\documentclass[pdflatex,sn-mathphys-num]{sn-jnl}

\usepackage{graphicx}%
\usepackage{multirow}%
\usepackage{amsmath,amssymb,amsfonts}%
\usepackage{amsthm}%
\usepackage{mathrsfs}%
\usepackage[title]{appendix}%
\usepackage{xcolor}%
\usepackage{textcomp}%
\usepackage{manyfoot}%
\usepackage{booktabs}%
\usepackage{algorithm}%
\usepackage{algorithmicx}%
\usepackage{algpseudocode}%
\usepackage{listings}%
\theoremstyle{thmstyleone}%
\theoremstyle{thmstyletwo}%

\theoremstyle{thmstylethree}%

\usepackage{physics}
\usepackage{centernot}
\usepackage{multirow}
\usepackage{moresize}
\usepackage{comment}
\usepackage{wasysym}
\usepackage{float}

\usepackage{makecell}

\usepackage{pdflscape}

\usepackage{mathtools}

\AtBeginDocument{\RenewCommandCopy\qty\SI} 

\newcommand{\appropto}{\mathrel{\vcenter{
  \offinterlineskip\halign{\hfil$##$\cr
    \propto\cr\noalign{\kern2pt}\sim\cr\noalign{\kern-2pt}}}}}

\usepackage{geometry}
\begin{document}

\title[The Group-IV-vacancy Color Center in Diamond]{The Group-IV-vacancy Color Center in Diamond}


\author[]{\fnm{Fenglei} \sur{Gu}}\email{f.gu@tudelft.nl}

\affil[]{\orgdiv{QuTech}, \orgname{Delft University of Technology}, \orgaddress{\street{Lorentzweg 1}, \city{Delft}, \postcode{2628 CJ}, \country{The Netherlands}}}


\abstract{Group-IV vacancy (G4V, or XV, where X = Si, Ge, Sn, Pb) color centers constitute a novel and promising class of defects in diamonds. This chapter reviews and refines the theoretical models for the XV systems, encompassing the intrinsic interactions, including spin-orbit coupling and electron-phonon interactions, and the external interactions involving strain, electric, light, and magnetic fields. Based on the refined model, we predict their properties, explain the experimental data, and suggest follow-up experiments. This article established a solid foundation for controlling the XV system, thus paving the way for quantum information processing.}

\keywords{Group-IV-vacancy Color Center, electron-phonon interaction, diamond defects, quantum system}



\maketitle

\section{Introduction}

Group-IV vacancy (XV, where X = Si, Ge, Sn, Pb) color centers in diamond~\cite{GS_SiV_sukachev2017silicon, GS_SiV_hepp2014electronic, GS_GeV_ekimov2015germanium, GS_GeV_siyushev2017optical, ZPL_GeV_palyanov2015,  GS_SnV_iwasaki2017tin, ZPL_SnV_gorlitz2020, Stark_de2021investigation, linear_shift_aghaeimeibodi2021electrical, GS_PbV_trusheim2019lead} have recently emerged as promising platforms for quantum information processing. Their appeal stems from their high zero-phonon line (ZPL) proportion—the spectral feature corresponding to phonon-free electronic transitions—as well as their robustness against charge noise and resistance to thermal excitation. These properties position XV centers as highly attractive candidates for applications in quantum technologies. The motivation for the work presented in this chapter is fourfold.

Firstly, despite the enormous potential of XV centers, these systems consist of complex interactions between the orbital and spin movements of a hole (the absence of an electron in a fully occupied system), phonons (vibrations of atomic nuclei), and their coupling with strain, as well as external electric and magnetic fields. These intricate interactions make understanding them difficult. Several models have been developed and used for numerical calculations and experimental data analysis~\cite{HeppThesis, trusheim2020transform, thiering2018ab, Meesala2018Strain, strain_guo2023microwave}; however, as detailed in Sec.~\ref{sec:comparison}, mutual discrepancies and missing interaction terms occasionally appear in these models. This emphasizes the necessity for a structured and comprehensible framework to elucidate the definitions of established concepts and to identify critical prerequisite conditions for using a specific approximate model in the characterization of XV centers. Starting from the foundational level, this article tackles these challenges by offering precise and comprehensive explanations of essential concepts and methodologies. In particular, this article elucidates the interactions arising from the fundamental force between a solitary electron and the atomic level, offering a detailed physical depiction of the XV systems. Additionally, it avoids employing sophisticated mathematical techniques like group theory, but ultimately arrives at the same conclusions regarding the configurations of electronic orbital eigenstates. This approach is intended to promote an accurate comprehension of XV centers, thereby minimizing errors in the application of concepts and models.

Secondly, this article addresses two key issues found in the previous modeling of the XV systems. To begin with, it contends that phonon modes should not be classified without considering the electronic orbital eigenstate configurations, as was done in Hepp's work~\cite{HeppThesis}. Instead, they should be classified with these configurations taken into account, as explained in Sec.~\ref{sec:ele-vib} and Sec.~\ref{sec:comparison}. This approach results in a simpler expression for the electron-phonon interaction Hamiltonian and validates the assumption of two-mode coupling employed in Thiering's calculations~\cite{thiering2018ab}. Moreover, this article highlights the discrepancy between the model formulated by Hepp~\cite{HeppThesis} and the one implemented in Thiering's calculations~\cite{thiering2018ab} regarding the Jahn-Teller effect (Sec.~\ref{sec:comparison}). Through meticulous analysis, this article endorses the model utilized in Thiering's calculations and introduces a refined model that builds upon it, achieving enhanced precision for the XV systems with higher spin-orbit coupling strengths (Sec.~\ref{sec:refined_def}).

Thirdly, despite abundant experimental measurements, those data lack sufficient theoretical explanation. This article explores possible theoretical frameworks behind these observations and obtains four main achievements. Firstly, given the fact that the electron-phonon interaction and strain interaction arise from the same mechanism, the interaction between the hole and the nuclear displacements, we explored the connection between the two interactions as presented in the quantitative relationship between their interaction strengths(Sec.~\ref{sec:strain_Hami}).  Besides, we explained the magnitude of the Stark shift (Sec.~\ref{fig:stark_shift}) and the intensity branching ratio in the light emission/absorption spectra (Sec.~\ref{sec:light_emission}). Finally, we anticipate that an unstrained XV system will preserve its degeneracy when subjected to a magnetic field perpendicular to the principal axis, a phenomenon that Kramer's theorem does not account for (Sec.~\ref{sec:degeneracy}). Our research indicates that this degeneracy originates from a novel symmetry associated with a joint reflection operation acting on the phonon, hole orbital, and hole spin movements (Sec.~\ref{sec:joint-reflection}).

Finally, with the relationship between electron-phonon interaction strength and strain susceptibilities now determined, this article identifies a discrepancy between the XV system parameters derived by Thiering \textit{et al.} and the strain susceptibilities reported by Meesala \textit{et al.}~\cite{Meesala2018Strain} as well as those calculated by Guo \textit{et al.}~\cite{strain_guo2023microwave}. These differences are observed in both signs and magnitude. This article offers possible explanations for these contradictions, attributing them to the intrinsic strain of the XV sample. Additionally, we introduce a technique for measuring the intrinsic strain in the XV sample (Sec.~\ref{sec:predictions}). This approach allows for mitigating any potential effects from the intrinsic strain.

\section{Intrinsic interactions}
\label{sec:intrinsic_H}

\subsection{XV Center Formation}
\begin{figure}
    \centering
    \includegraphics[width=0.7\linewidth]{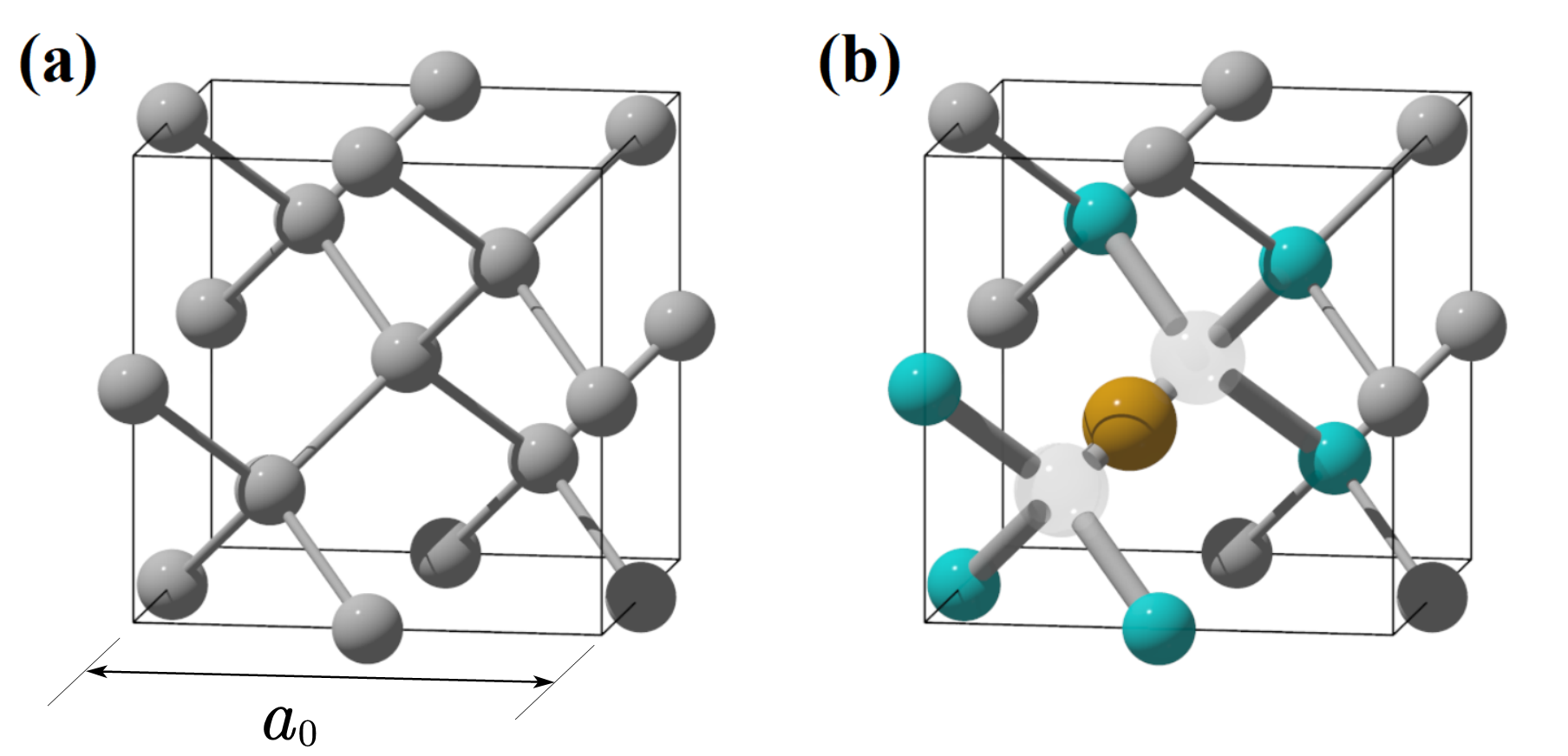}
    \caption{Atomic structure of the XV center in diamond. (a) The unit cell of the diamond crystal, where $a_0 = 3.567 \, \text{\r{A}}$ is the diamond lattice constant. (b) The atomic structure of the XV center. The two transparent spheres indicate the positions of the missing carbon atoms, while the solid brown sphere represents the X atom. The six solid blue spheres surrounding the X atom represent the six nearest carbon atoms. Adapted from Ref.~\cite{HeppThesis}.}
    \label{fig:XV_structure}
\end{figure}

A Group-IV vacancy (XV, where X=Si, Ge, Sn, Pb) center in diamond is a lattice defect formed by two stages. Firstly, two neighboring carbon atoms are removed and replaced with a single X (X=Si, Ge, Sn, Pb) atom positioned centrally between the two vacancies, as illustrated in Fig.~\ref{fig:XV_structure}. Secondly, an electron from a nearby donor is captured by the X atom and the six surrounding carbon atoms~\cite{nearby_donor_PhysRevB.84.245208}.

The X atom, together with six surrounding carbon atoms and the captured electron, constitutes an XV system—the central subject of this chapter. When two adjacent carbon atoms are removed from the diamond lattice, it leaves the six remaining neighboring carbon atoms each with an unsaturated covalent bond, commonly referred to as a dangling bond. Each dangling bond hosts an unpaired electron occupying an $sp^3$ hybrid orbital.

The introduction of the X atom, which has four valence electrons, alters this configuration. In the presence of two additional external electrons, the X atom could form covalent bonds with all six neighboring carbon atoms through its $sp^3d^2$ hybrid orbitals, fully saturating the bonding structure. However, in the actual XV center system, only a single external electron is added instead of two. As a result, one of the six X-carbon (X-C) covalent bonds remains incomplete, leaving behind a hole that exists in a quantum superposition over the six bonds. This delocalized hole is fundamental to the quantum properties of the XV center, as it enables the emergence of well-defined and controllable quantum states, which are central to its quantum applications. In the following, we will determine the eigenstates governing the hole's orbital motion and derive the corresponding interaction Hamiltonians that describe its coupling to both nuclear vibrations (phonons) and the hole’s spin.

\subsection{Electronic Orbital Eigenstates} 
\label{sec:electronic_orbital}

In order to determine the orbital eigenstates of the hole, we must first establish the orbital eigenstates of the 11 valence band electrons in the XV center system. To achieve this, we begin by determining the orbital eigenstates of a single electron in the Coulomb potential formed by the X atomic core and its six surrounding carbon atomic cores. Here, an atomic core refers to an atom with its valence-shell electrons removed, leaving only the nucleus and the fully occupied inner electron shells.

\begin{figure}
    \centering
    \includegraphics[width=0.8\linewidth]{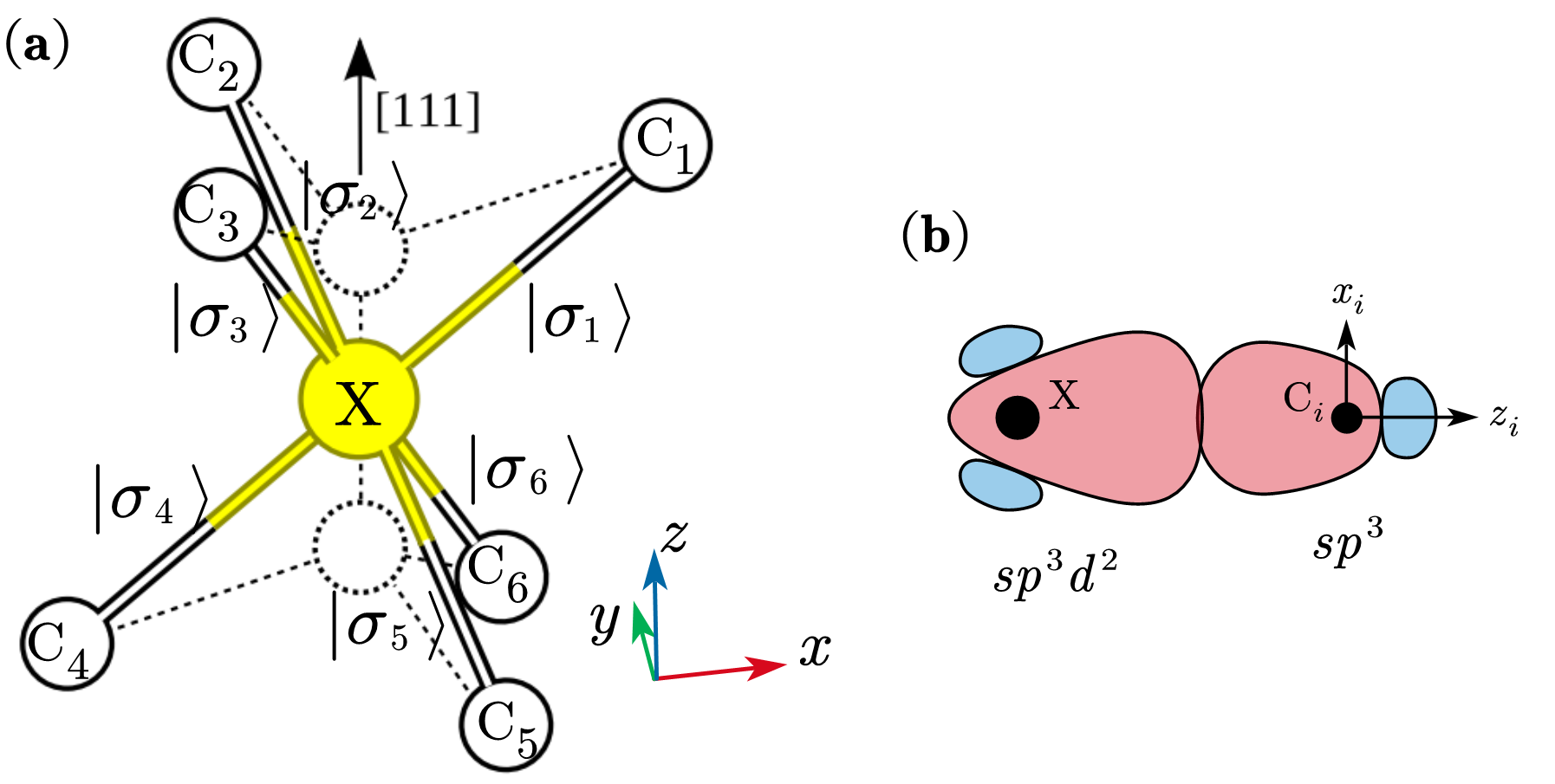}
    \caption{The X–C bonds as the basis states (\(\{\ket{\sigma_i}\}\)) for representing the electronic orbital states. (a) The arrangement of the X–C bonds and the Cartesian coordinate system for the XV center system. Adapted from Ref.~\cite{thiering2018ab}. (b) Sketch of the speculated shape of the electron orbital of an X–C bond. It is formed by combining an \(sp^3d^2\) hybridized orbital from the X atom with an \(sp^3\) hybridized orbital from the carbon atom, resembling a \(\sigma\) covalent bond. The black dots represent the X and carbon nuclei, while the red (blue) color in the electron orbital indicates the positive (negative) phase of the orbital components. A local Cartesian coordinate system \(\{x_i, y_i, z_i\}\) for the \(i\)-th carbon atomic core \(C_i\) is also shown. The \(z_i\)-axis is aligned along the X–C bond direction.}
    \label{fig:ele_bases}
\end{figure}

Fig.~\ref{fig:ele_bases}(a) shows the six covalent bonds in the XV center system, each bond resembles a \(\sigma\)-bond~\cite{sigma_bond_clayden2012organic} consisting of one $sp^3d^2$ hybrid orbital from the X atom and one $sp^3$ hybrid orbital from the nearby carbon atom as shown in Fig.~\ref{fig:ele_bases}(b). These six covalent bonds form a simplified set of basis states, nearly orthogonal and complete to present the eigenstate of the electron. We define the six covalent bonds as quantum states \(\left\{\ket{\sigma_i}\right\}\) (\(i = 1, 2, \dots, 6\)), ordered according to a Cartesian coordinate system as illustrated in Fig.~\ref{fig:ele_bases}(a). This coordinate system is established in the reference frame of the six covalent bonds, which are assumed to remain fixed relative to one another. Within this frame, the X atomic core and the six surrounding carbon atomic cores are free to vibrate around their equilibrium positions, which themselves are considered stationary.

The origin of the coordinate system is set at the equilibrium position of the X atomic core, effectively defining the center of the XV system. The \(z\)-axis is oriented along the line connecting the two vacant carbon sites, while the \(x\)-axis is chosen such that the \(x\)-\(z\) plane contains two of the X–C bonds—one located on the \(+z\) side and the other on the \(-z\) side. The \(y\)-axis is then uniquely determined to ensure a right-handed coordinate system.

Based on this coordinate system, we establish the spatial arrangement of the six basis states. The states \(\ket{\sigma_1}\), \(\ket{\sigma_2}\), and \(\ket{\sigma_3}\) reside on the \(+z\) side, with \(\ket{\sigma_1}\) lying in the \(x\)-\(z\) plane, while \(\ket{\sigma_2}\) and \(\ket{\sigma_3}\) are positioned counterclockwise when viewed from the \(-z\) direction. Similarly, the states \(\ket{\sigma_4}\), \(\ket{\sigma_5}\), and \(\ket{\sigma_6}\) are symmetrically positioned on the \(-z\) side, directly opposite to \(\ket{\sigma_1}\), \(\ket{\sigma_2}\), and \(\ket{\sigma_3}\), respectively, as depicted in Fig.~\ref{fig:ele_bases}(a).

Although group theory provides a rigorous framework for determining the eigenstates of an electron~\cite{HeppThesis}, the standard procedure is extensive and difficult to master quickly. Instead, we adopt a simplified approach based on a single foundational theorem: the symmetry of the eigenstates reflects the symmetry of the potential in a quantum system (detail shown below). By decomposing the XV center into two simpler subsystems with well-understood symmetries, this method enables a more intuitive and accessible derivation of the electronic eigenstates within the Coulomb potential of the XV center, referred to here as the XV potential.

The XV potential exhibits inversion symmetry, indicating that the $-z$ side of the potential mirrors the $+z$ side under an inversion operation. Consequently, the eigenstates of the XV system can be categorized into two groups, \(\{\ket{\psi_{+,s}}\}\) and \(\{\ket{\psi_{-,s}}\}\), with the subscript \(s\) denoting some indices. They take the following forms:
\begin{equation}
\begin{split}
    \ket{\psi_{+,s}} = & \frac{1}{\sqrt{2}}\left(\ket{\psi^{\rm upp}_{s}} + \ket{\psi^{\rm low}_{s}}\right)\;,\\
    \ket{\psi_{-,s}} = & \frac{1}{\sqrt{2}}\left(\ket{\psi^{\rm upp}_{s}} - \ket{\psi^{\rm low}_{s}}\right)\;.
\end{split}
\label{eq:Psi+-}
\end{equation}
Here, \(\left\{\ket{\psi^{\rm upp}_{s}}\right\}\) and \(\left\{\ket{\psi^{\rm low}_{s}}\right\}\) represent wavefunctions in the subspace spanned by \(\left\{\ket{\sigma_1}, \ket{\sigma_2}, \ket{\sigma_3}\right\}\) and \(\left\{\ket{\sigma_4}, \ket{\sigma_5}, \ket{\sigma_6}\right\}\), respectively. The states \(\ket{\psi^{\rm upp}_{s}}\) and \(\ket{\psi^{\rm low}_{s}}\) share the same coefficients according to the \(\ket{\sigma_i}\)–\(\ket{\sigma_{i+3}}\) correspondence (\(i=1, 2, 3\)). Consequently, determining the eigenstates of the XV system is reduced to identifying the three eigenstates of either subspace.

Considering the subspace spanned by $\left\{\ket{\sigma_1}, \ket{\sigma_2}, \ket{\sigma_3}\right\}$, the \(\frac{2\pi}{3}\) rotational symmetry of the $+z$ side of the XV system mandates that eigenstates must be invariant under cyclic permutation of the coefficients within the basis, except for a global phase factor. Only three distinct wavefunctions fulfill this criterion. They are:
\begin{equation}
    \begin{split}
        \ket{\psi_0^{\rm upp}} &= \frac{1}{\sqrt{3}}\left(1, 1, 1\right)^\text{T},\\
        \ket{\psi_{+1}^{\rm upp}} &= \frac{1}{\sqrt{3}}\left(1, \text{e}^{\text{i}2\pi/3}, \text{e}^{\text{i}4\pi/3}\right)^\text{T},\\
        \ket{\psi_{-1}^{\rm upp}} &= \frac{1}{\sqrt{3}}\left(1, 
        \text{e}^{-\text{i}2\pi/3}, 
        \text{e}^{-\text{i}4\pi/3}\right)^\text{T}.
    \end{split}
    \label{eq:eigen_rot}
\end{equation}

Notice that $\ket{\psi_{+1}^{\rm upp}}$ and $\ket{\psi_{-1}^{\rm upp}}$ correspond to clockwise and counterclockwise permutations, respectively, both acquiring the same global phase factor $\text{e}^{\text{i}2\pi/3}$. Due to the symmetry between these two rotational directions, these states possess the same energy eigenvalue and are thus degenerate.

As dictated by rotational symmetry, these states are closely related to the eigenstates of the $z$-component of the orbital angular momentum operator, \(L^{\rm o}_z\), making them particularly convenient for analyzing spin-orbit coupling interactions, as we will explore in Sec.~\ref{sec:SOC}. However, since these states exhibit uniform populations across the three basis orbitals, they interact identically with nuclear vibrational modes, regardless of the spatial direction of motion. For problems involving such interactions, it is advantageous to work with eigenstates that exhibit spatially polarized populations.

Fortunately, since \( \ket{\psi_{+1}^{\rm upp}} \) and \( \ket{\psi_{-1}^{\rm upp}} \) are degenerate in energy, any linear combination of these states is also an eigenstate of the Hamiltonian describing the upper subsystem of the XV center. This allows us to construct an alternative set of eigenstates with spatially polarized charge distributions.

\begin{figure}
    \centering
    \includegraphics[width=0.8\linewidth]{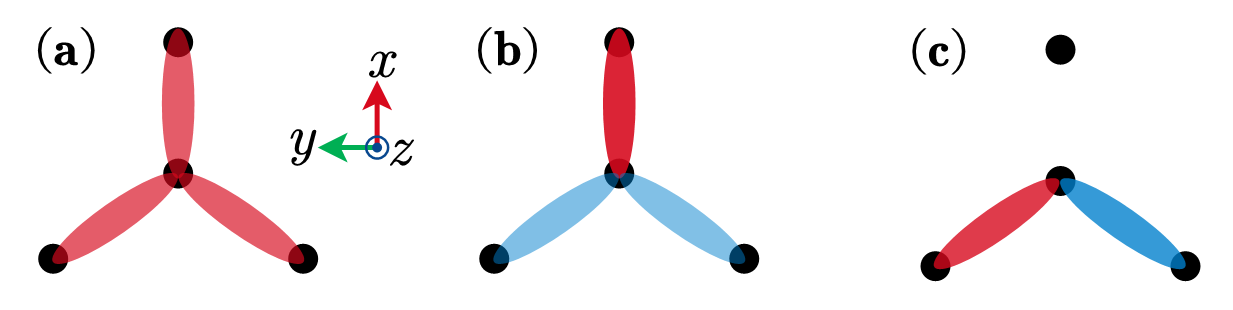}
    \caption{Illustrations of the eigenstates in the \( +z \) half of the XV center: (a) \( \ket{\psi_{0}^{\rm upp}} \), (b) \( \ket{\psi_{x}^{\rm upp}} \), and (c) \( \ket{\psi_{y}^{\rm upp}} \). Red (blue) shading represents the \( + \) (\( - \)) phase of the corresponding part of the eigenstate, and the color intensity reflects the wavefunction population. Black dots denote the X and carbon atomic cores.}
    \label{fig:ele_structure}
\end{figure}

The X–C bond orbitals are approximately linear in the radial directions (see Fig.~\ref{fig:ele_bases}), and the phase factors \( \{\phi_\pm^{\rm upp}\} = \left\{ 0, \pm \frac{2\pi}{3}, \pm \frac{4\pi}{3} \right\} \) in \( \ket{\psi_\pm^{\rm upp}} \) are spatially dependent, with \( \phi^{\rm upp} = \arctan(y/x) \), where \( x \) and \( y \) are the Cartesian coordinates in the XV system. This spatial dependence motivates the construction of the following linear combinations:
\begin{equation}
    \begin{split}
        \frac{1}{2} \left( \mathrm{e}^{\mathrm{i}\phi_+^{\rm upp}} + \mathrm{e}^{\mathrm{i}\phi_-^{\rm upp}} \right) &= x\;, \\
        \frac{-\mathrm{i}}{2} \left( \mathrm{e}^{\mathrm{i}\phi_+^{\rm upp}} - \mathrm{e}^{\mathrm{i}\phi_-^{\rm upp}} \right) &= y\;.
    \end{split}
    \label{eq:x_and_y}
\end{equation}
Using these relations, we define the spatially polarized eigenstates:
\begin{equation}
    \begin{split}
        \ket{\psi_{x}^{\rm upp}} &= \frac{1}{\sqrt{2}}\left( \ket{\psi_{+1}^{\rm upp}} + \ket{\psi_{-1}^{\rm upp}} \right) = \frac{1}{\sqrt{6}}\left( 2, -1, -1 \right)^\text{T}\;, \\
        \ket{\psi_{y}^{\rm upp}} &= \frac{-\mathrm{i}}{\sqrt{2}}\left( \ket{\psi_{+1}^{\rm upp}} - \ket{\psi_{-1}^{\rm upp}} \right) = \frac{1}{\sqrt{2}}\left( 0, 1, -1 \right)^\text{T}\;.
    \end{split}
    \label{eq:eigen_xy}
\end{equation}
These states can be interpreted as the state \( \ket{\psi_0^{\rm upp}} \) modulated by the spatial functions \( x \) and \( y \), respectively, as illustrated in Fig.~\ref{fig:ele_structure}. This polarization renders them particularly useful for describing interactions with nuclear vibrations.

Substituting the states~(\ref{eq:eigen_rot}) or their spatially polarized alternatives~(\ref{eq:eigen_xy}) into Eq.~(\ref{eq:Psi+-}), we obtain the six eigenstates of the XV center’s bare Hamiltonian. The first two take the form:
\begin{equation}
    \begin{split}
        \ket{\psi_{+,0}} &= \frac{1}{\sqrt{6}}(1,\,1,\,1,\,1,\,1,\,1)^\text{T},\\
        \ket{\psi_{-,0}} &= \frac{1}{\sqrt{6}}(1,\,1,\,1,\,-1,\,-1,\,-1)^\text{T}.
    \end{split}
    \label{eq:ele_eigen_0}
\end{equation}
The remaining four are given by:
\begin{equation}
    \begin{split}
        \ket{\psi_{+,+1}} &= \frac{1}{\sqrt{6}}(1,\; 
        \text{e}^{\text{i}2\pi/3},\; 
        \text{e}^{\text{i}4\pi/3},\;1,\; 
        \text{e}^{\text{i}2\pi/3},\; 
        \text{e}^{\text{i}4\pi/3})^\text{T},\\
        \ket{\psi_{+,-1}} &= \frac{1}{\sqrt{6}}(1,\; 
        \text{e}^{-\text{i}2\pi/3},\; 
        \text{e}^{-\text{i}4\pi/3},\;1,\; 
        \text{e}^{-\text{i}2\pi/3},\; 
        \text{e}^{-\text{i}4\pi/3})^\text{T},\\
        \ket{\psi_{-,+1}} &= \frac{1}{\sqrt{6}}(1,\; 
        \text{e}^{\text{i}2\pi/3},\; 
        \text{e}^{\text{i}4\pi/3},\;-1,\; 
        -\text{e}^{\text{i}2\pi/3},\; 
        -\text{e}^{\text{i}4\pi/3})^\text{T},\\
        \ket{\psi_{-,-1}} &= \frac{1}{\sqrt{6}}(1,\; 
        \text{e}^{-\text{i}2\pi/3},\; 
        \text{e}^{-\text{i}4\pi/3},\;-1,\; 
        -\text{e}^{-\text{i}2\pi/3},\; 
        -\text{e}^{-\text{i}4\pi/3})^\text{T}.
    \end{split}
    \label{eq:ele_eigen_+1-1}
\end{equation}
Alternatively, using spatially polarized bases, the four states can be rewritten as:
\begin{equation}
    \begin{split}
        \ket{\psi_{+,x}} &= \frac{1}{\sqrt{12}}(2, -1, -1, 2, -1, -1)^\text{T},\\
        \ket{\psi_{+,y}} &= \frac{1}{2}(0, 1, -1, 0, 1, -1)^\text{T},\\
        \ket{\psi_{-,x}} &= \frac{1}{\sqrt{12}}(2, -1, -1, -2, 1, 1)^\text{T},\\
        \ket{\psi_{-,y}} &= \frac{1}{2}(0, 1, -1, 0, -1, 1)^\text{T}.
    \end{split}
    \label{eq:ele_eigen_xy}
\end{equation}
For problems involving spin-orbit interactions, we employ the eigenstates in Eq.~(\ref{eq:ele_eigen_+1-1}), as they are expressed in terms of eigenstates of the orbital angular momentum. For electron-phonon interactions, we use the spatially polarized states in Eq.~(\ref{eq:ele_eigen_xy}), which facilitate analysis of interactions with nuclear vibrations.

\subsection{Eigenenergy Ordering}

Evidently, \(\ket{\psi_{+,x}}\) and \(\ket{\psi_{+,y}}\) are degenerate in energy, as are \(\ket{\psi_{-,x}}\) and \(\ket{\psi_{-,y}}\). Consequently, the XV bare Hamiltonian has only four distinct eigenenergies, denoted as \(E_{+,0}\), \(E_{-,0}\), \(E_{+,1}\), and \(E_{-,1}\), corresponding to four energetic manifolds: \(\left\{ \ket{\psi_{+,0}} \right\}\), \(\left\{ \ket{\psi_{-,0}} \right\}\), \(\left\{ \ket{\psi_{+,x}}, \ket{\psi_{+,y}} \right\}\), and \(\left\{ \ket{\psi_{-,x}}, \ket{\psi_{-,y}} \right\}\), respectively.

Notice that there are 11 outer-shell electrons in an XV system. They tend to occupy the six eigenstates in the order of their eigenenergies from low to high. Therefore, it is crucial to determine the ordering of these four eigenenergies. 

In the case of a single electron, we can employ a heuristic method to determine the order of the four eigenenergies: eigenstates with more nodes tend to have higher eigenenergies. Here, a node refers to a region where the wavefunction—and hence the probability density—vanishes, such as a plane in three dimensions, a line in two, or a point in one. This rule originates from the uncertainty principle: the presence of nodes confines the spatial extent of the wavefunction, reducing position uncertainty and increasing momentum uncertainty, which typically raises the energy~\cite{nodes_and_energy}. 

\begin{figure}
    \centering
    \includegraphics[width=0.9\linewidth]{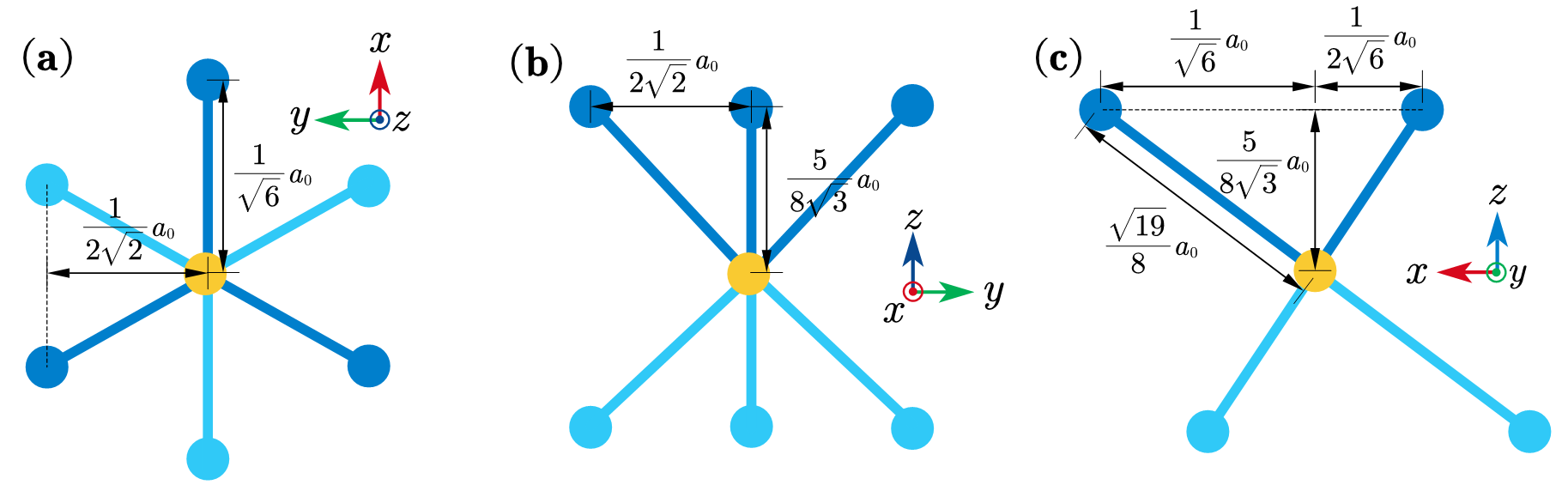}
    \caption{Dimensions of the XV system. The three carbon atoms and their corresponding dangling bonds on the $+z$ ($-z$) sides are depicted in dark blue (light blue). The central X atom is shown in yellow. Here, \( a_0 \) denotes the diamond lattice constant, same as shown in Fig.~\ref{fig:XV_structure}.}
\label{fig:XV_size}
\end{figure}

\begin{figure}
    \centering
    \includegraphics[width=0.75\linewidth]{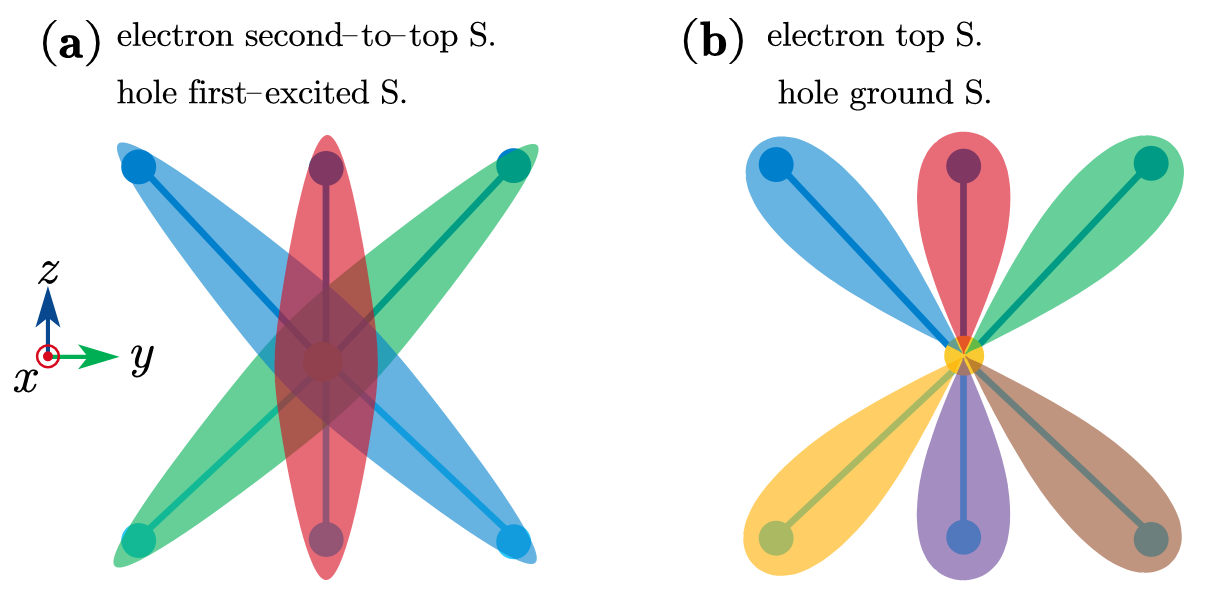}
    \caption{Shape of the orbitals of eigenstates (a) \(\ket{\psi_{+,+1}}\) and (b) \(\ket{\psi_{-,+1}}\). The state \(\ket{\psi_{+,+1}}\) has no node on the \(z=0\) plane, while the state \(\ket{\psi_{-,+1}}\) does. The state \(\ket{\psi_{+,+1}}\) corresponds to the second-to-top energy state of an electron or the first-excited state of the hole. The state \(\ket{\psi_{-,+1}}\) corresponds to the top energy state of an electron or the ground state of the hole. Different colors represent different phases of the orbital components.}
    \label{fig:WF_node}
\end{figure}

According to this principle, \(\ket{\psi_{+,0}}\), having no nodes, is the ground state. The states \(\ket{\psi_{-,0}}\), \(\ket{\psi_{+,x}}\), and \(\ket{\psi_{+,y}}\) each contain a single nodal plane: \(z=0\), \(x=0\), and \(y=0\), respectively [see Eq.~(\ref{eq:Psi+-}), Fig.~\ref{fig:WF_node}, and Eq.~(\ref{eq:x_and_y})]. However, as shown in Fig.~\ref{fig:XV_size}, since the XV center’s spatial extent is smaller along \(x\) and \(y\) than along \(z\), reducing position uncertainty, \(\ket{\psi_{-,0}}\) with node plane \(z=0\) has lower energy corresponding to the first excited state, while \(\ket{\psi_{+,x}}\) and \(\ket{\psi_{+,y}}\) have node planes \(x=0\) and \(y=0\), respectively, have higher energies forming the degenerate second excited states. Finally, \(\ket{\psi_{-,x}}\) and \(\ket{\psi_{-,y}}\), each possessing two nodal planes (\(z=0\) and either \(x=0\) or \(y=0\)), correspond to the highest-energy degenerate states. This argument is initially introduced in the work of Hepp~\cite{HeppThesis}.

Nonetheless, this rule may not apply to systems with multiple electrons as Coulomb interaction and/or exchange interaction take effect. The smaller the spatial extent of confinement, the larger the interaction energy. In the large interaction energy case, when occupied by two electrons, the orbitals with more nodes may have lower energies as allowing for larger spacing between the two electrons. Therefore, a more accurate determination of the order of the eigenenergies relies on the DFT (Density Functional theory) simulation. Coincidentally, the DFT result~\cite{thiering2018ab, ordering_PhysRevB.88.235205} aligns with the ordering of the single-electron case. 

The presence/absence of the central ($z=0$) node affects the partition factor as introduced in Sec.~\ref{sec:ele-vib} and the screening factor as introduced in Sec.~\ref{sec:SOC}. Therefore, by calculating these factors based on the DFT calculations, one can infer the node configurations. We present these calculations based on Thiering's DFT simulation~\cite{thiering2018ab} in in Sec.~\ref{sec:comparison}.

\subsection{Hole Orbital Eigenstate}
\label{sec:hole_eigen}

\begin{figure}
    \centering
    \includegraphics[width=0.8\linewidth]{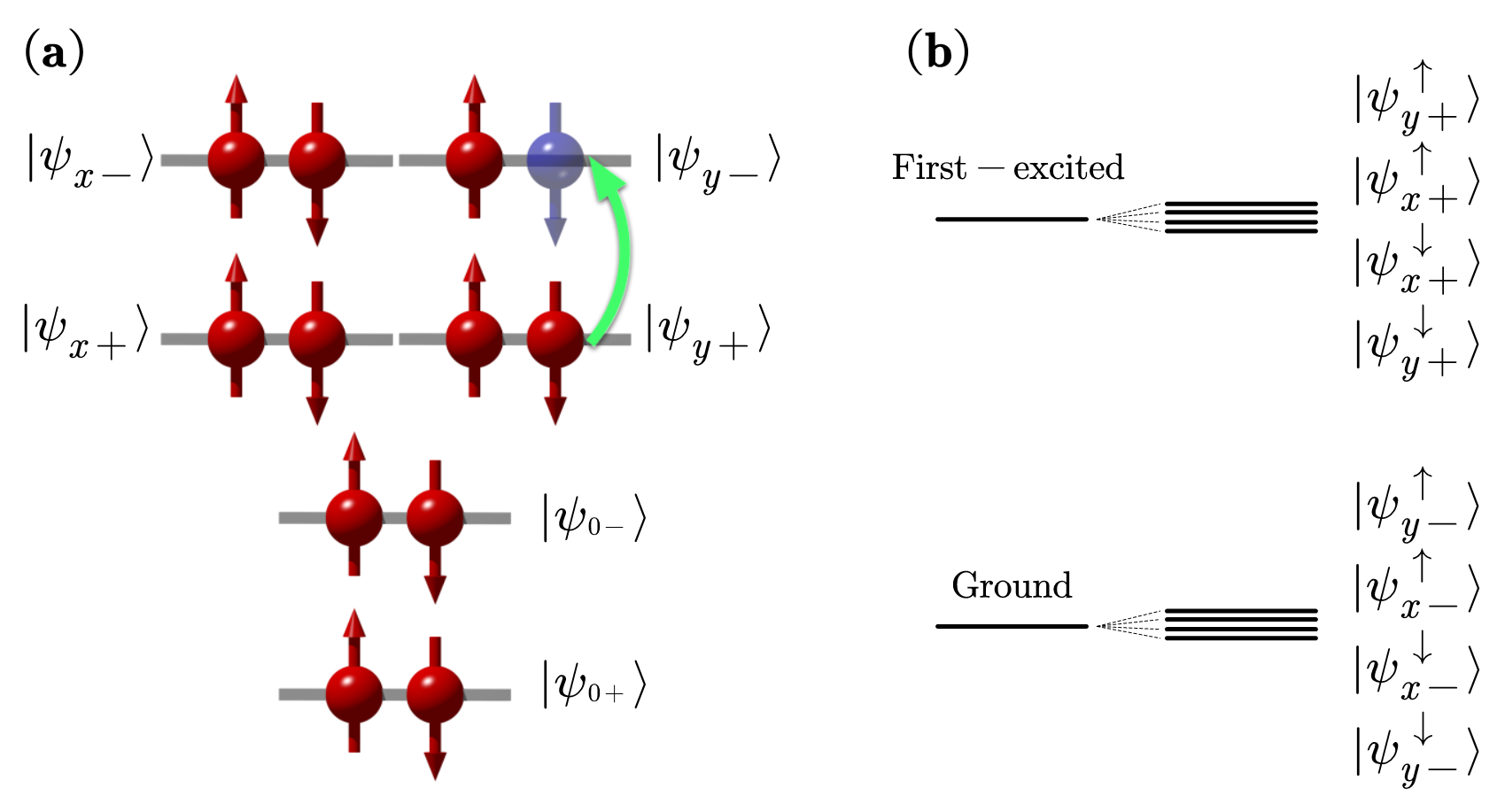}
    \caption{Hole orbital states of the XV system. (a) Occupation of the six electronic states by 11 free electrons (solid red spheres). The transparent blue sphere represents the hole, while up and down arrows indicate electron or hole spin. The green arc denotes a possible excitation involving an electron transition. The written states denote the hole orbitals of the corresponding energy level. Adapted from Ref.~\cite{HeppThesis}. (b) Degeneracy of the four ground states and four first-excited states of the hole, including spin degrees of freedom.}
    \label{fig:ele_arrangement}
\end{figure}

With the eigenstates and energy ordering of the XV system established, we now determine the orbital states of the 11 free electrons. Due to the low electron density, Coulomb interactions have a negligible effect on the electronic orbital eigenstates. Following the Pauli exclusion principle, electrons sequentially occupy the six eigenstates from lowest to highest energy, with each level accommodating up to two electrons of opposite spin. Consequently, the lowest-energy configuration leaves a single hole in the \(\{\ket{\psi_{-,x}}, \ket{\psi_{-,y}}\}\) manifold. The first excited configuration arises when an electron is promoted from the \(\{\ket{\psi_{+,x}}, \ket{\psi_{+,y}}\}\) manifold to the \(\{\ket{\psi_{-,x}}, \ket{\psi_{-,y}}\}\) manifold, as illustrated in Fig.~\ref{fig:ele_arrangement}(a).

From the hole’s perspective, the \(\{\ket{\psi_{-,x}}, \ket{\psi_{-,y}}\}\) manifold constitutes the ground states, while \(\{\ket{\psi_{+,x}}, \ket{\psi_{+,y}}\}\) forms the first excited manifold. Higher-energy manifolds can be constructed similarly but lie beyond the scope of this thesis.

Including spin, each manifold is four-fold degenerate in the absence of a magnetic field, as shown in Fig.~\ref{fig:ele_arrangement}(b). The energy gap between the ground and first-excited manifolds is significantly larger than that of other interactions, such as electron-phonon, static electric and magnetic field, and strain interactions. Thus, without influences (e.g., light) of high enough energy, inter-manifold couplings are suppressed, allowing us to treat these manifolds as independent subspaces when analyzing additional interactions.

For convenience, when specifying the manifold is unnecessary, we may use \( \ket{\psi_{x}} \) and \( \ket{\psi_{y}} \), or equivalently \( \ket{\psi_{+1}} \) and \( \ket{\psi_{-1}} \), to denote states within either manifold. These eigenbases are related by
\begin{equation}
    \begin{split}
        \ket{\psi_{x}} &= \frac{1}{\sqrt{2}}\left(\ket{\psi_{+1}} + \ket{\psi_{-1}}\right), \\
        \ket{\psi_{y}} &= \frac{-\text{i}}{\sqrt{2}}\left(\ket{\psi_{+1}} - \ket{\psi_{-1}}\right).
    \end{split}
    \label{eq:basis_transition}
\end{equation}
Thus, the Pauli-\(Z\) operator in the \(\{ \ket{\psi_{+1}}, \ket{\psi_{-1}}\} \) basis corresponds to the Pauli-\(Y\) operator in the \(\{\ket{\psi_{x}}, \ket{\psi_{y}} \}\) basis. Since these two sets of bases are convertible, in the following sections, when introducing various interaction Hamiltonians, we will adopt the convention of expressing all the interaction Hamiltonians in the \(\{\ket{\psi_{x}}, \ket{\psi_{y}} \}\) basis.

\subsection{Electron-phonon interaction}
\label{sec:ele-vib}

\subsubsection{Localized phonon}

As studied in solid-state physics, phonons, the quantized vibrations of a crystal lattice, come in various types. In the context of group-IV vacancy color centers in diamond, the most relevant phonons are localized phonons—vibrations of nuclei near the defect that are largely decoupled from the rest of the crystal lattice.

In contrast, bulk phonons, which propagate throughout the entire crystal, are macroscopically numerous but have limited influence on the electronic (or hole) orbitals localized at the defect. This is because bulk phonons have long wavelengths, especially at low temperatures, and primarily induce a global translation of the defect rather than energy shifts.

Localized phonons arise due to the unique properties of the defect site. The nuclei near the defect have different masses and bond stiffness compared to those in the bulk crystal, causing them to vibrate at distinct frequencies. This frequency mismatch limits their coupling to the bulk phonons, effectively confining the vibrations to the defect region. Localized or quasi-localized photons have been theoretically predicted and/or experimentally observed in GeV~\cite{local_GeV_boldyrev2022localized}, SiV~\cite{local_SiV_PhysRevB.98.035306}, and nitrogen-vacancy (NV) color centers in diamond~\cite{Local_NV_PhysRevB.84.035211}.

In this section, we assume that these vibrations are effectively localized, so we confine the involved nuclei to the scope of the central X nucleus and its surrounding six carbon nuclei.

\subsubsection{nuclear Vibrations}
\label{sec:nuclear_vib}

To begin with, we choose the rest frame of the equilibrium positions of the seven nuclei as the reference frame for analyzing the nuclear movement. Furthermore, for the free movement of the seven nuclei, we neglect the hole's influence and treat all bonds as saturated. Following the concept of localized photons, we model each nucleus in the XV system as moving independently within a potential created by the covalent bonds surrounding it.

\begin{figure}
    \centering
    \includegraphics[width=0.98\linewidth]{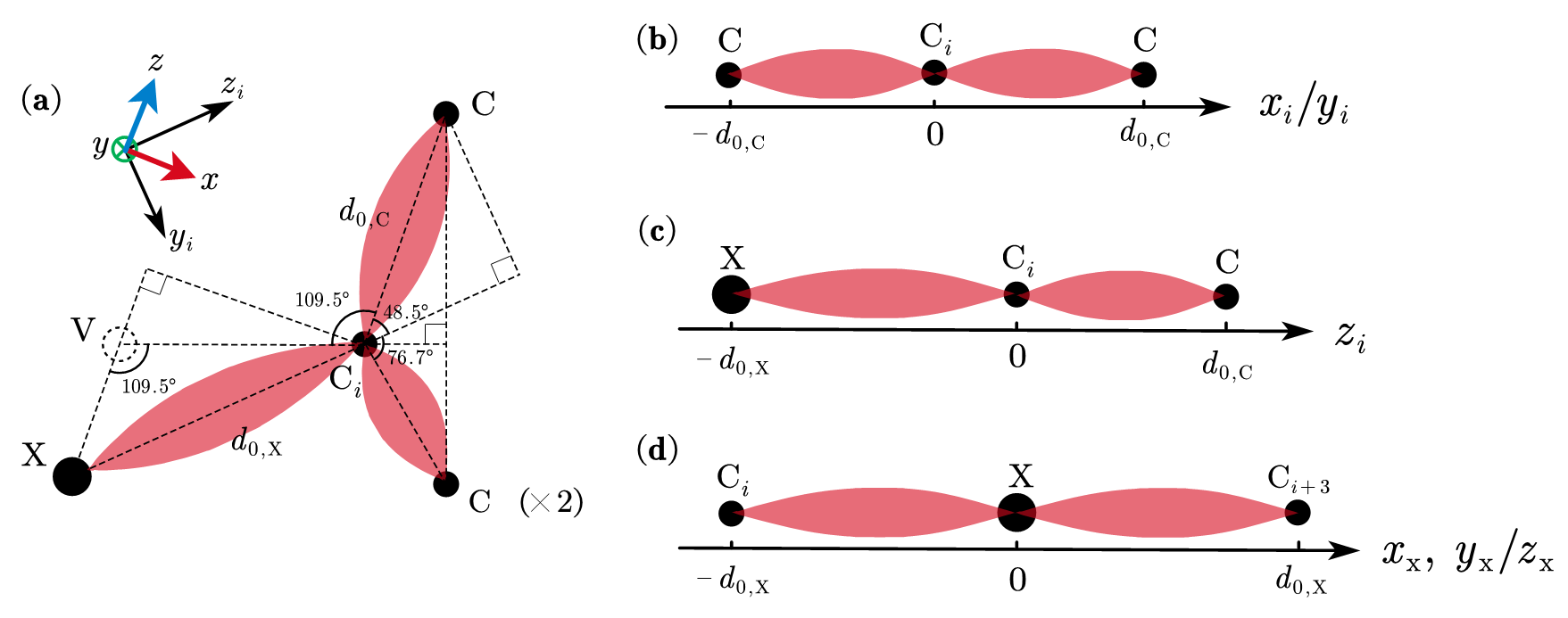}
    \caption{
        Surrounding bonds of the $i$-th carbon nucleus (\(C_i\)). 
        (a) View of the bonds attached to \(C_i\) along the \(x_i\) direction (perpendicular to the \(z\)-axis [see Fig.~\ref{fig:ele_bases}(a)] and the \(i\)-th X--C bond). 
        Note: Carbon nuclei and vacancies are plotted based on an ideal diamond crystal, not DFT simulations. 
        (b)--(d) Simplified potential models for \(C_i\) along: 
        (b) \(x_i\) or \(y_i\), (c) \(z_i\), and (d) X nucleus along all directions (\(x_X\), \(y_X\), \(z_X\)). 
        Black dots denote nuclei; red leaf-like shapes represent covalent bonds.
    }
    \label{fig:X-C-X}
\end{figure}

For the six carbon nuclei, each is attached to one X--C bond on one side and three C--C bonds on the other. Fig.~\ref{fig:XV_structure}(b) shows the 3D bonding structure, while Fig.~\ref{fig:X-C-X}(a) details the bond orientations. Due to the distinct X--C bond, the potential profile along the $z_i$ direction differs from those along the perpendicular $x_i$ and $y_i$ directions. 

For simplicity, we model the potential along $x_i$ or $y_i$ as a fictitious 1D structure where the nucleus is confined between two C--C bonds (one on each side), as shown in Fig.~\ref{fig:X-C-X}(b). Conversely, the potential along $z_i$ is modeled as a 1D structure with the carbon nucleus confined between an X--C bond on one side and a C--C bond on the other (Fig.~\ref{fig:X-C-X}(c)).

For the X nucleus, the six surrounding X--C bonds are approximately evenly spaced (Fig.~\ref{fig:ele_bases}(a)). We thus model its potential along all directions as a 1D structure with the X nucleus confined between two X--C bonds (Fig.~\ref{fig:X-C-X}(d)).

For each covalent bond, we model the bond potential energy $V(r_{\rm a})$ as a function of internuclear distance $r_{\rm a}$ using the Morse potential~\cite{morse1929diatomic, costa2013morse}:
\begin{equation}
    V(r_{\rm a}) = E_{\rm 0} \left[1 - \mathrm{e}^{-\alpha (r_{\rm a} - d_{\rm 0})}\right]^2 - E_{\rm 0},
\end{equation}
where $E_{\rm 0}$ is the bond dissociation energy, $d_{\rm 0}$ the equilibrium bond length, and $\alpha$ the potential-width parameter, representing the potential's depth, minimum position, and curvature, respectively.

\begin{table}[h]
    \centering
    \begin{tabular}{|c|c|c|c|}
        \hline
        Bond & $d_{\rm 0, C/X}$ (\AA) & $E_{\rm 0, C/X}$ (kJ/mol) & $\alpha_{\rm C/X}$ (\AA$^{-1}$) \\
        \hline
        C--C & 1.54  & 346 & 2.2 \\
        Si--C & 1.85 & 318 & 1.7 \\
        Ge--C & 1.95 & 238 & 1.2 \\
        Sn--C & 2.16 & 192 & 0.87 \\
        Pb--C & 2.30 & 130 & 0.55 \\
        \hline
    \end{tabular}
    \caption{Parameters for C--C and X--C bonds. The C--C bond length ($d_{\rm 0, C}$) and dissociation energy ($E_{\rm 0, C}$) are taken from Ref.~\cite{Bond_data}, as are the X--C bond parameters ($d_{\rm 0, X}$, $E_{\rm 0, X}$). The potential-width parameter $\alpha_{\rm C}$ is determined from both experiment~\cite{C_C_alpha_nguyen2021high} and DFT calculations~\cite{C-C_alpha_nickabadi2021derivation}, while $\alpha_{\rm X}$ values are estimated via Eq.~\ref{eq:stiff_guess}. Note: These values correspond to molecular systems and should be interpreted accordingly in crystalline contexts.}
    \label{tab:bond_data}
\end{table}

The potential-width parameter $\alpha_{\rm C}$ for the C--C bond has been well-characterized through experimental measurements~\cite{C_C_alpha_nguyen2021high} and theoretical calculations~\cite{C-C_alpha_nickabadi2021derivation}. For X--C bonds, we estimate $\alpha_{\rm X}$ using the empirical relation:
\begin{equation}
    \alpha_{\rm X} \approx \frac{E_{\rm 0, X} d_{\rm 0, C}}{E_{\rm 0, C} d_{\rm 0, X}} \alpha_{\rm C},
    \label{eq:stiff_guess}
\end{equation}
where subscripts C and X denote C--C and X--C bonds, respectively.

The Morse potential has a characteristic concave shape that can be well approximated by a harmonic oscillator potential when the vibration amplitude satisfies
\begin{equation}
    \Delta D \ll 1/\alpha,
    \label{eq:Morse_condi}
\end{equation}
where $\Delta D$ represents the typical nuclear displacement. The approximating harmonic oscillator potential is expressed as
\begin{equation}
    V(r_{\rm a}) \approx -E_{\rm 0} + \frac{1}{2} k (r_{\rm a} - d_{\rm 0})^2,
\end{equation}
where the stiffness $k$ is given by
\begin{equation}
    k = 2 E_{\rm 0} \alpha^2.
    \label{eq:poten_stiff}
\end{equation}

Based on the simplified models in Fig.~\ref{fig:X-C-X}(b)--(d), where each carbon or X nucleus is confined between two covalent bonds, the effective potential can be approximated as a superposition of two Morse potentials. The resulting effective stiffnesses are:
\begin{align}
    K_{\rm C,\,x/y} &= 2k_{\rm C}, \\
    K_{\rm C,\,z}   &= k_{\rm C} + k_{\rm X}, \\
    K_{\rm X}       &= 2k_{\rm X},
\end{align}
where $K_{\rm C,\,x/y}$, $K_{\rm C,\,z}$, and $K_{\rm X}$ represent the stiffnesses for carbon nuclei along the $x/y$ and $z$ directions, and for X nuclei, respectively.

The phonon energy of the X nucleus, derived from the quantum harmonic oscillator model, is:
\begin{equation}
    \hbar\omega_{\rm X} =
    \begin{cases}
        74.7\,\text{meV} & \text{(SiV)} \\
        28.5\,\text{meV} & \text{(GeV)} \\
        14.5\,\text{meV} & \text{(SnV)} \\
        5.8\,\text{meV} & \text{(PbV)}
    \end{cases},
    \label{eq:hbo_X}
\end{equation}
where $\omega_{\rm X} = \sqrt{K_{\rm X}/\mu_{\rm X}}$ is the vibrational angular frequency and $\mu_{\rm X} \in \{28.0, 72.6, 118.7, 207.2\}\,\text{Da}$ (1 Da = \(1.66 \times 10^{-27}\) kg) is the mass of the X nucleus.

Similarly, the carbon nucleus phonon energies in the XV system are:
\begin{align}
    \hbar \omega_{\rm{C},x/y} &= 156\,\text{meV}, \\
    \hbar\omega_{\rm{C}, z} &=
    \begin{cases}
        136\,\text{meV} & \text{(SiV)} \\
        120\,\text{meV} & \text{(GeV)} \\
        114\,\text{meV} & \text{(SnV)} \\
        111\,\text{meV} & \text{(PbV)}
    \end{cases},
    \label{eq:hbo_Cz}
\end{align}
with $\omega_{\rm{C},x/y} = \sqrt{K_{\rm{C},x/y}/\mu_{\rm C}}$ and $\omega_{\rm{C},z} = \sqrt{K_{\rm{C},z}/\mu_{\rm C}}$.

To verify the condition~(\ref{eq:Morse_condi}) for XV systems, we calculate the root-mean-square displacement of nuclei using the ground state variance of a 1D quantum harmonic oscillator:
\begin{equation}
    \Delta D = \sqrt{\frac{\hbar}{2\mu \omega}},
\end{equation}
where $\mu$ is the reduced mass and $\omega$ the vibrational angular frequency.

For different species of the X nuclei, we obtain the following typical displacements:
\begin{equation}
    \Delta D_{\rm X} = 
    \begin{cases}
        0.032\,\text{\AA} & \text{(SiV)} \\
        0.032\,\text{\AA} & \text{(GeV)} \\
        0.035\,\text{\AA} & \text{(SnV)} \\
        0.042\,\text{\AA} & \text{(PbV)}
    \end{cases}.
    \label{eq:DeltaD_X}
\end{equation}

For carbon nuclei, the displacements are:
\begin{align}
    \Delta D_{\rm{C}, x/y} &= 0.034\,\text{\AA}, \label{eq:DeltaD_Cxy} \\
    \Delta D_{\rm{C}, z} &= 
    \begin{cases}
        0.036\,\text{\AA} & \text{(SiV)} \\
        0.038\,\text{\AA} & \text{(GeV)} \\
        0.039\,\text{\AA} & \text{(SnV)} \\
        0.040\,\text{\AA} & \text{(PbV)}
    \end{cases}. \label{eq:DeltaD_Cz}
\end{align}

Comparing these displacements with typical $\alpha$ values (2.2 \AA$^{-1}$ for C--C and 0.55-1.7 \AA$^{-1}$ for X--C bonds from Table~\ref{tab:bond_data}), we find $1/\alpha$ ranges from 0.59 \AA \; to 1.8 \AA. Since all calculated $\Delta D$ values are $\ll 1/\alpha$, the harmonic approximation is justified for all cases considered.

\subsubsection{Interaction Hamiltonian}
\label{sec:phonon_int_Hami}

As discussed, by employing the picture of a hole, the system originally comprising 11 covalent electrons is simplified to one single hole within a structure of six fully filled X–C bonds (totaling 12 electrons). The electron-phonon interaction problem thus results in the interaction between a single hole and the seven nuclei within the XV system. In this context, the interaction species under investigation is the Coulomb interaction.

Within this model, the nearest-neighbor approximation is applicable: When a hole is located in the $i$-th X–C covalent bond orbital ($i = 1, 2, \dots, 6$), it primarily interacts with (1) the central X nucleus and (2) the $i$-th carbon nucleus through the Coulomb potential. The other five carbon nuclei are disregarded because their longer distances make their influence negligible. The symmetry of the six X--C bonds permits the interaction analysis to be focused on a single X--C bond unit, as depicted in Fig.~\ref{fig:ele_bases}(b).

We adopt the rest frame defined by the equilibrium positions of the nuclei. In this frame, the nuclei are allowed to move, and the X--C bond can deform accordingly. To first-order approximation, the hole-nucleus interaction is considered separately for each nucleus. That is, when evaluating the interaction between the hole and the motion of the carbon nucleus, the X nucleus is assumed to remain fixed, and vice versa.

The interaction Hamiltonian between a hole and a nucleus is characterized by the potential energy of the hole-nucleus setup, which depends on the positions of the hole and the nuclei. However, similar to the idea of an electronic cloud, the hole is spread in space across the X--C bonding orbital. Additionally, its configuration and density alter in response to nuclear movement. This non-local and adaptable feature of the hole makes evaluating the Coulomb energy between the hole and the nucleus challenging. For a rough magnitude estimate, we employ a simplified model. To begin with, the portion of the hole population adjacent to the X and carbon nuclei can be considered stationary relative to their respective nuclei and thus has minimal impact on the Coulomb potential variation. It is primarily the segment of the population farther from the nuclei, specifically the middle portion of the X--C bond, that alters the distance to the two nuclei and consequently the Coulomb potential energy. Consequently, we define a partition factor $\alpha$ ($0<\alpha<1$), such that the hole's effective charge is $\alpha e$, with $e$ representing the elementary charge.

Since the middle portion of the hole cloud is distant from the X and carbon nuclei, we can approximate it as a point charge. In this context, the effective position of the point charge is the average position of the middle portion of the hole cloud. In Sec.~\ref{sec:electronic_orbital}, we showed that there is a node present (absent) in the ground (first-excited) manifold of the XV system as shown in Fig.~\ref{fig:WF_node}. The presence or absence of the node modifies the shape of the hole cloud, consequently affecting the average position of the effective point charge. In the ground (first-excited) manifold, the average position is deviated from the average position as in an isolated X--C bond toward the carbon (X) nucleus.  That is the reason why the electron-phonon interaction strengths differ between the two manifolds. Here, to give an order-of-magnitude estimation, we neglect this difference, examining an isolated X--C bond and assuming that the average position of the middle portion of the hole cloud is at the midpoint of the X--C bond.

The X and carbon nuclei can also be approximated as point charges. Regarding the effective charge of the nuclei, we assume that the middle portion of the hole cloud is distant enough from the two nuclei. Consequently, the inner-shell electrons provide full screening for the charges of the two nuclei, resulting in effective charges of $4e$ for each nucleus.

The Coulomb energy between the hole and a nucleus depends on their relative distance. Being independent of gauge choices, only changes in potential energy are physically meaningful. Therefore, we investigate only the energy variation induced by the displacements of the nuclei.

We first examine the displacement of the carbon nucleus, assuming a stationary X nucleus. For small displacements, as justified in Sec.~\ref{sec:nuclear_vib}, the displacements in the \( x_i \) and/or \( y_i \) directions approximately result in rotations of the X--C bond, thus no energy variations. In contrast, the displacements in the $z_i$ direction cause changes in the relative distance between the hole and the two nuclei. Assuming that the effective position of the hole remains at the midpoint of the X--C bond, the variation in distance between the point charge of the hole and either the X or the carbon nucleus is expressed by
\begin{equation}
    \Delta d = \frac{1}{2} z_i.
    \label{eq:delta_d}
\end{equation}
Additionally, the derivative of the sum of the hole-X and hole-carbon  potential energies with respect to $\Delta d$ is given by
\begin{equation}
    \frac{\Delta E}{\Delta d} \approx - 2F_{\rm d0} \coloneqq - \frac{8\alpha e^2}{\pi \epsilon_0 d_{\rm 0, X}^2}.
\end{equation}
In classical terms, this corresponds to the Coulomb force between the hole and the screened X or carbon nucleus. Here, the minus sign comes from the fact that the hole is of positive charge.

For the time being, we consider the interaction term related to the displacement of the X nucleus, \( H_{\rm h-x} \), as an undefined entity.

In conclusion, the expression of the hole-nucleus interaction Hamiltonian as
\begin{equation}
    H_{\rm vo} = - F_{\rm d0} \sum_{i=1}^{6} z_i \ket{\sigma_i} \bra{\sigma_i} + H_{\rm h-x},
    \label{eq:_H_vo_sigma}
\end{equation}

As the final step, we convert the above interaction Hamiltonian~\eqref{eq:_H_vo_sigma} into the hole basis $\{\ket{\psi_x}, \ket{\psi_y}\}$ as specified in Eq.~\eqref{eq:ele_eigen_xy}. Recognizing that $\{z_i\}$ forms a linear space, choosing an optimal set of basis vectors can simplify the expression of $H_{\rm vo}$. Also, noticing that the coefficients in the spatially polarized eigenstates of the hole, $\ket{\psi_j}$ ($j \in \{+,0;\; -,0;\;$ $ +,x;\; -,x;\;$ $ +,y;\; -,y\}$) in the bases of $\ket{\sigma_i}$ are all real (see Sec.~\ref{sec:electronic_orbital}), we may choose the set of basis vectors ($\{\mathbf{V}_{j}\}$) of the displacements of the carbon nuclei following the coefficients of $\ket{\psi_j}$, \textit{i.e.},
\begin{equation}
    \mathbf{z}_i^{\mathrm{T}}\mathbf{V}_{j} \coloneqq -\bra{\sigma_i}\ket{\psi_j},
    \label{eq:vib_ele_coef}
\end{equation}
where $\mathbf{z}_i$ is the unit vector in the $z_i$ direction and the minus sign is intentionally included.

In this displacement basis, $\{\mathbf{V}_{j}\}$, the interaction Hamiltonian in Eq.~\eqref{eq:_H_vo_sigma} becomes
\begin{equation}
    H_{\rm vo} = - F_{\rm d0} \sum_j Q_j \; \mathrm{diag}(\mathbf{V}_j) + H_{\rm h-x},
    \label{eq:_H_vo_Vxy}
\end{equation}
where variables $\{Q_j\}$ and $\{z_i\}$ are related via
\begin{equation}
    Q_j = \sum_{i=1}^{6} \mathbf{V}_j[i] \cdot z_i\;,
\end{equation}
and \( \mathrm{diag}(\cdot) \) denotes a function that maps a vector to a diagonal matrix with its components placed on the diagonal in order. 

\begin{figure}[h]
    \centering
    \includegraphics[width=0.98\linewidth]{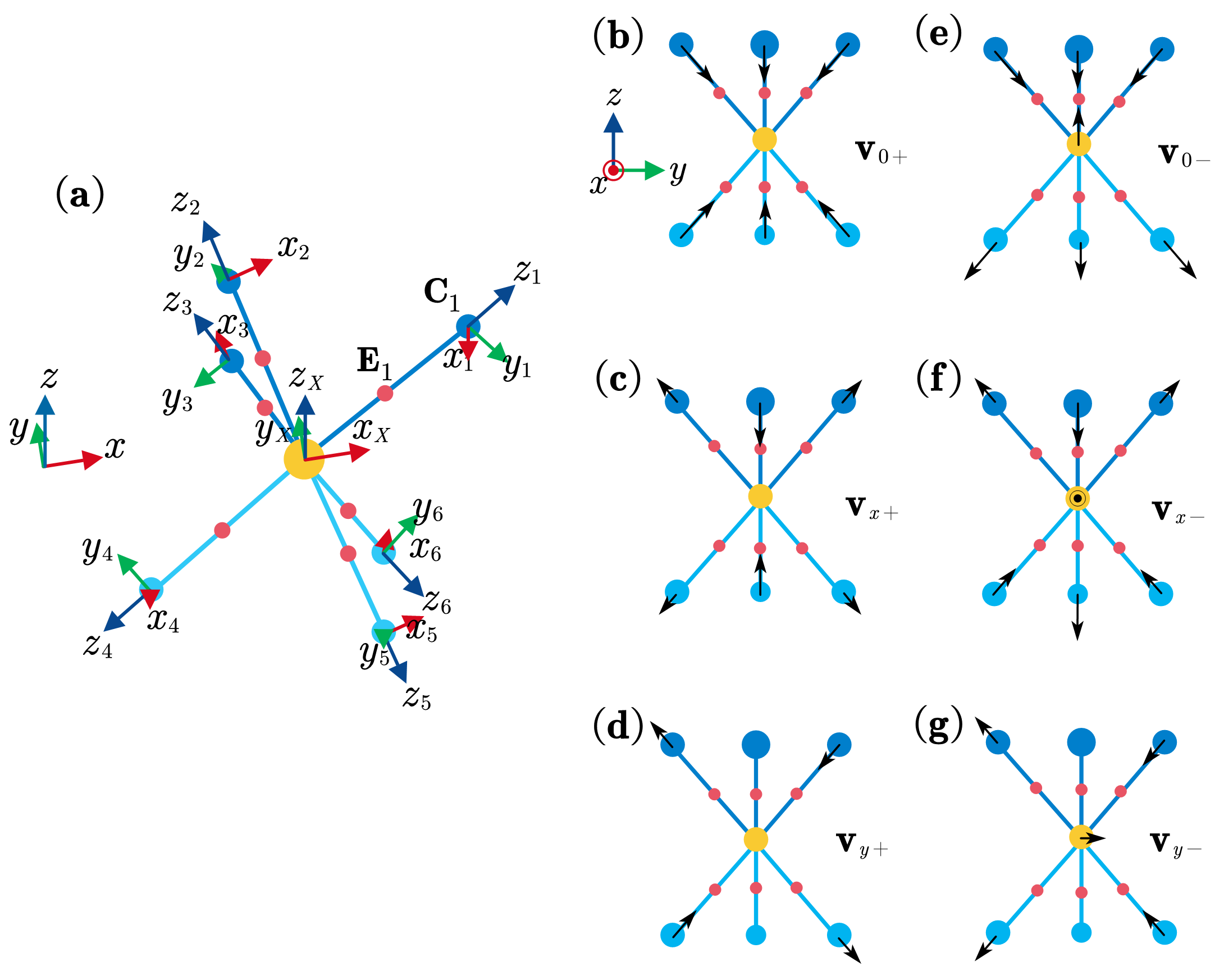}
    \caption{nuclear vibrations. (a) Cartesian coordinate systems $(x_{\rm X}, y_{\rm X}, z_{\rm X})$ and $\{\mathbf{x}_i, \mathbf{y}_i, \mathbf{z}_i\}$ for the central X nucleus and the $i$-th peripheral carbon nucleus, respectively. (b)--(g) Orthogonal vibration modes involving the six carbon nuclei. The black arrows indicate instantaneous nuclear velocities at equilibrium positions. Each carbon nucleus oscillates along the marked directions. Red dots indicate the effective hole positions under the point-charge model.}
    \label{fig:vib_bases}
\end{figure}

Upon transforming the interaction Hamiltonian into the hole orbital basis $\{\ket{\psi_{x}}, \ket{\psi_{y}}\}$, we observe that all terms with $j \in \{-,0; -,x; -,y\}$ vanish. These correspond to collective displacement modes with odd inversion symmetry—i.e., the nuclear displacements reverse sign under spatial inversion (see Fig.~\ref{fig:vib_bases}). Likewise, the term $H_{\rm h-x}$ vanishes, as the displacement of the central X nucleus also exhibits odd inversion symmetry. As a result, the interaction Hamiltonian simplifies to
\begin{equation}
    H_\text{vo} = F_0 Q_0 I^{\text{o}} + F \left( Q_x \sigma_z^\text{o} - Q_y \sigma_x^\text{o} \right),
    \label{eq:H_vo}
\end{equation}
where the \( + \) subscript on \( Q_j \) is dropped for brevity, and
\begin{equation}
    \begin{split}
        I^{\text{o}} &= \ket{\psi_{x}} \bra{\psi_{x}} + \ket{\psi_{y}} \bra{\psi_{y}}, \\
        \sigma_z^{\text{o}} &= \ket{\psi_{x}} \bra{\psi_{x}} - \ket{\psi_{y}} \bra{\psi_{y}}, \\
        \sigma_x^{\text{o}} &= \ket{\psi_{y}} \bra{\psi_{x}} + \ket{\psi_{x}} \bra{\psi_{y}}.
    \end{split}
\end{equation}

The coefficients \( F_0 \) and \( F \) are given by
\begin{equation}
    \begin{split}
        F_0 &= \frac{1}{\sqrt{3}} F_{\rm d0} 
        =\begin{cases}
            39\,\alpha_{\rm Si}\; \text{eV/\AA} & \text{(SiV)}, \\
            35\,\alpha_{\rm Ge}\; \text{eV/\AA} & \text{(GeV)}, \\
            28\,\alpha_{\rm Sn}\; \text{eV/\AA} & \text{(SnV)}, \\
            25\,\alpha_{\rm Pb}\; \text{eV/\AA} & \text{(PbV)},
        \end{cases} \\
        F &= \frac{1}{\sqrt{6}} F_{\rm d0} 
        =\begin{cases}
            27\,\alpha_{\rm Si}\; \text{eV/\AA} & \text{(SiV)}, \\
            25\,\alpha_{\rm Ge}\; \text{eV/\AA} & \text{(GeV)}, \\
            20\,\alpha_{\rm Sn}\; \text{eV/\AA} & \text{(SnV)}, \\
            18\,\alpha_{\rm Pb}\; \text{eV/\AA} & \text{(PbV)}.
        \end{cases}
    \end{split}
    \label{eq:F_range}
\end{equation}
where $\alpha_{\rm X}$ is the partition factor $\alpha$ of the X nucleus. And, we have
\begin{equation}
    F_0=\sqrt{2}F\;.
\end{equation}

We have thus derived the electron-phonon interaction Hamiltonian in the hole basis \( \left\{\ket{\psi_{x}}, \ket{\psi_{y}}\right\} \). Note that the full Hamiltonian of the XV system also includes the free Hamiltonian for the vibrational variables \( Q_x \) and \( Q_y \), while the one for \( Q_0 \) may be neglected, as it induces only a global energy shift. The free Hamiltonian takes the form of a two-dimensional harmonic oscillator:
\begin{equation}
    H_{\rm vib} = -\frac{\hbar^2}{2 \mu} \left(\frac{\partial^2}{\partial Q_x^2} + \frac{\partial^2}{\partial Q_y^2}\right) + \frac{\mu}{2} \omega^2 \left(Q_x^2 + Q_y^2\right),
    \label{eq:H_vib}
\end{equation}
where \( \mu = \mu_{\rm C} \) and \( \omega = \omega_{\rm{C}, z} \). This Hamiltonian relates to those of the individual carbon displacements \( z_i \) through
\begin{equation}
    H_{\rm vib} = \sum_{i=1}^{6} \left[\left(V_{+,x}[i]\right)^2 + \left(V_{+,y}[i]\right)^2\right] H_{{\rm C}, i},
\end{equation}
with
\begin{equation}
    H_{{\rm C}, i} = -\frac{\hbar^2}{2 \mu} \frac{\partial^2}{\partial z_i^2} + \frac{\mu}{2} \omega^2 z_i^2.
    \label{eq:H_har_i}
\end{equation}

In summary, we proposed an approximate model characterizing the electron-phonon interaction in the XV systems with undetermined parameters $\alpha_{\rm X}$ ($0<\alpha_{\rm X}<1$). We list the approximations used in App.~\ref{app:list_of_approx}. We compare this model with previous theoretical works~\cite{HeppThesis}, reported experimental measurements~\cite{Meesala2018Strain}, and DFT simulations~\cite{thiering2018ab, strain_guo2023microwave} in Sec.~\ref{sec:comparison}.

\subsection{Spin-Orbit Interaction}
\label{sec:SOC}

The origin of spin-orbit interaction describes the interaction between the spatial (orbital) movement of the hole and its own spin. It can be explained by the principles of classical physics. According to classical electromagnetism, even when no magnetic field is present in the laboratory frame, an object moving in an electric field experiences a magnetic field in its own rest frame. In the non-relativistic limit (\(|\mathbf{v}| \ll c\)), this induced magnetic field is given by
\begin{equation}
    \mathbf{B} = -\frac{\mathbf{v} \times \mathbf{E}}{c^2},
    \label{eq:B_vE}
\end{equation}
where \(\mathbf{v}\) is the object’s velocity, \(\mathbf{E}\) is the electric field at its position, and \(c\) is the speed of light. If the object possesses spin, the motion of the spin couples to this effective magnetic field generated by its orbital motion, giving rise to the spin-orbit interaction.

In quantum mechanics, the velocity \(\mathbf{v}\) is not a well-defined operator and is replaced by the momentum \(\mathbf{p}\). The association between them is given by \(\mathbf{p} = m_{\rm e} \mathbf{v}\), where \(m_{\rm e}\) is the object’s mass. Additionally, the electric field can be expressed as the gradient of the electrostatic potential, \(\mathbf{E} = -\nabla V\). Substituting these relations into Eq.~\eqref{eq:B_vE}, the spin-orbit interaction Hamiltonian takes the form~\cite{SOC_peter2010fundamentals}
\begin{align}
    H_{\rm os} &= - \frac{g_{\rm s} \mu_{\rm B}}{\hbar} \mathbf{B} \cdot \mathbf{S} \\
    &= \frac{g_{\rm s} \mu_{\rm B}}{m_{\rm e} \hbar c^2} \left( \nabla V \times \mathbf{p} \right) \cdot \mathbf{S},
\end{align}
where \(\mathbf{S}=\frac{\hbar}{2}\left[\sigma_x^{\rm s}, \sigma_y^{\rm s}, \sigma_z^{\rm s}\right]\) is the spin operator with $\{\sigma_i^{\rm s}\}$ ($i=x, y, z$) being the Pauli matrices, \(g_{\rm s}\) is the particle’s \(g\)-factor, and \(\mu_{\rm B} = \frac{e \hbar}{2 m_{\rm e}}\) is the Bohr magneton.

Return to the problem of the spin-orbit coupling of the hole in the XV system, the hole is delocalized over the six X–-C bonds, which surround the central X atom and are approximately confined by the outer six carbon nuclei of the diamond lattice (see Fig.~\ref{fig:ele_bases}). This particular arrangement leads to the consequence that the majority of the spin-orbit interaction is predominantly determined by the electric field generated by the central X nucleus and its inner-shell electrons, while the effect from the electric field generated by the carbon nuclei and their inner-shell electrons can be neglected. To explain, we take the classical picture. For a state with non-zero orbital momentum, the hole can be seen as orbiting around the central X nucleus~\cite{SOC_griffiths2018introduction}. Only the electric field generated by the X nucleus and its inner-shell electrons can have non-zero divergence within the area enclosed by the hole's trajectory, giving rise to non-zero spin-orbit coupling. In contrast, the electric field stemming from any carbon nucleus and its inner-shell electrons has zero divergence within that area. The corresponding magnetic field derived from Eq.~\eqref{eq:B_vE} reverses with the hole's motion, resulting in a vanished net spin-orbit coupling effect. Consequently, the spin-orbit interaction is primarily governed by the electric field of the central X nucleus, as in the case of an isolated X atom.

According to the central field approximation~\cite{central_feild_approx_fox2018student}, the electric field of the central X atom is spherically symmetric and can be expressed as
\begin{equation}
    \nabla V = \beta_{\rm X} \frac{Z_{\rm X} e}{ 4\pi \epsilon_0 r^3} \mathbf{r},
\end{equation}
where \(Z_{\rm X}\) is the atomic number of the X atom, and \(\beta_{\rm X}\) ($0<\beta_{\rm X}<1$) is the screening factor (unspecified) accounting for the reduction of the nuclear electric field by inner-shell electrons.

Substituting this form of \(\nabla V\) into the spin-orbit Hamiltonian $H_{\rm os}$ yields
\begin{equation}
    H_{\rm os} = \beta_{\rm X} \frac{\mu_0 \mu_{\rm B}^2}{2\pi \hbar^2} \frac{Z_{\rm X}}{r^3} \mathbf{L} \cdot \mathbf{S},
    \label{eq:H_os_LS}
\end{equation}
where \(\mathbf{L} = \mathbf{r} \times \mathbf{p}\) is the orbital angular momentum of the hole, and \(\mu_0 = \frac{1}{\epsilon_0 c^2}\) is the vacuum permeability. Here we have used \(g_{\rm s} \approx 2\) and included the \(\frac{1}{2}\) factor arising from Thomas precession~\cite{SOC_griffiths2018introduction, Thomas_procession_namias1989electrodynamics, thomas1926motion, SOC_spavieri2015origin}.

The hole's orbital angular momentum is defined in three-dimensional space, \textit{e.g.}, \(\mathbf{L} = \left[L_x^{\rm o}, L_y^{\rm o}, L_z^{\rm o}\right]\), where each component \(L_i^{\rm o}\) (with \(i \in \{x, y, z\}\)) corresponds to rotation about the respective axis in the XV system [see Fig.~\ref{fig:ele_bases}(a)].

As previously discussed, each X--C bond consists of an \(sp^3d^2\) hybridized orbital from the X atom and an \(sp^3\) hybridized orbital from the carbon atom. These bonds may be considered as extended \(sp^3d^2\) orbitals oriented along the radial directions. As a result, the orbital angular momentum of the electronic state (see Sec.~\ref{sec:hole_eigen}) is determined by the orbital angular momentum associated with the \(sp^3d^2\) component.

Since \(d\)-orbitals have the largest angular momentum quantum number \(l = 2\), the magnetic quantum number \(m_z\) of the \(sp^3d^2\) orbital takes on values in \(\{-2, -1, 0, 1, 2\}\). Consequently, the \(z\)-component of the orbital angular momentum can be written as
\begin{equation}
    L_z^{\rm o} = \sum_{m_z = -2}^{2} m_z \hbar \ket{m_z} \bra{m_z},
    \label{eq:Lz_mz}
\end{equation}
where \(m_z\) is defined with respect to the \(z\)-axis of the XV system, taken as the quantization axis. The corresponding non-normalized eigenstate of \(L_z^{\rm o}\) is
\begin{equation}
    \ket{m_z} = \mathrm{e}^{\mathrm{i} m_z \phi^{\rm o}},
    \label{eq:m_z}
\end{equation}
with \(\phi^{\rm o} = \arctan\left(\frac{y}{x}\right)\) denoting the azimuthal angle of the hole's position.

\begin{figure}
    \centering
    \includegraphics[width=0.98\linewidth]{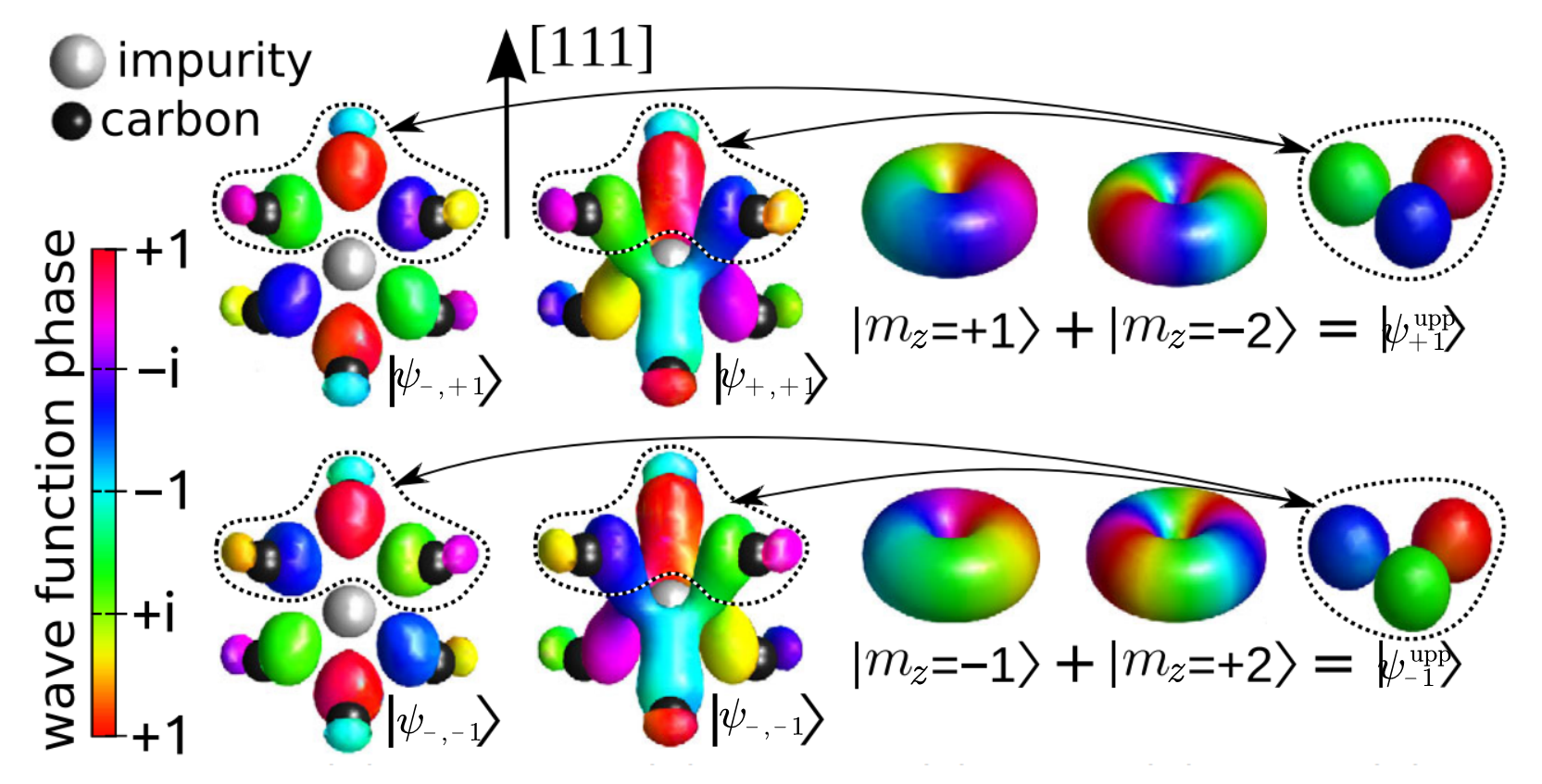}
    \caption{Visualization of the linear combinations of the \(\ket{\psi_{\pm1}}\) hole orbital states as described in Eq.~\eqref{eq:mz}. Adapted from Ref.~\cite{thiering2018ab}.}
    \label{fig:combi_visul}
\end{figure}

Analyzing the electronic orbital eigenstates as represented by Eqs.~\eqref{eq:ele_eigen_+1-1}, the state \(\ket{\psi_{+1}}\) (\(\ket{\psi_{-1}}\)) exhibits a non-zero overlap exclusively with \(\ket{m_z=+1}\) and \(\ket{m_z=-2}\) (\(\ket{m_z=-1}\) and \(\ket{m_z=+2}\)). Additionally, since the longitudinal extent of the X--C bond is small compared to the bond length, the magnitudes of the two overlaps are approximately equal. Thus,
\begin{equation}
\begin{split}
    \ket{\psi_{+1}} &\appropto \frac{1}{\sqrt{2}}\left(\ket{m_z=+1} + \ket{m_z=-2} \right), \\
    \ket{\psi_{-1}} &\appropto \frac{1}{\sqrt{2}}\left(\ket{m_z=-1} + \ket{m_z=+2} \right).
\end{split}
\label{eq:mz}
\end{equation}
A visualization of these composition of the \(\ket{\psi_{\pm1}}\) states is shown in Fig.~\ref{fig:combi_visul}.

Utilizing Eqs.~\eqref{eq:mz} and \eqref{eq:basis_transition}, the operator $L_z^{\rm o}$ can be represented in the basis \(\{\ket{\psi_x}, \ket{\psi_y}\}\) as
\begin{equation}
    L_z^{\rm o} = -\frac{\hbar}{2} \sigma_y^{\rm o},
    \label{eq:L_z_sigma_y}
\end{equation}
where
\begin{equation}
\begin{split}
    \sigma_y^{\rm o} &= \ket{\psi_{+}} \bra{\psi_{+}} - \ket{\psi_{-}} \bra{\psi_{-}} \\
    &= \mathrm{i} \ket{\psi_y} \bra{\psi_x} - \mathrm{i} \ket{\psi_x} \bra{\psi_y}.
\end{split}
\end{equation}

Regarding the $y$-component of the orbital angular momentum, $L_y^{\rm o}$, the XV system has two X--C bonds ($\ket{\sigma_2}$ and $\ket{\sigma_6}$) on the $+y$ side, two ($\ket{\sigma_3}$ and $\ket{\sigma_5}$) on the $-y$ side, and the remaining two ($\ket{\sigma_1}$ and $\ket{\sigma_4}$) on the $y=0$ plane. Thus, all states within the XV system exhibit no preference towards either the positive or negative orientation of $L_y^{\rm o}$, leading to the conclusion that
\begin{equation}
    L_y^{\rm o} = 0 \;.
\end{equation}

Concerning the $x$-component of the orbital angular momentum, $L_x^{\rm o}$, similar to the case of $L_z^{\rm o}$, the XV system has three X--C bonds ($\ket{\sigma_1}$, $\ket{\sigma_5}$, and $\ket{\sigma_6}$) on the $+x$ side and the other three ($\ket{\sigma_4}$, $\ket{\sigma_2}$, and $\ket{\sigma_3}$) on the $-x$ side [see Fig.~\ref{fig:XV_size}]. The six X--C bonds can be regarded as presenting a \underline{distorted} $\frac{2\pi}{3}$ symmetry about the $x$ axis. Therefore, $L_x^{\rm o}$ is smaller than $L_z^{\rm o}$ in magnitude but not completely vanish, \textit{i.e.},
\begin{equation}
    L_x^{\rm o} \ne 0 \;.
    \label{eq:Lx_ne0}
\end{equation} 
The magnitude and form of the operator $L_x^{\rm o}$ are left for future examination. Due to its smaller magnitude, this term is often neglected in most of the thesis; nevertheless, it affects the energy degeneracy as discussed in Sec.~\ref{sec:degeneracy}.

Therefore, considering only the $z$-component of orbital angular momentum, \(L_z^{\rm o}\), Eq.~\eqref{eq:H_os_LS} is simplified to
\begin{equation}
    H_{\rm os} = \beta_{\rm X} \frac{\mu_0 \mu_{\rm B}^2}{8\pi} \frac{Z_{\rm X}}{r^3} \sigma_y^{\rm o} \sigma_z^{\rm s},
    \label{eq:H_os_LzSz}
\end{equation}
where \(\frac{\hbar}{2} \sigma_z^{\rm s}\) is the \(z\)-component of the spin operator of the hole.

Below, we estimate the expectation value, \(\left\langle r^{-3} \right\rangle\). The \(\ket{\psi_{+1}}\) states can be regarded as comprising half \(sp^3\) orbitals from the six adjacent carbon atoms, along with half \(sp^3d^2\) hybrid orbitals from the X atom. We neglect the contribution from the \(sp^3\) half, \textit{i.e.},
\begin{equation}
    \left\langle r^{-3} \right\rangle_{sp^3} \approx 0 \;.
\end{equation}
The reason is twofold: (1) the spin-orbit coupling strength is reciprocal cubic decay in distance from the X nucleus; (2) the population in the \(sp^3\) orbitals suffers a more substantial screening effect on the electric field of the X nucleus than that in the \(sp^3d^2\) orbitals.

For the \(sp^3d^2\) half, we estimate the expectation value \(\left\langle r^{-3} \right\rangle\) based on a single atom as given by~\cite{SOC_griffiths2018introduction}
\begin{equation}
    \left\langle \frac{1}{r^3} \right\rangle_{\rm sgl} = \frac{Z_{\rm X}^3}{a_{\rm B}^3 n_{\rm X}^3 l(l+1)(l+0.5)} \ket{l} \bra{l},
    \label{eq:r3_atomic_states}
\end{equation}
where \(n_{\rm X} = 3, 4, 5, 6\) for Si, Ge, Sn, and Pb, respectively, is the principle quantum number, $l$ is the angular momentum quantum number and \(a_{\rm B} = 5.29 \times 10^{-11}\text{m}\) is the Bohr radius. As shown in Eq.~\eqref{eq:mz}, for the \(\ket{\psi_{+1}}\) states, the \(sp^3d^2\) half has only two equal-weight components of \(\ket{m_z = +1}\) and \(\ket{m_z = -2}\). According to the Aufbau principle, the \(\ket{m_z = +1}\) component arises from the \(l = 1\) (i.e., \(p\)) state, while the \(\ket{m_z = -2}\) component comes from the \(l = 2\) (i.e., \(d\)) state. As such, the expectation value of \(r^{-3}\) for the \(\ket{\psi_{\pm1}}\) state can be approximated as
\begin{equation}
    \left\langle \frac{1}{r^3} \right\rangle \approx \frac{Z_{\rm X}^3}{10 a_{\rm B}^3 n_{\rm X}^3}\;,
\end{equation}

Hence, the spin-orbit interaction Hamiltonian becomes
\begin{equation}
    H_{\rm os} \approx \frac{\lambda_{\rm X}}{2} \sigma_y^{\rm o} \sigma_z^{\rm s},
    \label{eq:H_os_sigma}
\end{equation}
where
\begin{equation}
    \lambda_{\rm X} = \beta_{\rm X} \frac{\mu_0 \mu_{\rm B}^2}{40\pi a_{\rm B}^3} \frac{Z_{\rm X}^4}{n_{\rm X}^3} =
    \begin{cases}
        52\, \beta_{\rm Si} \; \text{meV} & \text{(SiV)} \\
        603\, \beta_{\rm Ge} \; \text{meV} & \text{(GeV)} \\
        1841\, \beta_{\rm Sn} \; \text{meV} & \text{(SnV)} \\
        7706\, \beta_{\rm Pb} \; \text{meV} & \text{(PbV)}
    \end{cases}.
    \label{eq:spin_orbit_strength}
\end{equation}

In summary, we proposed an approximate model characterizing the spin-orbit interaction in the XV systems with undetermined parameters $\beta_{\rm X}$ ($0<\beta_{\rm X}<1$). We list the approximations used in App.~\ref{app:list_of_approx}. We compare this model with DFT simulations~\cite{thiering2018ab} and measured ground splittings~\cite{GS_SiV_hepp2014electronic, GS_SiV_sukachev2017silicon, GS_GeV_ekimov2015germanium, GS_GeV_siyushev2017optical, GS_SnV_iwasaki2017tin, GS_PbV_trusheim2019lead} in Sec.~\ref{sec:comparison}.


\section{Intrinsic Properties} 
\label{sec:eigenstate}

In the previous section, we have shown the modeling of the intrinsic interactions of the XV system and derived the intrinsic Hamiltonian. It includes terms of free motions of hole orbitals, atomic vibrations, hole spin, and the interaction between them. The ground and first-excited energy manifolds are analyzed separately with their respective Hamiltonians sharing the same form but with different parameters. Ignoring the term $F_0 Q_0 I^{\text{o}}$ in Eq.~\eqref{eq:H_vo}, as it is irrelevant to the splitting of the hole orbital states, the intrinsic Hamiltonian for each manifold is given by
\begin{equation}
\begin{split}
    H_0 &= -\frac{\hbar^2}{2 \mu} \left(\frac{\partial^2}{\partial Q_x^2} + \frac{\partial^2}{\partial Q_y^2}\right) + \frac{\mu}{2} \omega^2 \left(Q_x^2 + Q_y^2\right) \\
    &\quad + F \left(Q_x \sigma_{z}^{\text{o}} - Q_y \sigma_{x}^{\text{o}}\right) + \frac{\lambda}{2} \sigma_y^\text{o} \sigma_z^\text{s},
\end{split}
    \label{eq:H_0}
\end{equation}
where \(Q_x\) and \(Q_y\) are linear combinations of the movements on the six carbon atomic core in the $\{z_i\}$ directions and can be interpreted as the displacements of the center of the mass of the upper three carbon atomic core in the $x$ and $y$ directions, respectively. In this section, we analyze the properties of \(H_0\).

Before we proceed, to simplify the analysis, we first make a unitary transformation $U_c$ for the hole orbital operators $\sigma^\text{o}_i$ \(i = x, y, z\) in Hamiltonian $H_0$. The unitary operator $U_c$ is defined as
\begin{equation}
    U_{\rm c} = R_y\left(\frac{\pi}{2}\right) R_z\left(-\frac{\pi}{2}\right),
\end{equation}
where
\begin{equation}
    R_i\left(\theta\right) = \cos\left(\frac{\theta}{2}\right) I^\mathrm{o} + \mathrm{i} \sin\left(\frac{\theta}{2}\right) \sigma^\text{o}_i\;.
\end{equation}
It transforms the hole-orbital Pauli matrices as
\begin{equation}
    \begin{split}
        \sigma^\text{o}_z &\rightarrow \tau_{x}^{\text{o}} = U_{\rm c} \, \sigma^\text{o}_z \, U_{\rm c}^\dagger, \\
        \sigma^\text{o}_x &\rightarrow -\tau_{y}^{\text{o}} = U_{\rm c} \, \sigma^\text{o}_x \, U_{\rm c}^\dagger, \\
        \sigma^\text{o}_y &\rightarrow -\tau_{z}^{\text{o}} = U_{\rm c} \, \sigma^\text{o}_y \, U_{\rm c}^\dagger.
    \end{split}
    \label{eq:pauli_convert}
\end{equation}

Consequently, in the new hole orbital bases \(\{U_{\rm c} \ket{\psi_{x}}, U_{\rm c} \ket{\psi_{y}}\}\), the intrinsic Hamiltonian \(H_0\)~(\ref{eq:H_0}) is given by
\begin{equation}
    H_0 = H_{\rm vib} + F \left(Q_x \tau_{x}^{\text{o}} + Q_y \tau_{y}^{\text{o}}\right) - \frac{\lambda}{2} \tau_z^\text{o} \sigma_z^\text{s},
    \label{eq:H_0_tau}
\end{equation}
where the notations of $\tau_{x}^{\text{o}}$ and $\tau_{y}^{\text{o}}$ have been made uniform with $Q_x$ and $Q_y$ in terms of subscripts and signs.

\subsection{Conserved Quantities}
\label{sec:cons_quan}

In addition to the total energy, we identify two other conserved quantities in the intrinsic Hamiltonian $H_0$. The first is the hole spin, as evidenced by the following commutation relation:
\begin{equation}
    \left[H_0, \tau_z^{\rm s}\right] = 0.
\end{equation}
The second conserved quantity is the $z$-component of the orbital-phonon angular momentum, or vibronic angular momentum, defined as
\begin{equation}
    L_{z}^{\rm vo} = -\mathrm{i}\hbar \frac{\partial}{\partial \phi^{\rm v}} + \frac{\hbar}{2} \tau_z^{\rm o},
\end{equation}
[see Eqs.~(\ref{eq:Lz_mz}), (\ref{eq:m_z}), (\ref{eq:L_z_sigma_y}), and (\ref{eq:pauli_convert})]. Here, $\phi^{\rm v} = \arctan\frac{Q_y}{Q_x}$ represents the azimuthal angle in the $Q_x$-$Q_y$ plane.

The conservation of $L_{z}^{\rm vo}$ is confirmed by the commutation relation
\begin{equation}
    \left[H_0, L_{z}^{\rm vo}\right] = 0,
\end{equation}
which can be verified using the following auxiliary relations:
\begin{gather}
    H_{\rm vo} = F Q \left( M_+ \tau^\text{o}_- + M_- \tau^\text{o}_+ \right), \label{eq:H_vo_pm}\\
    \left[ L^\mathrm{v}_{z}, \; M_{\pm} \right] = \pm \hbar\, M_{\pm}, \label{eq:raise_lower_M}\\
    \left[ \frac{\hbar}{2} \tau_z^{\rm o}, \; \tau^\text{o}_{\pm} \right] = \pm \hbar \, \tau^\text{o}_{\pm}, \label{eq:raise_lower_tau}
\end{gather}
where \(Q = \sqrt{Q_x^2 + Q_y^2} \), \(M_{\pm} = \mathrm{e}^{\pm \mathrm{i} \phi^\mathrm{v}}\), and
\begin{equation}
    \tau^\text{o}_{\pm} \coloneqq \frac{1}{2} \left( \tau^\text{o}_x \pm \mathrm{i} \tau^\text{o}_y \right).
\end{equation}

These conserved quantities partition the Hilbert space of the XV center system into multiple independent subspaces. As a result, states in different subspaces do not interact. A state initially in one subspace remains confined to that subspace throughout its evolution. Furthermore, the intrinsic Hamiltonian can be block-diagonalized, meaning that each of its eigenstates can be constructed exclusively from basis states within a single subspace.

The conservation of the spin ($\tau_z^{\rm s}$) implies that states with spin-up (eigenstates of $\tau_z^{\rm s}$ with eigenvalue $+1$) form one subspace, while states with spin-down (eigenvalue $-1$) form another distinct subspace.

Similarly, the conservation of the $z$-component of the vibronic angular momentum ($L_z^{\rm vo}$) further divides the spin-up and spin-down subspaces into smaller subspaces, each characterized by a unique value of the $z$-component vibronic angular momentum. A detailed presentation and visualization of these subspaces will be provided in Sec.~\ref{sec:interaction_diag}.

\subsection{Symmetries}
\label{sec:symmetries}

The XV center system exhibits both temporal and spatial symmetries: the well-known time-reversal symmetry (associated with motion directions) and a newly defined spatial symmetry termed "joint-reflection symmetry".

\subsubsection{Time-Reversal Symmetry}
The XV center system possesses half-integer spin, and thus the Kramers' degeneracy theorem~\cite{time_reversal_abragam2012electron} applies. This theorem states that the system is at least doubly degenerate due to time-reversal symmetry.

Time-reversal symmetry, or intuitively, motion-reversal symmetry, is based on the time-reversal operation. This operation transforms the momentum and all spins of an object at a given time into their respective opposite directions while keeping the spatial positions unchanged. The time-reversal operator takes different forms in different representations. In the spatial representation (not the momentum representation), the time-reversal operator can be expressed as
\begin{equation}
    \mathcal{T} = \mathrm{i}\sigma_y^{\rm s} K_0,
    \label{eq:time_rev_oper}
\end{equation}
where \(K_0\) represents the complex conjugation operation, which transforms all imaginary numbers in the quantum states or operators it acts on into their complex conjugates. This operation reverses the momentum operators and the Pauli-$y$ spin operators, as they are imaginary in the spatial representation. The \(\sigma_y^{\rm s}\) operator reverses the Pauli-$x$ and Pauli-$z$ spin operators. The factor \(\mathrm{i}\) is included to make the operator \(\mathrm{i}\sigma_y^{\rm s}\) real, ensuring it commutes with \(K_0\).

With the mathematical form of the time-reversal operator, the time-reversal symmetry of Hamiltonian~(\ref{eq:H_0}) can also be verified by the commutation relation
\begin{equation}
    \left[H_0, \mathcal{T}\right] = 0,
\end{equation}
or equivalently \(\mathcal{T}H_0\mathcal{T}^{-1} = H_0\). This can be easily proved, as the time-reversal operator transforms only the two operators \(\sigma_y^{\rm o}\) and \(\sigma_z^{\rm s}\) in \(H_0\) by simply flipping their signs, \textit{i.e.},
\begin{equation}
\begin{split}
    \mathcal{T} \sigma_y^{\rm o} \mathcal{T}^{-1} = -\sigma_y^{\rm o}, \\
    \mathcal{T} \sigma_z^{\rm s} \mathcal{T}^{-1} = -\sigma_z^{\rm s}. \\
\end{split}
\end{equation}

\subsubsection{Joint-Reflection Symmetry}
\label{sec:joint-reflection}
The joint-reflection symmetry is based on the operation of spatial reflections on both the phonon and orbital states, which flips both their positions and momenta, combined with a $\pi$-rotation that flips the $z$-component of the hole spin.

The $Q_x-Q_y$ plane and the $\tau_{x}^{\text{o}}-\tau_{y}^{\text{o}}$ "plane" are reflected about the axes \(Q_y = Q_x \tan \theta\) and \(\cos \theta \tau^\text{o}_{x} + \sin \theta \tau^\text{o}_{y}\), respectively, with the same angle $\theta$. The transformations can be presented as
\begin{equation}
    \begin{bmatrix}
        Q_x\\
        Q_y
    \end{bmatrix}
    \rightarrow
    \begin{bmatrix}
        Q_{x,\theta}\\
        Q_{y,\theta}
    \end{bmatrix}
    =
    F_\theta
    \begin{bmatrix}
        Q_x\\
        Q_y
    \end{bmatrix}\;,
    \quad \text{and} \quad
    \begin{bmatrix}
        \tau^\text{o}_{x} \\
        \tau^\text{o}_{y}
    \end{bmatrix}
    \rightarrow
    \begin{bmatrix}
        \tau^\text{o}_{x,\theta} \\
        \tau^\text{o}_{y,\theta}
    \end{bmatrix}
    =
    F_\theta
    \begin{bmatrix}
        \tau^\text{o}_{x} \\
        \tau^\text{o}_{y}
    \end{bmatrix}\;.
\end{equation}
with a reflection matrix
\begin{equation}
    F_\theta = \begin{bmatrix}
        \cos(2\theta) & \sin(2\theta) \\
        \sin(2\theta) & -\cos(2\theta)
    \end{bmatrix}.
\end{equation}
The transformation is not unique with valid angle $\theta \in [0, 2\pi)$.

Or alternatively, these transformations can be presented as operators $F^\mathrm{v}_\theta$ and $F^\mathrm{o}_\theta$, where
and operator $F^\mathrm{v}_\theta$ applies the following transformation to a quantum state $\ket{\psi(Q_x, Q_y)}$:
\begin{equation}
    F^\mathrm{v}_\theta\ket{\psi(Q_x, Q_y)}=\ket{\psi(Q_{x,\theta}, Q_{y, \theta})}\;.
\end{equation}
and
\begin{equation}
    F^\mathrm{o}_\theta =\mathrm{i}\left( \cos(2\theta)\, \tau^\text{o}_{x} + \sin(2\theta)\, \tau^\text{o}_{y}\right).
\end{equation}

The joint-reflection operation also involves a $\pi$-rotation on the hole spin, given by the operator
\begin{equation}
    F_{\phi}^{\rm s} = \cos \phi \, \sigma_x^{\rm s} + \sin \phi \, \sigma_y^{\rm s},
    \label{eq:F_z_s}
\end{equation}
whose definition is also not unique, with valid \(\phi \in [0, 2\pi)\).

The joint-reflection operation is then represented by the operator
\begin{equation}
    F^\mathrm{vos}_{\theta \phi} = F^\mathrm{v}_\theta F^\mathrm{o}_\theta F^\mathrm{s}_\phi.
\end{equation}

One can verify the commutation relation of the intrinsic Hamiltonian with the joint-reflection operator:
\begin{equation}
    \left[H_0, F^\mathrm{vos}_{\theta \phi} \right] = 0\;.
\end{equation}
This follows from the symmetry properties of the individual terms in \( H_0 \). The phonon bare Hamiltonian \( H_{\rm v} \) possesses reflection symmetry, ensuring its invariance under \( F^\mathrm{vos}_{\theta \phi} \). In the electron-phonon interaction Hamiltonian \( H_{\rm vo} \), the relevant operators appear in pairs and transform in the same manner under the symmetry operation. Finally, in the spin-orbit interaction Hamiltonian \( H_{\rm os} \), both contributing operators undergo sign changes, preserving the overall commutation relation. Consequently, the intrinsic Hamiltonian exhibits joint-reflection symmetry.

\subsection{Degeneracy}
\label{sec:degeneracy}

Above, we see that the intrinsic Hamiltonian \(H_0\) commutes with four operators: 
\begin{itemize}
    \item the \(z\)-component vibronic angular momentum operator \(L_z^{\rm vo}\), 
    \item the spin-\(z\) operator \(\tau_z^{\rm s}\), 
    \item the time-reversal operator \(\mathcal{T}\), and
    \item the flip-all operator \(F_{\theta\phi}^{\rm vos}\). 
\end{itemize}
However, operators among the four do not always commute with each other. Specifically,
\begin{gather}
    \left[\tau_z^{\rm s}, \mathcal{T}\right] \ne 0,\label{eq:un_commute_z_T}\\
    \left[\tau_z^{\rm s}, F_{\theta\phi}^{\rm vos}\right] \ne 0,\label{eq:un_commute_z_F}\\
    \left[L_z^{\rm vo}, F_{\theta\phi}^{\rm vos}\right] \ne 0. \label{eq:un_commute_L_F}
\end{gather}

These non-commuting relationships imply two-fold degeneracy in the Hamiltonians $H_0$. To show this, we use the following lemma.

\textbf{Lemma.} Let $H$, $A$, and $B$ be operators acting on a Hilbert space $\mathcal{H}$, satisfying:
\begin{enumerate}
    \item $[H, A] = [H, B] = 0$,
    \item The commutator $[A, B]$ acts as a full-rank operator on $\mathcal{H}$.
\end{enumerate}
Then every eigenvalue of $H$ has at least two linearly independent eigenfunctions.

\textbf{Proof.} Let $\lambda$ be an eigenvalue of $H$, and denote its corresponding eigenspace by
\[
\mathcal{H}_\lambda = \{ \psi \in \mathcal{H} \mid H \psi = \lambda \psi \}.
\]
Since $[H, A] = [H, B] = 0$, both $A$ and $B$ preserve $\mathcal{H}_\lambda$, i.e., $A\psi \in \mathcal{H}_\lambda$ and $B\psi \in \mathcal{H}_\lambda$ for all $\psi \in \mathcal{H}_\lambda$.

Suppose, for contradiction, that $\mathcal{H}_\lambda$ is one-dimensional. Then any vector $\psi \in \mathcal{H}_\lambda$ must satisfy $A\psi = a\psi$ and $B\psi = b\psi$ for some scalars $a$ and $b$. It follows that
\[
[A, B] \psi = (AB - BA)\psi = (ab - ba)\psi = 0,
\]
which contradicts the assumption that $[A, B]$ acts as a full-rank operator on $\mathcal{H}_\lambda$. Therefore, $\dim \mathcal{H}_\lambda \geq 2$.

Once an eigenstate of \(H\) is given, the proof above also provides a method to find its degenerate counterpart.

Based on this lemma, any one of the relationships~(\ref{eq:un_commute_z_T}), (\ref{eq:un_commute_z_F}), and (\ref{eq:un_commute_L_F}) can demonstrate the double degeneracy of the intrinsic Hamiltonian \(H_0\). It also reveals that, for an eigenstate with a \(z\)-component vibronic angular momentum expectation value \(l\) and spin expectation value \(s\), there must be a degenerate state with \(-l\) and \(-s\). The interaction diagram in Fig.~\ref{fig:interaction_dia} visualizes this symmetry and degeneracy.

In the latter sections, we show that the XV system may also couple to lattice strain (Sec.~\ref{sec:strain}) and magnetic fields (Sec.~\ref{sec:Zeeman}). Below, we examine the degeneracy situations when the XV system is exposed to strain and/or the $x$- and/or $y$-components of magnetic fields.

The strain interaction Hamiltonian takes the form
\begin{equation}
    H^{\rm str} \propto  \left(\cos\theta_0 \sigma^\text{o}_{z} + \sin\theta_0 \sigma^\text{o}_{x}\right),
    \label{eq:strain_form}
\end{equation}
with a parameter \(\theta_0\) depending on the stress direction, while the interaction Hamiltonian with the \(x\)- and/or \(y\)-components of the magnetic field takes the form
\begin{equation}
    H_{xy}^{\rm mag} \propto  \left(\cos\phi_0 \sigma^\text{s}_{x} + \sin\phi_0 \sigma^\text{s}_{y}\right),
    \label{eq:magnetic_xy_form}
\end{equation}
with a parameter \(\phi_0\) depending on the magnetic field direction.

The XV center system remains doubly degenerate in energy when strain is applied because the Hamiltonian \(H_0 + H^{\rm str}\) commutes with both \(\sigma^\text{s}_{z}\) and \(\mathcal{T}\), but the two of them do not commute.

Similarly, the XV center system remains doubly degenerate in energy when the the magnetic field in the \(x\)- and/or \(y\)-direction are applied because the Hamiltonian \(H_0 + H_{xy}^{\rm mag}\) commutes with both \(L_z^{\rm vo}\) and \(F_{\theta\phi}^{\rm vos}(\phi = \phi_0)\), but the two of them do not commute. Please note that this result is based on the assumption of a vanished $x$-component of the orbital angular momentum. However, as discussed in Sec.~\ref{sec:SOC}, it does not strictly vanish [Eq.~\eqref{eq:Lx_ne0}], which may lead to observable splittings.

However, when both the strain and the magnetic field in the \(x\)- and/or \(y\)-direction are present, this double degeneracy will be lifted. This observation suggests that when a constant \(x\)- and/or \(y\)-direction magnetic field is applied, the XV system's splitting depends on the strain's magnitude. This phenomenon leads to a potential application as follows.

Currently, the in situ strain of the XV center is estimated by numerical calculation of the finite element analysis of the bulk sample, which is computationally costly and not always precise. Since the splitting under the magnetic field in the \(x\)- and/or \(y\)-direction depends on strain, it may serve as a means to probe local strain with improved accuracy. We provide the relationship between the splitting, magnetic field, and strain based on the improved quench model (see Sec.~\ref{sec:quench_f}) in Sec.~\ref{sec:predictions}.

\subsection{Interaction Diagram}
\label{sec:interaction_diag}

The interaction diagram provides a visual representation of the interactions between quantum eigenstates in a certain representation. To construct this diagram, we first select a set of complete basis states for the atomic motions, hole orbitals, and hole spin. 

For atomic motions, we use the basis set \(\{\ket{n}\ket{m}\}\), obtained by solving the 2D harmonic oscillator in polar coordinates~\cite{pauling2012introduction}. Specifically, $\ket{n}$ represents the eigenstate of $H_{\rm vib}$ [Eq.~\eqref{eq:H_vib}]:
\begin{equation}
    H_\mathrm{v} \ket{n} = \left(n + 1\right)\hbar\omega \ket{n}\;,
    \label{eq:n_eigen}
\end{equation}
and $\ket{m} = e^{\mathrm{i} m \phi^{\rm v}}$ is the eigenstate of $L_z^{\rm v}$:
\begin{equation}
    L_z^{\rm v} \ket{m} = m\hbar \ket{m}\;.
    \label{eq:m_eigen}
\end{equation}
Here, \(n \in \{0, 1, 2, \ldots\}\) denotes the total phonon number, and \(m \in \{-n, -n+1, \ldots, n-1, n\}\) is the quantum number corresponding to the $z$-component of angular momentum for the atomic motion.

For the hole orbitals, we select the eigenstates of the operator $\tau_z^{\rm o}$, denoted as $\ket{+}_z^{\rm o}$ and $\ket{-}_z^{\rm o}$, as the basis set, \textit{i.e.},
\begin{equation}
    \tau_z^{\rm o}\,\ket{\pm}_z^{\rm o} = \pm \ket{\pm}_z^{\rm o}\;.
\end{equation}
Similarly, for the hole spin, we use the eigenstates of the operator $\sigma_z^{\rm s}$, labeled as $\ket{\uparrow}_z^{\rm s}$ and $\ket{\downarrow}_z^{\rm s}$, \textit{i.e.},
\begin{equation}
    \begin{split}
        \sigma_z^{\rm s}\,\ket{\uparrow}_z^{\rm s} &= + \ket{\uparrow}_z^{\rm s}\;, \\
        \sigma_z^{\rm s}\,\ket{\downarrow}_z^{\rm s} &= - \ket{\downarrow}_z^{\rm s}\;.
    \end{split}
\end{equation}

In this basis set, the states $\ket{n}$ and $\ket{n+1}$, corresponding to the same hole orbital and spin, are split by an energy $\hbar\omega$. Similarly, the states $\ket{+}_z^{\rm o}$ and $\ket{-}_z^{\rm o}$, associated with the same atomic motion and hole spin, are split by an energy $\lambda$. Furthermore, as analyzed in Sec.~\ref{sec:cons_quan}, the interactions in $H_0$ occur only between states $\ket{m}\ket{+}_z^{\rm o}$ and $\ket{m+1}\ket{-}_z^{\rm o}$ with the same hole spin because of the conservation of the $z$-component of the total angular momentum.

The interaction strengths can be obtained by expressing the operators
\begin{equation}
    Q_\pm = Q M_{\pm}
\end{equation}
[see Eq.~(\ref{eq:H_vo_pm})] in the $\{\ket{n}\ket{m}\}$ basis~\cite{longuet1958studies}:
\begin{equation}
    Q_+ = \sqrt{\frac{\hbar}{2\mu\omega}} \sum_{nm}
    \bordermatrix{
        ~ & \langle nm | \cr
        |n-1 \, m-1 \rangle & 0 \cr
        |n-1 \, m+1 \rangle & \sqrt{n-m} \cr
        |n+1 \, m-1 \rangle & 0 \cr
        |n+1 \, m+1 \rangle & \sqrt{n+m+2}
    }\;;
    \label{Q+}
\end{equation}
\begin{equation}
    Q_- = \sqrt{\frac{\hbar}{2\mu\omega}} \sum_{nm}
    \bordermatrix{
        ~ & \langle nm | \cr
        |n-1 \, m-1 \rangle & \sqrt{n+m} \cr
        |n-1 \, m+1 \rangle & 0 \cr
        |n+1 \, m-1 \rangle & \sqrt{n-m+2} \cr
        |n+1 \, m+1 \rangle & 0
    }\;.
    \label{Q-}
\end{equation}
Based on this, the minimum interaction strength, which occurs between the two states, 
\begin{align}
    &\ket{n=0,\; m=0,\; -_z^{\rm o},\;\uparrow_z^{\rm s}}\; \text{and} \\
    &\ket{n=1,\; m=-1,\; +_z^{\rm o},\;\uparrow_z^{\rm s}}\;,
\end{align}
is given by
\begin{equation}
    J = F\sqrt{\frac{\hbar}{2\mu\omega}}\;.
    \label{eq:J_esti}
\end{equation}

\begin{figure}
    \centering
    \includegraphics[width=0.95\linewidth]{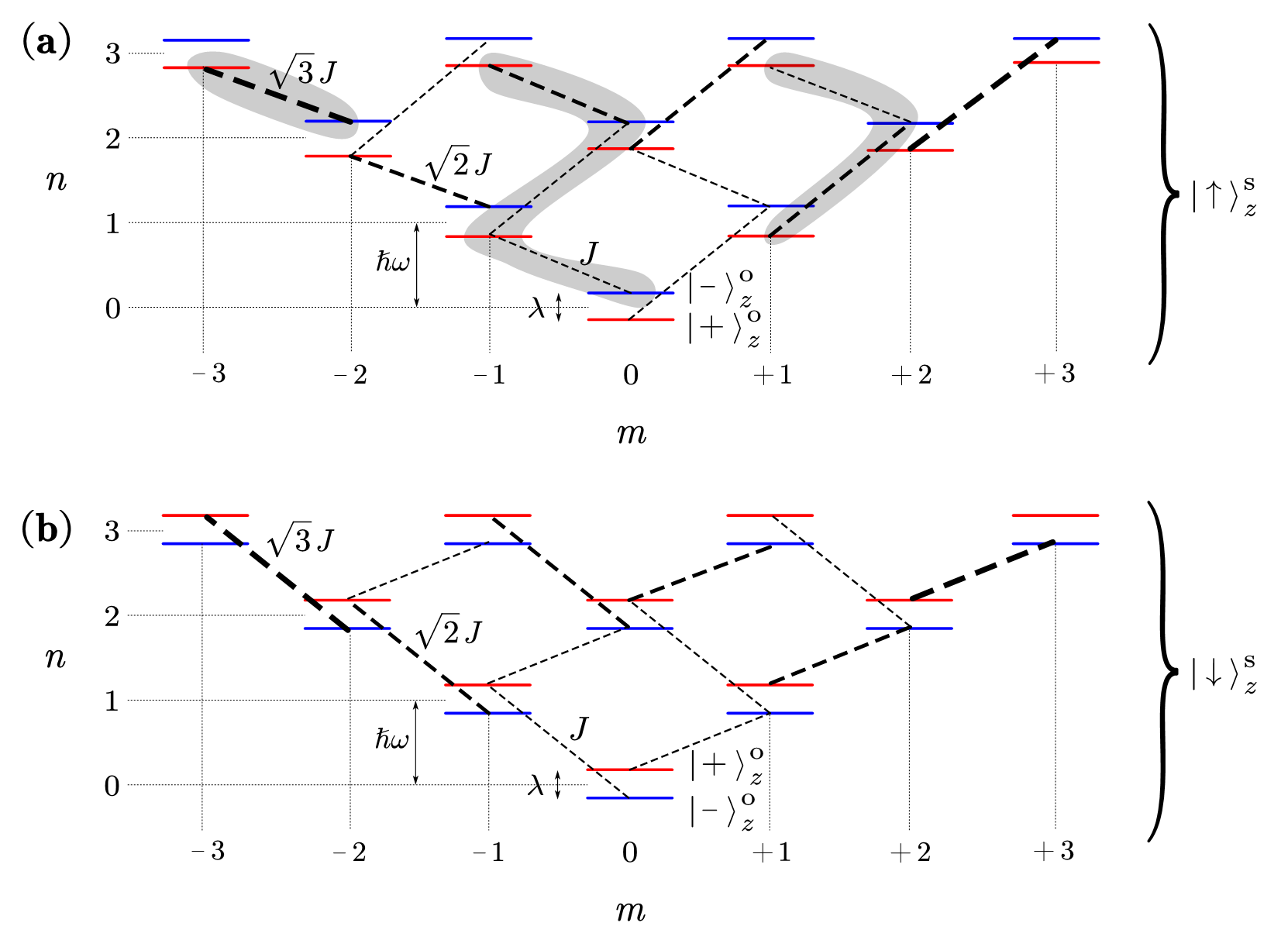}
    \caption{Interaction diagram of Hamiltonian $H_0$. It is separated into spin-up (a) and spin-down (b) subspaces. The horizontal red and blue lines represent the basis states. The tilted dashed lines connecting the horizontal lines denote interactions between the corresponding states, with their thickness representing the coupling strengths. The energy splittings and coupling strengths are also indicated. The shaded interactions indicate the smallest subspace of the XV system.}
    \label{fig:interaction_dia}
\end{figure}

\begin{figure}
    \centering
    \includegraphics[width=0.8\linewidth]{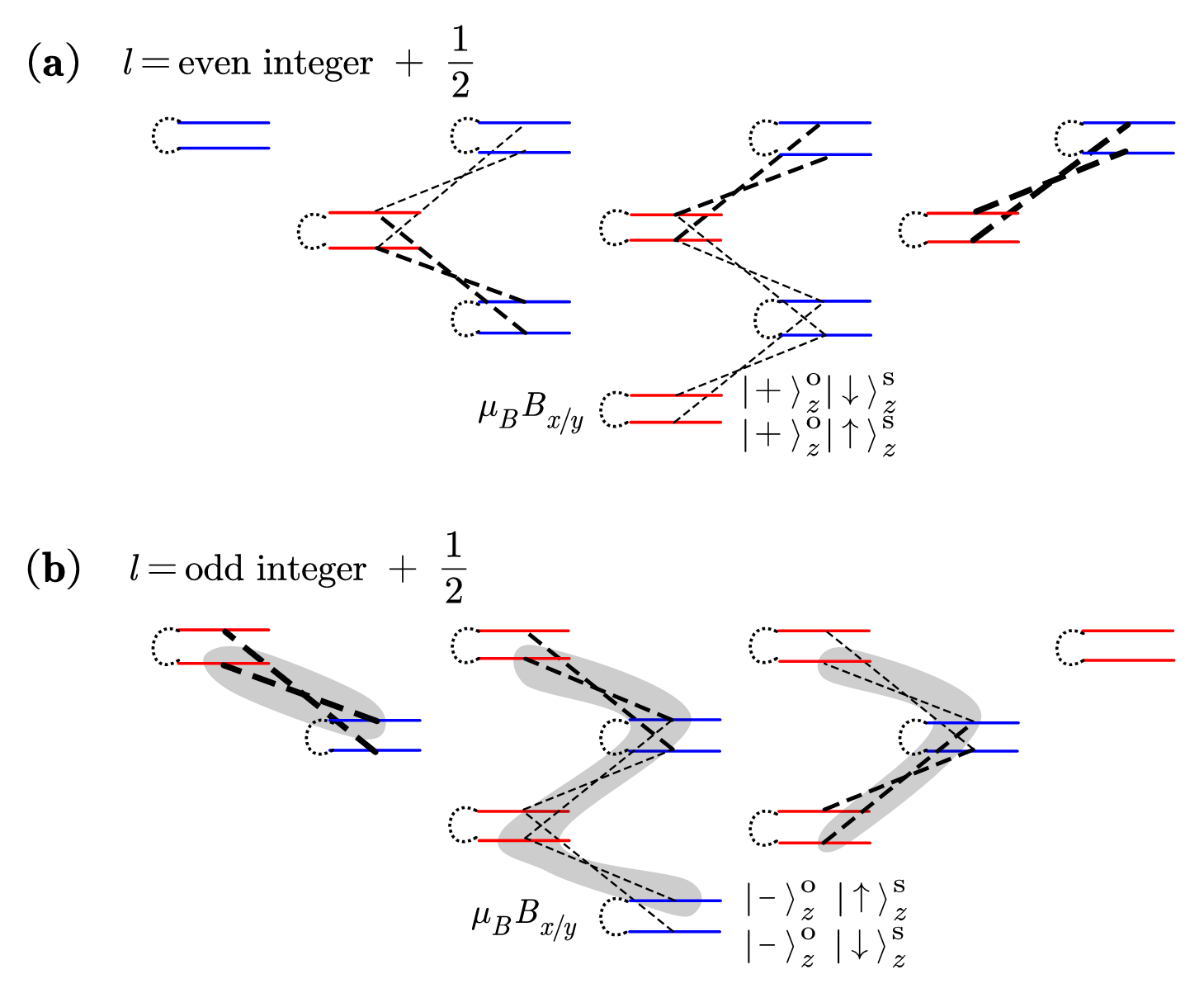}
    \caption{Interaction diagram of the Hamiltonian \( H_0 + H_{xy}^{\rm mag} \). The diagram is separated into two subspaces defined by the vibronic angular momentum number \( l \): (a) even integer \( + \frac{1}{2} \), and (b) odd integer \( - \frac{1}{2} \). The energy levels (red and blue lines) are rearranged from Fig.~\ref{fig:interaction_dia}. The c-shape dashed curve indicates the interaction induced by a magnetic field in the $x$- and/or $y$-directions, whose interaction strength is $\mu_B B_{x/y}$. The shaded levels are the subspaces of $H_0$, corresponding to the ones shown in Fig.~\ref{fig:interaction_dia}.}
    \label{fig:interaction_dia_Bx}
\end{figure}

Using these results, we derive the interaction diagram of $H_0$, as shown in Fig.~\ref{fig:interaction_dia}. From the interaction diagram, one sees that the whole Hilbert space of the XV system can be decomposed into independent subspaces, each characterized by a distinct \(z\)-component vibronic angular momentum number $l$. The Hamiltonian $H_0$ has non-zero matrix elements only between states belonging to the same subspace.

In the latter sections, we show that the XV system may also couple to lattice strain (Sec.~\ref{sec:strain}) and magnetic fields (Sec.~\ref{sec:Zeeman}). The strain interaction Hamiltonian takes the form
\begin{equation}
    H^{\rm str} \propto  \left(\cos\theta_0 \sigma^\text{o}_{z} + \sin\theta_0 \sigma^\text{o}_{x}\right),
\end{equation}
with a parameter \(\theta_0\) depending on the stress direction. The Hamiltonian commutes with the operator $F^\mathrm{vos}_{\theta \phi}(\theta=\theta_0)$ and $\tau_z^{\rm s}$, but not $L_z^{\rm vo}$.

Additionally, the interaction Hamiltonian with a magnetic field in the \(x\)- and/or \(y\)-directions takes the form
\begin{equation}
    H_{xy}^{\rm mag} \propto  \left(\cos\phi_0 \sigma^\text{s}_{x} + \sin\phi_0 \sigma^\text{s}_{y}\right),
\end{equation}
with a parameter \(\phi_0\) depending on the magnetic field direction. The Hamiltonian commutes with the operator $F^\mathrm{vos}_{\theta \phi}(\phi=\phi_0)$ and $L_z^{\rm vo}$, but not $\tau_z^{\rm s}$.

The strain interaction Hamiltonian $H^{\rm str}$ has non-zero matrix elements only between the hole orbital states $\ket{+}_z^o$ and $\ket{+}_z^o$ with the same phonon number $n$, phononic angular momentum number $m$, and hole spin. One may easily draw the interaction diagram of Hamiltonian $H_0+H^{\rm str}$ based on Fig.~\ref{fig:interaction_dia}. 

On the other hand, the interaction Hamiltonian $H_{xy}^{\rm mag}$ has non-zero matrix elements only between the spin states $\ket{\uparrow}_z^s$ and $\ket{\downarrow}_z^s$ of the same phonon number $n$,  phononic angular momentum number $m$, and hole orbital state. The cross-spin interaction in $H_{xy}^{\rm mag}$ makes the construction of the interaction diagram for the Hamiltonian $H_0+H_{xy}^{\rm mag}$ based on Fig.~\ref{fig:interaction_dia} difficult. A simple way to present it is by a rearrangement of the independent subspaces of $H_0$, grouping the states with the same hole orbital movement and different spins close to each other, as shown in Fig.~\ref{fig:interaction_dia_Bx}.

From the interaction diagrams Fig.~\ref{fig:interaction_dia} and~\ref{fig:interaction_dia_Bx}, each with two panels, we observed symmetric interaction configurations between the two panels, \textit{i.e.}, for every bare state (eigenstate of Hamiltonian $H_{\rm vib}+H_{\rm os}$, represented by the red and blue solid lines) in the upper panel, there exists a corresponding bare state in the lower panel with the same energy and the same corresponding interaction strengths. As a result, all eigenstates of the Hamiltonians $H_0$, $H_0+H^{\rm str}$, and $H_0+H_{xy}^{\rm mag}$ are two-fold degenerate. In particular, we also observe that the two degenerate states, $\ket{\psi_{l/s}}$ and $\ket{\psi'_{l/s}}$, of certain $z$-direction vibronic angular momentum numbers ($l$) or certain $z$-direction hole spin ($s$) are related to each other as
\begin{equation}
    \ket{\psi_{l/s}}=F^\mathrm{vos}_{\theta \phi}(\theta=\theta_0, \phi=\phi_0)\ket{\psi'_{l/s}}\;.
\end{equation}

\subsection{Numerical diagonalization}
\label{sec:num_diag}
\begin{figure}
    \centering
    \includegraphics[width=0.7\linewidth]{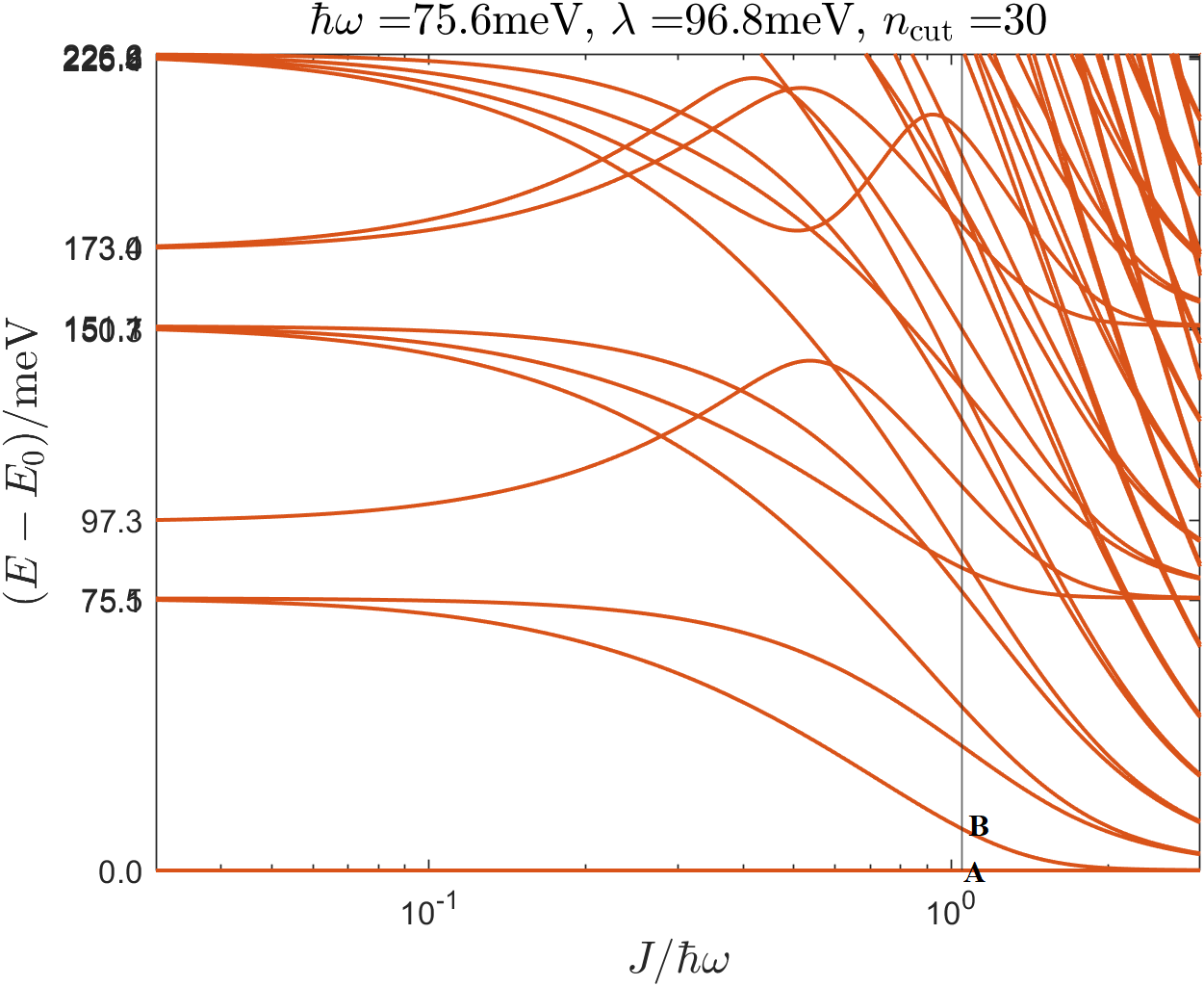}
    \caption{The relative energies ($E - E_0$) of the Hamiltonian $H_0$ as functions of the electron-phonon coupling strength $J$. The eigenenergies are from numerically diagonalizing the Hamiltonian $H_0$, where the parameters used are $\hbar\omega = 75.6~\mathrm{meV}$ and $\lambda = 96.8~\mathrm{meV}$. The vertical black line corresponds to $J = 79.3~\mathrm{meV}$. The spin-up wavefunctions corresponding to points A and B are visualized in Fig.~\ref{fig:state_visual}.}
    \label{fig:energy_lines}
\end{figure}

\begin{figure}
    \centering
    \includegraphics[width=0.9\linewidth]{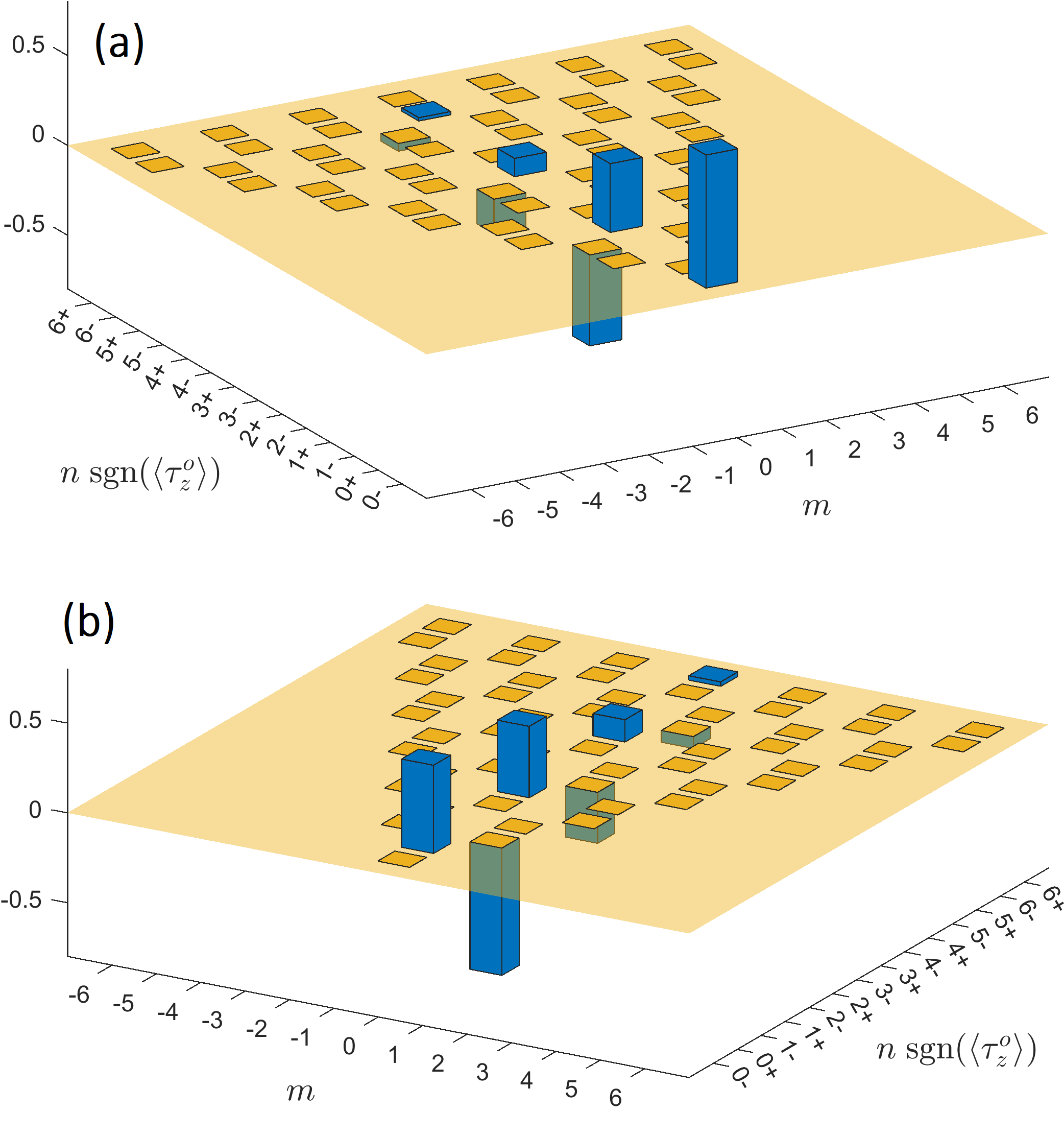}
    \caption{The probability amplitudes of the spin-up (+ state of operator $\sigma_z^s$) eigenstates corresponding to eigenenergies of the points (a) A and (b) B in Fig.~\ref{fig:energy_lines}. The arrangement of the bases follows the one in Fig.~\ref{fig:interaction_dia}(a). The probability amplitudes are from numerically diagonalizing the Hamiltonian $H_0$, where the parameters used include $J = 79.3~\mathrm{meV}$, with other values identical to those in Fig.~\ref{fig:energy_lines}.}
    \label{fig:state_visual}
\end{figure}

The Hamiltonian \( H_0 \) can be diagonalized numerically. However, this approach is limited to finite photon numbers \( n \), requiring a cut-off such that \( n \in [0, n_{\rm cut}] \). The cut-off value \( n_{\rm cut} \) must be sufficiently large to ensure the accuracy of the derived eigenstates and eigenenergies. To assess this accuracy, we examine their precision as functions of \( n_{\rm cut} \). The precision of an eigenenergy \( E \) and an eigenstate \( \ket{\psi} \) is defined as:
\begin{gather}
    \epsilon_E(n_{\rm cut}) = \frac{\left|E(n_{\rm cut}) - E(n_{\rm cut}=40)\right|}{E(n_{\rm cut}=40)}\;, \\
    \epsilon_\psi(n_{\rm cut}) = \left| \bra{\psi(n_{\rm cut}=40)}\ket{\psi(n_{\rm cut})} \right|\;.
\end{gather}
Here, the reference value \( n_{\rm cut} = 40 \) can be replaced with any sufficiently large number to ensure convergence. 

For the lowest 10 eigenenergies (each corresponding to two degenerate eigenstates), when \( \{J, \lambda\} < 2\hbar\omega \), we find that the precision satisfies \( \{\epsilon_E, \epsilon_\psi\} < 10^{-4} \; (10^{-15}) \) for \( n_{\rm cut} > 10 \; (20) \).

Figure~\ref{fig:energy_lines} shows the spectrum for the parameters $\hbar\omega = 75.6~\mathrm{meV}$ and $\lambda = 96.8~\mathrm{meV}$ as $J$ is varied from $0.03\hbar\omega$ to $3\hbar\omega$. Figure~\ref{fig:state_visual} illustrates the two spin-up eigenstates with the lowest energies for $J = 79.3~\mathrm{meV}$, while using the same parameter set.


\section{Stress Effect \label{sec:stress}}

\subsection{Stress} \label{subsec:stress}
The stress experienced by an infinitesimal rectangular prism within a solid material is represented by the stress tensor, a second-order tensor expressed as
\begin{equation}
    \left(\sigma_{ij}\right) =
    \begin{pmatrix}
        \sigma_{xx} & \sigma_{xy} & \sigma_{xz} \\
        \sigma_{yx} & \sigma_{yy} & \sigma_{yz} \\
        \sigma_{zx} & \sigma_{zy} & \sigma_{zz}
    \end{pmatrix},
\end{equation}
where each tensor component, such as \(\sigma_{xx}\), is defined as
\begin{equation}
    \sigma_{xx} = \frac{F_x}{\Delta_y \Delta_z}.
    \label{eq:stress_def}
\end{equation}
Here, \(F_x\) is the force in the \(x\)-direction acting on the \(y\)-\(z\) surface of the prism, and \(\Delta_y\) and \(\Delta_z\) are the dimensions of this surface. The other components of the tensor are defined analogously. For static stress, the stress tensor is symmetric, satisfying \(\sigma_{ij} = \sigma_{ji}\).

\begin{figure}
    \centering
    \includegraphics[width=0.5\linewidth]{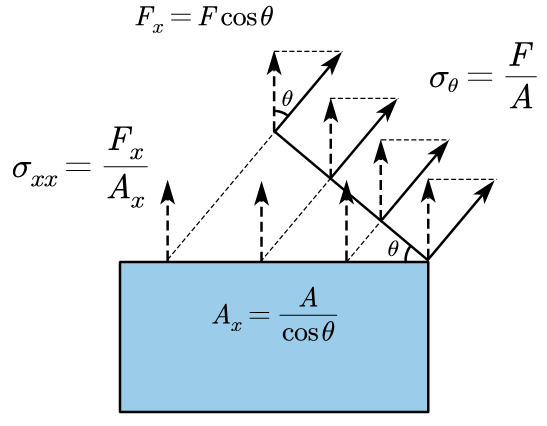}
    \caption{Projection of uniaxial stress in a 2D material onto the \(x\)-direction and the edge aligned with the \(x\)-direction of a rectangle. Both the force \(F\) and the cross-sectional length \(A\) are rescaled to account for the projection.}
    \label{fig:double_proj}
\end{figure}

Uniaxial stress is a type of stress where the force acts along a single direction, resulting in either stretching or compression along that axis. For uniaxial stress with a magnitude \(S\) applied along the direction \(\Vec{e} = \alpha_x \Vec{e}_x + \alpha_y \Vec{e}_y + \alpha_z \Vec{e}_z\), where \(\Vec{e}_i\) are unit vectors in the \(i\)-direction and \(\alpha_x^2 + \alpha_y^2 + \alpha_z^2 = 1\), the stress tensor elements are given by~\cite{hughes1967uniaxial}
\begin{equation}
    \sigma_{ij} = S \alpha_i \alpha_j.
\end{equation}
This expression follows from the definition of stress as internal pressure, where the stress \(\sigma\) is the force divided by the cross-sectional area: \(\sigma = \frac{F}{A}\). Both the force and the cross-sectional area must be properly rescaled, as illustrated in Fig.~\ref{fig:double_proj}, which provides an example for a 2D material.

\subsection{Strain \label{sec:strain}}
\begin{figure}
    \centering
    \includegraphics[width=0.7\linewidth]{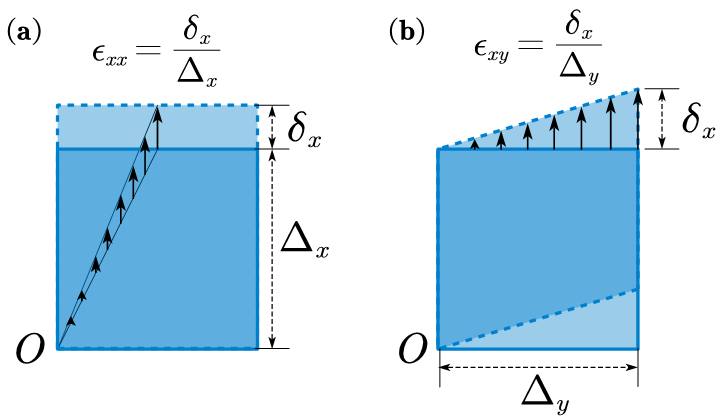}
    \caption{Deformation of a 2D square subjected to strains (a) \(\epsilon_{xx}\) and (b) \(\epsilon_{xy}\).}
    \label{fig:strain}
\end{figure}

The strain in an infinitesimal rectangular prism within a solid material is described by the strain tensor, a second-order tensor expressed as
\begin{equation}
    \left(\epsilon_{ij}\right) =
    \begin{pmatrix}
        \epsilon_{xx} & \epsilon_{xy} & \epsilon_{xz} \\
        \epsilon_{yx} & \epsilon_{yy} & \epsilon_{yz} \\
        \epsilon_{zx} & \epsilon_{zy} & \epsilon_{zz}
    \end{pmatrix}.
\end{equation}
Each tensor component is defined as
\begin{equation}
    \epsilon_{ij} = \frac{\delta_i}{\Delta_j},
    \label{eq:strain_def}
\end{equation}
where \(\delta_i\) is the \(i\)-th component of the displacement of a point accumulated along the \(j\)-th direction relative to an arbitrary reference point, and \(\Delta_j\) is the \(j\)-th component of the distance from the examined point to the reference point. Fig.~\ref{fig:strain} illustrates the deformation of a 2D square subjected to strains \(\epsilon_{xx}\) and \(\epsilon_{xy}\), respectively. Notably, for uniaxial strain, the strain tensor is symmetric, i.e., \(\epsilon_{ij} = \epsilon_{ji}\).

Based on the definition of the strain tensor elements, consider a material point initially located at \(\left(\Delta_x, \Delta_y, \Delta_z\right)^\text{T}\) that is displaced to \(\left(\Delta_x', \Delta_y', \Delta_z'\right)^\text{T}\) due to the application of a strain described by the tensor \(\left(\epsilon_{ij}\right)\). The relationship between the initial and strained positions is given by
\begin{equation}
    \begin{pmatrix}
        \Delta_x' \\
        \Delta_y' \\
        \Delta_z'
    \end{pmatrix}
    =
    \begin{pmatrix}
        \Delta_x \\
        \Delta_y \\
        \Delta_z
    \end{pmatrix}
    +
    \begin{pmatrix}
        \epsilon_{xx} & \epsilon_{xy} & \epsilon_{xz} \\
        \epsilon_{yx} & \epsilon_{yy} & \epsilon_{yz} \\
        \epsilon_{zx} & \epsilon_{zy} & \epsilon_{zz}
    \end{pmatrix}
    \begin{pmatrix}
        \Delta_x \\
        \Delta_y \\
        \Delta_z
    \end{pmatrix}.
\end{equation}

\subsection{Generalized Hooke's Law \label{sec:Hooke_s_law}}
The generalized Hooke's law describes the relationship between the stress tensor and the strain tensor:
\begin{equation}
    \left(\epsilon_{ij}\right) = \left(E_{ijkl}\right) \left(\sigma_{kl}\right),
\end{equation}
where \(\left(E_{ijkl}\right)\) is the fourth-order elasticity tensor, and the expression involves a tensor product. While \(\left(E_{ijkl}\right)\) contains 81 components, both \(\left(\epsilon_{ij}\right)\) and \(\left(\sigma_{kl}\right)\) are symmetric, i.e., \(\epsilon_{ij} = \epsilon_{ji}\) and \(\sigma_{kl} = \sigma_{lk}\), reducing each tensor to six independent terms. Consequently, \(\left(E_{ijkl}\right)\) can be represented by a \(6 \times 6\) elastic matrix \(C\). 

For cubic lattices, such as diamond, the generalized Hooke's law simplifies to:
\begin{equation}
    \begin{pmatrix}
        \epsilon_{xx} \\
        \epsilon_{yy} \\
        \epsilon_{zz} \\
        \epsilon_{xy} \\
        \epsilon_{yz} \\
        \epsilon_{zx}
    \end{pmatrix}
    =
    \begin{pmatrix}
        C_{11} & C_{12} & C_{12} & 0 & 0 & 0 \\
        C_{12} & C_{11} & C_{12} & 0 & 0 & 0 \\
        C_{12} & C_{12} & C_{11} & 0 & 0 & 0 \\
        0 & 0 & 0 & C_{44} & 0 & 0 \\
        0 & 0 & 0 & 0 & C_{44} & 0 \\
        0 & 0 & 0 & 0 & 0 & C_{44}
    \end{pmatrix}
    \begin{pmatrix}
        \sigma_{xx} \\
        \sigma_{yy} \\
        \sigma_{zz} \\
        \sigma_{xy} \\
        \sigma_{yz} \\
        \sigma_{zx}
    \end{pmatrix},
\end{equation}
where the matrix \(C\) is characterized by three independent parameters: \(C_{11}\), \(C_{12}\), and \(C_{44}\). For diamond, these parameters are \(C_{11} = 1075 \, \mathrm{GPa}\), \(C_{12} = 139 \, \mathrm{GPa}\), and \(C_{44} = 567 \, \mathrm{GPa}\).

\subsection{Strain Engineering}
\begin{figure}
    \centering
    \includegraphics[width=0.95\linewidth]{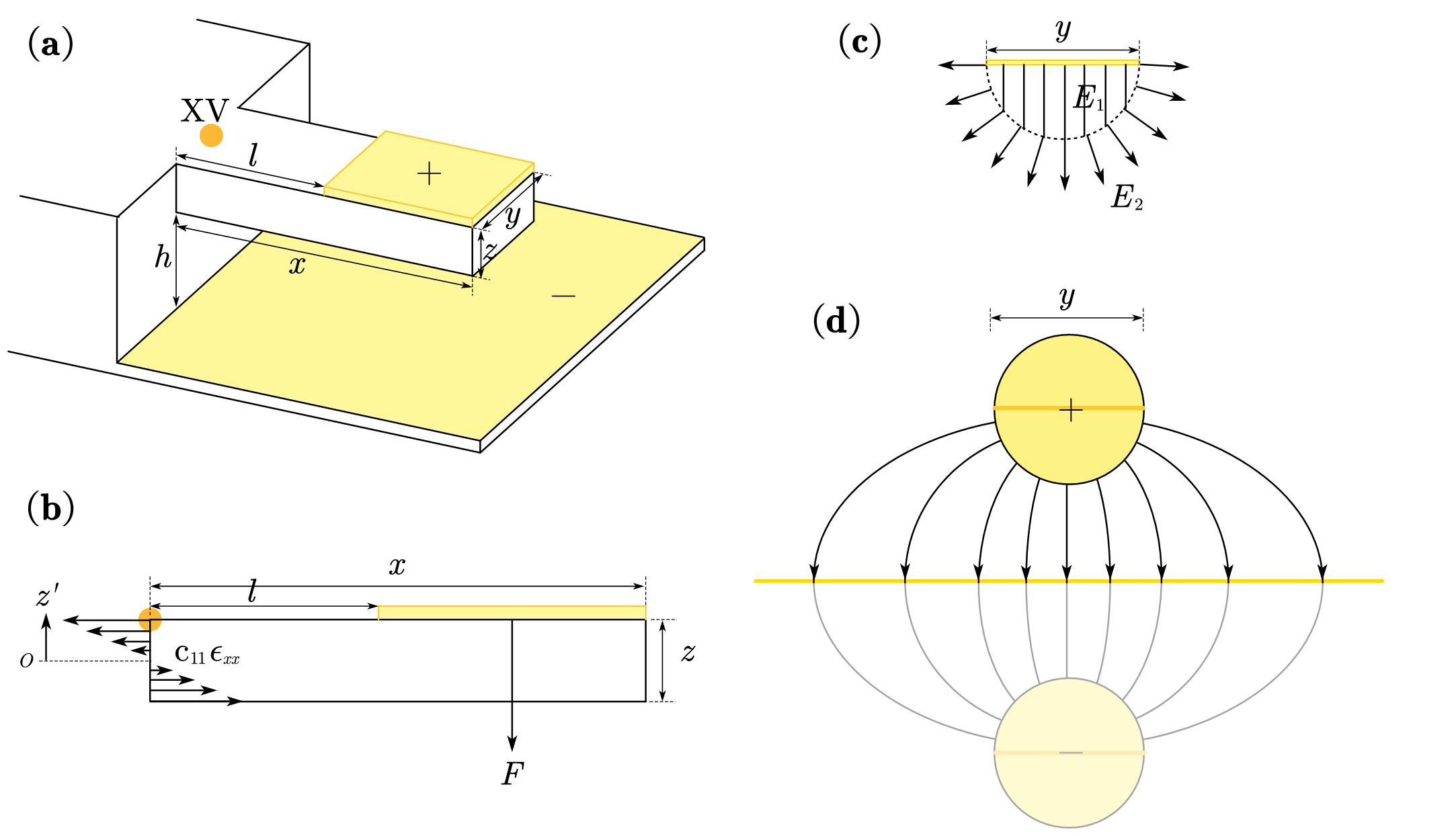}
    \caption{Artificial strain implementation and estimation. (a) A cantilever etched from a diamond substrate for use in an integrated optical system. The XV center (yellow dot) is embedded on the top surface at the joint of the cantilever. (b) Force analysis for the cantilever: stress $\mathrm{c}_{11}\epsilon_{xx}$ acts on the left $y$-$z$ surface, and an electric force $F$ acts on the top $x$-$y$ surface. (c) Approximation of the electric field near the upper electrode: inside the semi-cylinder, the field $E_1$ is uniform, while outside the semi-cylinder, the field $E_2$ resembles that produced by an evenly charged cylinder. (d) Method of images used to calculate the voltage between the upper electrode and the lower electrode plane.}
    \label{fig:artificial_strain}
\end{figure}

Natural strains can arise from various lattice defects near the XV center within the diamond bulk. Artificial strain, on the other hand, can be induced by applying engineered external forces to the diamond. One method to implement artificial strain in an integrated optical system involves using electric forces by applying a voltage to electrodes positioned on a cantilever etched from a substrate, as shown in Fig.~\ref{fig:artificial_strain}(a).

Given the size of the cantilever and electrodes shown in Fig.~\ref{fig:artificial_strain}(a), along with the voltage \(V\) applied between the electrodes on the cantilever and the substrate, we derive a theoretical formula to estimate the magnitude of the induced strain.

First, we perform a force analysis for the cantilever. As shown in Fig.~\ref{fig:artificial_strain}(b), the cantilever experiences two forces: the stress from the substrate and the electric force \(F\). We assume that the stress has a maximum value of \(\mathrm{c}_{11}\epsilon_{xx}^{\rm R}\) at the upper surface of the cantilever, decreases linearly to zero at the middle plane, reverses direction, and increases linearly to \(-\mathrm{c}_{11}\epsilon_{xx}^{\rm R}\) at the bottom surface. For static equilibrium, the torques caused by these two forces must balance, yielding:
\begin{equation}
    2\int_{0}^{\frac{Z}{2}} \mathrm{c}_{11} \epsilon_{xx} y z' \, dz' = \frac{1}{2}F\left(l + x\right),
    \label{eq:str1}
\end{equation}
where \(z'\) is the coordinate along the \(z\)-direction with the origin at the middle plane of the cantilever, and \(F\) is the electric force, assumed to act at the center of the upper electrode.

The electric force \(F\) is given by the product of the electric field strength \(E_1\) (assumed uniform near the upper electrode) and the charge \(Q\) on the upper electrode:
\begin{equation}
    F = E_1 Q.
    \label{eq:str2}
\end{equation}

Due to the high relative permittivity of diamond (\(5.5\)–\(10\)), the electric field within the diamond bulk is nearly zero. For simplicity, we consider an equivalent case where the upper electrode is placed at the bottom of the cantilever. Next, we approximate the electric field induced by the upper electrode. As shown in Fig.~\ref{fig:artificial_strain}(c), we assume the electric field inside the semi-cylinder covering the electrode is uniform, given by:
\begin{equation}
    E_1 = \frac{Q}{2\varepsilon_0 y (x - l)},
    \label{eq:str3}
\end{equation}
where \(\varepsilon_0 = 8.85 \times 10^{-12} \, \mathrm{F/m}\) is the vacuum permittivity.

For the case \(h > \frac{y}{2}\), we approximate the electric field \(E_2\) outside the semi-cylinder as that induced by a charged cylinder:
\begin{equation}
    E_2(r) = \frac{Q}{2\pi\varepsilon_0 r (x - l)}, \quad r > \frac{y}{2}.
    \label{eq:str4}
\end{equation}
Note that \(E_1\) and \(E_2\) are the electric fields induced by the upper electrode alone, without accounting for the presence of the electrode placed on the lower substrate [see Fig.~\ref{fig:artificial_strain}(a)].

Given that the electrode on the lower substrate is a large plane, the method of images can be used to calculate the voltage between the two electrodes. The main idea is that the electric field induced by the upper electrode and the substrate electrode is equivalent to the field induced by the upper electrode and an image electrode, located on the opposite side of the substrate plane and carrying an opposite charge, as shown in Fig.~\ref{fig:artificial_strain}(d). Using this approach, the voltage between the upper and substrate electrodes can be calculated as
\begin{equation}
    V = E_1 \cdot \frac{y}{2} + \int_{\frac{y}{2}}^h \big[E_2(r) + E_2(2h - r)\big] \, dr.
    \label{eq:str5}
\end{equation}

From Eqs.~(\ref{eq:str1})--(\ref{eq:str5}), the strain-voltage relationship is derived as
\begin{equation}
    \epsilon_{xx} = \frac{4\varepsilon_0}{\mathrm{c}_{11}} \frac{x^2 - l^2}{y^2 z^2 \left[\frac{1}{2} + \frac{1}{\pi} \ln{\left(\frac{4h}{y} - 1\right)}\right]^2} V^2.
    \label{eq:str_result}
\end{equation}

The above derivation assumes that the cantilever remains stationary. In reality, the cantilever bends toward the substrate electrode. For this case, \(h\) can be replaced with \(h' = h - \frac{1}{2}\epsilon_{xx}(l + x)\) in Eq.~(\ref{eq:str_result}) to obtain a more accurate relationship. However, this substitution introduces additional complexities in solving the equation.

\begin{figure}
    \centering
    \includegraphics[width=0.9\linewidth]{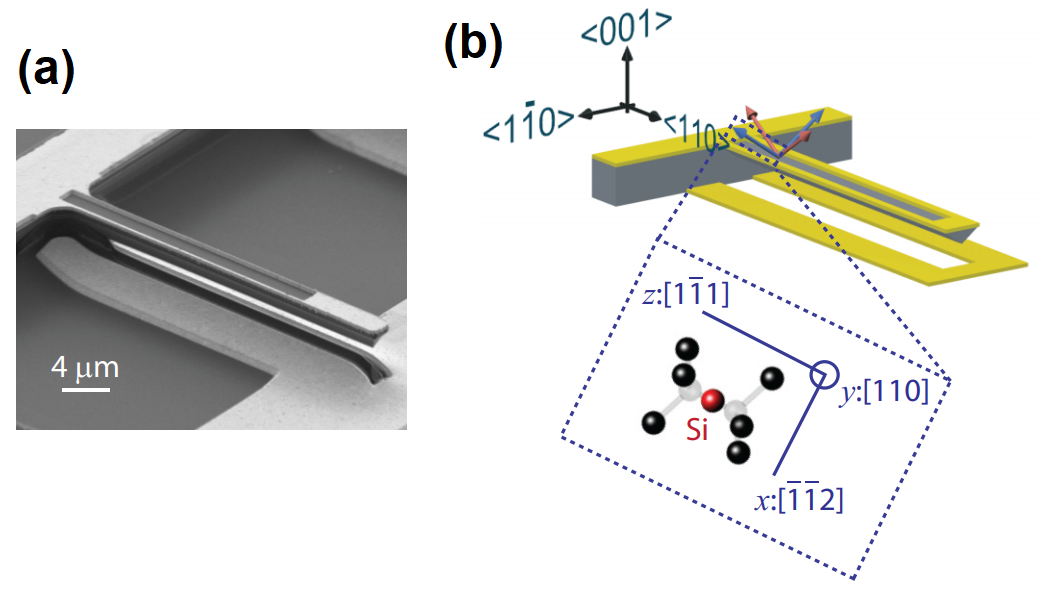}
    \caption{Triangular prism cantilever for inducing artificial strain with electric force. (a) Image of the experimental sample. (b) Model of the sample showing the diamond lattice direction and the orientation of the SiV center. The SiV center can have four possible orientations: two axial orientations (red arrows) and two transverse orientations (blue arrows). The analyzed SiV sample adopts the left transverse orientation, circled with dashed lines. Note that the \(y\)-axis of the SiV coordinate system follows the [110] crystal direction in diamond and aligns with the cantilever’s length. Adapted from Fig.~\cite{Meesala2018Strain}.}
    \label{fig:tri_prism}
\end{figure}

Meesala \textit{et al.}~\cite{Meesala2018Strain} demonstrated artificial strain experimentally using a triangular-prism cantilever, as shown in Fig.~\ref{fig:tri_prism}.

From Fig.~\ref{fig:tri_prism}(a), the triangular prism is estimated to have the dimensions listed in Tab.~\ref{tab:prism_size}.
\begin{table}[h]
    \centering
    \begin{tabular}{c|ccccc}
        \hline
        Dimension & \(x\) & \(l\) & \(y\) & \(z\) & \(h\) \\
        \hline
        Size (\(\mu\)m) & 29.0 & 22.6 & 2.0 & 1.6 & 2.4 \\
        \hline
    \end{tabular}
    \caption{Dimensions of the triangular prism cantilever and electrodes, extracted from Fig.~\ref{fig:tri_prism}.}
    \label{tab:prism_size}
\end{table}

Using the finite element method (FEM), Meesala \textit{et al.}~\cite{Meesala2018Strain} calculated the strain at the position of the SiV center near the joint of the cantilever, as shown in Fig.~\ref{fig:tri_prism}(b), under different electric voltages. Tab.~\ref{tab:Volt-Str} presents their calculated results for the strain \( |\epsilon_{xx}^{\rm M} - \epsilon_{yy}^{\rm M}| \), where
\begin{equation}
    \epsilon_{xx}^{\rm M} = -\frac{\rm c_{12}}{\rm c_{11} + c_{12}} \epsilon_{yy}^{\rm M} \approx -0.11 \epsilon_{yy}^{\rm M}
\end{equation}
in the SiV coordinate system.

Separately, the setup of Meesala \textit{et al.} can be approximated using the rectangular-prism cantilever model discussed above. Using this model and Eq.~(\ref{eq:str_result}), Table~\ref{tab:Volt-Str} also lists the calculated strain \( |\epsilon_{xx}^{\rm R} - \epsilon_{yy}^{\rm R}| \), where \( \epsilon_{yy}^{\rm R} \approx -0.11 \epsilon_{xx}^{\rm R} \). Since the \(x\)-\(y\) plane of the cantilever coordinate system aligns with that of the SiV coordinate system, the values of \( |\epsilon_{xx}^{\rm R} - \epsilon_{yy}^{\rm R}| \) and \( |\epsilon_{xx}^{\rm M} - \epsilon_{yy}^{\rm M}| \) are expected to match. As shown in Table~\ref{tab:Volt-Str}, the close agreement between the FEM results and the rectangular-prism cantilever model demonstrates the effectiveness of Eq.~(\ref{eq:str_result}) as a practical tool for guiding cantilever design.

\begin{table}[h]
    \centering
    \begin{tabular}{ccccccc}
        \hline
        Voltage (V) & 25 & 50 & 75 & 100 & 125 & 150 \\
        \hline
        \( |\epsilon_{xx}^{\rm M} - \epsilon_{yy}^{\rm M}| \) (\(10^{-4}\)) & 0.0200 & 0.0500 & 0.110 & 0.200 & 0.300 & 0.430 \\
        \( |\epsilon_{xx}^{\rm R} - \epsilon_{yy}^{\rm R}| \) (\(10^{-4}\)) & 0.0154 & 0.0615 & 0.138 & 0.246 & 0.384 & 0.553 \\
        \hline
    \end{tabular}
    \begin{tabular}{cccccccc}
        \hline
        Voltage (V) & 175 & 200 & 220 & 240 & 260 & 270 & 280 \\
        \hline
        \( |\epsilon_{xx}^{\rm M} - \epsilon_{yy}^{\rm M}| \) (\(10^{-4}\)) & 0.600 & 0.790 & 0.960 & 1.17 & 1.38 & 1.50 & 1.62 \\
        \( |\epsilon_{xx}^{\rm R} - \epsilon_{yy}^{\rm R}| \) (\(10^{-4}\)) & 0.753 & 0.984 & 1.19 & 1.42 & 1.66 & 1.79 & 1.93 \\
        \hline
    \end{tabular}
    \caption{Comparison of FEM results for \( |\epsilon_{xx}^{\rm M} - \epsilon_{yy}^{\rm M}| \) extracted from Ref.~\cite{Meesala2018Strain} and results for \( |\epsilon_{xx}^{\rm R} - \epsilon_{yy}^{\rm R}| \) calculated using Eq.~(\ref{eq:str_result}) with dimensions from Table~\ref{tab:prism_size} as inputs.}
    \label{tab:Volt-Str}
\end{table}

\begin{figure}[h]
    \centering
    \includegraphics[width=0.98\linewidth]{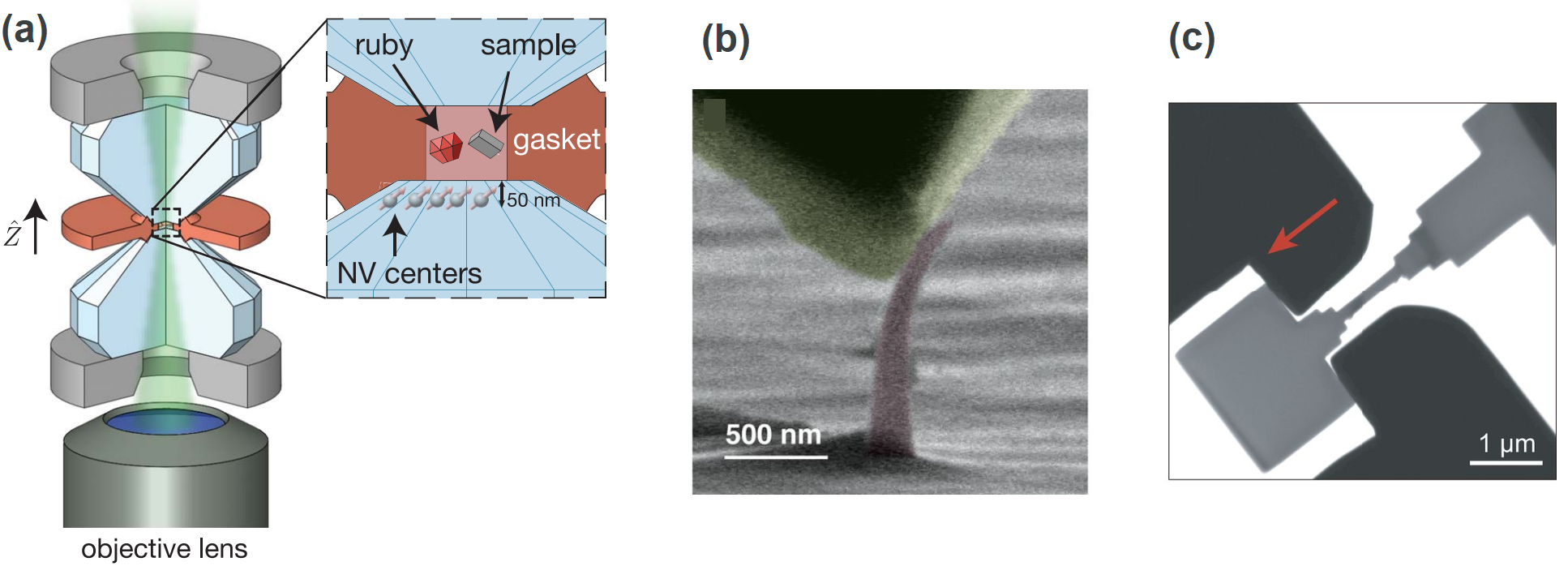}
    \caption{Methods of inducing strain in diamond using mechanical forces. (a) Compression of a diamond sample with two anvils (adapted from Ref.~\cite{anvils_hsieh2019imaging}). (b) Bending of a diamond needle with an indenter (adapted from Ref.~\cite{indenter_banerjee2018ultralarge}). (c) Stretching of a diamond bridge with a gripper (adapted from Ref.~\cite{gripper_dang2021achieving}).}
    \label{fig:mechanical_strain}
\end{figure}

Finally, it is worth noting that mechanical forces have also been employed to induce strain in diamonds. Examples include compressing a diamond sample with two anvils~\cite{anvils_hsieh2019imaging}, bending a diamond needle with an indenter~\cite{indenter_banerjee2018ultralarge}, and stretching a diamond bridge with a gripper~\cite{gripper_dang2021achieving}, as illustrated in Fig.~\ref{fig:mechanical_strain}(a), (b), and (c), respectively.

\subsection{Strain Interaction Hamiltonian}
\label{sec:strain_Hami}

In Sec.~\ref{sec:ele-vib}, we derived the electron-phonon interaction Hamiltonian, which describes the interaction between the hole orbital and the atomic displacements. The interaction between the hole orbitals and strain arises from the same mechanism, as strain can also induce atomic displacements, similar to thermal vibrations. Therefore, the strain interaction Hamiltonian can be derived by expressing the atomic displacement magnitudes \( Q_0 \), \( Q_x \), and \( Q_y \) (as in Eq.~\ref{eq:H_vo}) in terms of the strain tensor elements \(\epsilon_{k}\) (\( k \in \{zz, xx, yy, xy, yx, xz, zx, yz, zy\} \)) using the dimensions of the XV system.

\begin{figure}
    \centering
    \includegraphics[width=0.96\linewidth]{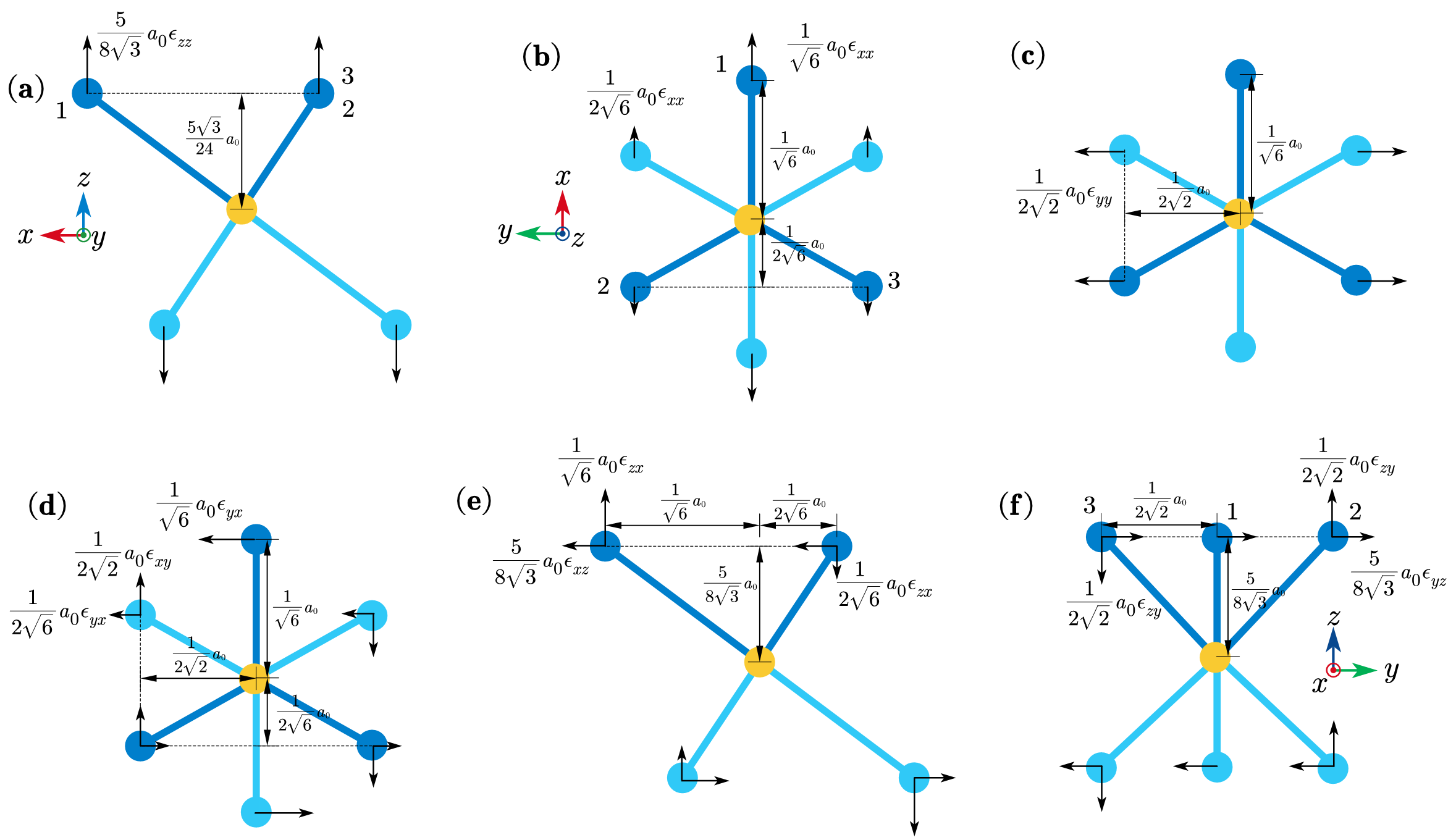}
    \caption{Directions and magnitudes of the displacements of the six carbon atomic cores under the strain components (a) \(\epsilon_{zz}\), (b) \(\epsilon_{xx}\), (c) \(\epsilon_{yy}\), (d) \(\epsilon_{xy}\) and \(\epsilon_{yx}\), (e) \(\epsilon_{xz}\) and \(\epsilon_{zx}\), and (f) \(\epsilon_{yz}\) and \(\epsilon_{zy}\).}
    \label{fig:XV_strain}
\end{figure}

As defined in Sec.~\ref{sec:strain}, each of the nine elements in the strain tensor represents a distinct type of lattice distortion. Given the dimensions of the XV system (Fig.\ref{fig:XV_size}), one can determine the atomic displacements induced by each strain component. These displacements, corresponding to the nine strain tensor elements, are illustrated in Fig.~\ref{fig:XV_strain}.

With the displacements of each atom under strain known for each strain tensor element, we can derive the generalized displacement vectors -- concatenations of the displacement vectors of multiple atoms -- for each strain tensor element. Since the displacement of the central X atom decouples from the hole orbitals, and the displacements of the lower (\(-z\) side) three carbon atoms are symmetric to the upper (\(+z\) side) three, we decompose the generalized displacement vector of \(\epsilon_{k}\) as
\begin{equation}
    \mathbf{D}_{k} = \mathbf{D}_{k}^{\rm upp} + \mathbf{D}_{k}^{\rm low},
\end{equation}
where \(\mathbf{D}_{k}^{\rm upp}\) and \(\mathbf{D}_{k}^{\rm low}\) are the generalized displacement vectors of the upper and lower three carbon atoms, respectively. The expressions for the set \(\{\mathbf{D}_{k}^{\rm upp}\}\) are given by
\begin{equation}
    \begin{split}
        \mathbf{D}_{zz}^{\rm upp} &= \frac{5}{8\sqrt{3}}a_0\epsilon_{zz}\left[ 0, 0, 1, 0, 0, 1, 0, 0, 1 \right], \\
        \mathbf{D}_{xx}^{\rm upp} &= \frac{1}{2\sqrt{6}}a_0\epsilon_{xx}\left[ 2, 0, 0, -1, 0, 0, -1, 0, 0 \right], \\
        \mathbf{D}_{yy}^{\rm upp} &= \frac{1}{2\sqrt{2}}a_0\epsilon_{yy}\left[ 0, 0, 0, 0, 1, 0, 0, -1, 0 \right], \\
        \mathbf{D}_{xy}^{\rm upp} &= \frac{1}{2\sqrt{2}}a_0\epsilon_{xy}\left[ 0, 0, 0, 1, 0, 0, -1, 0, 0 \right], \\
        \mathbf{D}_{yx}^{\rm upp} &= \frac{1}{2\sqrt{6}}a_0\epsilon_{yx}\left[ 0, 2, 0, 0, -1, 0, 0, -1, 0 \right], \\
        \mathbf{D}_{xz}^{\rm upp} &= \frac{5}{8\sqrt{3}}a_0\epsilon_{xz}\left[ 1, 0, 0, 1, 0, 0, 1, 0, 0 \right], \\
        \mathbf{D}_{zx}^{\rm upp} &= \frac{1}{2\sqrt{6}}a_0\epsilon_{zx}\left[ 0, 0, 2, 0, 0, -1, 0, 0, -1 \right], \\
        \mathbf{D}_{yz}^{\rm upp} &= \frac{5}{8\sqrt{3}}a_0\epsilon_{yz}\left[ 0, 1, 0, 0, 1, 0, 0, 1, 0 \right], \\
        \mathbf{D}_{zy}^{\rm upp} &= \frac{1}{2\sqrt{2}}a_0\epsilon_{zy}\left[ 0, 0, 0, 0, 0, 1, 0, 0, -1 \right],
    \end{split}
\end{equation}
expressed in the \(x^1, y^1, z^1, x^2, y^2, z^2, x^3, y^3, z^3\) basis, where \(x^i\), \(y^i\), and \(z^i\) are the coordinates of the \(i\)-th carbon atom in the XV Cartesian coordinate system (see Fig.~\ref{fig:ele_bases}).

On the other hand, in the same basis, the unit vectors in the three generalized directions \(\mathbf{V}_{j}\) (\( j \in \{0+, x+, y+\} \)) as in Eq.~(\ref{eq:vib_ele_coef}) can also be decomposed as
\begin{equation}
    \mathbf{V}_{j} = \mathbf{V}_{j}^{\rm upp} + \mathbf{V}_{j}^{\rm low},
\end{equation}
with \(\{\mathbf{V}_{j}^{\rm upp}\}\) given by:
\begin{equation}
    \begin{split}
        \mathbf{V}_{0+}^{\rm upp} &= -\frac{1}{\sqrt{6}}\left[\mathbf{e}_1, \mathbf{e}_2, \mathbf{e}_3\right], \\
        \mathbf{V}_{x+}^{\rm upp} &= -\frac{1}{2\sqrt{3}}\left[2\mathbf{e}_1, -\mathbf{e}_2, -\mathbf{e}_3\right], \\
        \mathbf{V}_{y+}^{\rm upp} &= -\frac{1}{2}\left[0, \mathbf{e}_2, -\mathbf{e}_3\right],
    \end{split}
\end{equation}
where
\begin{equation}
    \begin{split}
        \mathbf{e}_1 &= \frac{1}{\sqrt{57}}\left( 4\sqrt{2}, 0, 5 \right), \\
        \mathbf{e}_2 &= \frac{1}{\sqrt{57}}\left( -2\sqrt{2}, 2\sqrt{6}, 5 \right), \\
        \mathbf{e}_3 &= \frac{1}{\sqrt{57}}\left( -2\sqrt{2}, -2\sqrt{6}, 5 \right)
    \end{split}
\end{equation}
are the unit vectors in the \(z_i\) (\( i = 1, 2, 3 \)) directions, respectively.

Due to the symmetry between the displacements of the upper and lower three carbon atoms, when projecting the generalized displacement vector \(\mathbf{V}_{j}\) onto the generalized direction \(\mathbf{V}_{j}\), we can use the relationship
\begin{equation}
    \mathbf{V}_{j} \cdot \mathbf{D}_{k} = 2\mathbf{V}_{j}^{\rm upp} \cdot \mathbf{D}_{k}^{\rm upp}.
\end{equation}
The projections (inner products) are shown in Tab.~\ref{tab:inner_prod}.

\begin{table}[h]
    \centering
    \begin{tabular}{c|c|c|c|c|c|c|c|c|c}
        \hline
        $\frac{\mathbf{V}_j \cdot \mathbf{D}_k}{a_0\epsilon_k}$
        & $\mathbf{D}_{zz}$ 
        & $\mathbf{D}_{xx}$ 
        & $\mathbf{D}_{yy}$ 
        & $\mathbf{D}_{xy}$ 
        & $\mathbf{D}_{yx}$ 
        & $\mathbf{D}_{xz}$ 
        & $\mathbf{D}_{zx}$ 
        & $\mathbf{D}_{yz}$ 
        & $\mathbf{D}_{zy}$ \cr
                     \hline
        $\mathbf{V}_{0+}$ & \footnotesize $-\frac{25}{4\sqrt{114}}$ & \footnotesize $-\frac{4}{\sqrt{114}}$ & \footnotesize $-\frac{4}{\sqrt{114}}$ & 0 & 0 & 0 & 0 & 0 & 0 \cr
        \hline
        $\mathbf{V}_{x+}$ & 0 & \footnotesize $-\frac{2}{\sqrt{57}}$ & \footnotesize $\frac{2}{\sqrt{57}}$ & 0 & 0 & \footnotesize $-\frac{5}{\sqrt{114}}$ & \footnotesize $-\frac{5}{\sqrt{114}}$ & 0 & 0 \cr
        \hline
        $\mathbf{V}_{y+}$ & 0 & 0 & 0 & \footnotesize $\frac{2}{\sqrt{57}}$ & \footnotesize $\frac{2}{\sqrt{57}}$ & 0 & 0 & \footnotesize $-\frac{5}{\sqrt{114}}$ & \footnotesize $-\frac{5}{\sqrt{114}}$ \cr
        \hline
    \end{tabular}
    \caption{Inner products between the normalized vectors \(\mathbf{V}_j\) (\( j \in \{0, x, y\} \)) and vectors \(\mathbf{D}_k/(a_0\epsilon_k)\) (\( k \in \{zz, xx, yy, xy, yx, xz, zx, yz, zy\} \)).}
    \label{tab:inner_prod}
\end{table}

With the projection coefficients, and noting that \(\epsilon_{xy} = \epsilon_{yx}\), \(\epsilon_{xz} = \epsilon_{zx}\), and \(\epsilon_{yz} = \epsilon_{zy}\), we can express the atomic displacement magnitudes \( Q_0 \), \( Q_x \), and \( Q_y \) in terms of the strain tensor elements \(\epsilon_{k}\) as
\begin{equation}
    \begin{split}
        Q_0 &= -\frac{25}{4\sqrt{114}}a_0 \epsilon_{zz} - \frac{4}{\sqrt{114}}a_0 \left(\epsilon_{xx} + \epsilon_{yy}\right), \\
        Q_x &= -\frac{2}{\sqrt{57}}a_0 \left(\epsilon_{xx} - \epsilon_{yy}\right) - \frac{10}{\sqrt{114}}a_0\epsilon_{xz}, \\
        Q_y &= \frac{4}{\sqrt{57}}\epsilon_{xy} - \frac{10}{\sqrt{114}}a_0\epsilon_{yz}.
    \end{split}
\end{equation}
Substituting these equations into Eq.~(\ref{eq:H_vo}), we obtain the strain interaction Hamiltonian as
\begin{equation}
    \begin{split}
        H^{\text{str}} =& \left[t_{\parallel} \epsilon_{zz} + t_{\perp} \left(\epsilon_{xx} + \epsilon_{yy}\right)\right] I^{\text{o}} - \left[d \left(\epsilon_{xx} - \epsilon_{yy}\right) + f\epsilon_{xz}\right] \sigma_z^{\text{o}} \\
        &+ \left(-2d \epsilon_{xy} + f\epsilon_{yz}\right) \sigma_x^{\text{o}},
    \end{split}
    \label{eq:H_strain}
\end{equation}
where
\begin{equation}
    \begin{split}
        t_{\parallel} &= -\frac{25}{4\sqrt{114}}F_0a_0, \\
        t_{\perp} &= -\frac{4}{\sqrt{114}}F_0a_0, \\
        d &= \frac{2}{\sqrt{57}}Fa_0, \\
        f &= \frac{10}{\sqrt{114}}Fa_0,
    \end{split}
    \label{eq:d_F_relation}
\end{equation}
are the electron-strain coupling strengths or strain susceptibilities.

The ratio between \( d \) and \( f \) is given by
\begin{equation}
    \frac{d}{f} = \frac{\sqrt{2}}{5} = 0.28.
    \label{eq:ratio}
\end{equation}

\subsection{Stress-Strain-Spectrum Converter}
With the Hamiltonian~(\ref{eq:H_strain}) describing the strain effect, the energy shifts for the states in both the ground and excited manifolds can be calculated. Furthermore, a transition from a state in the excited manifold to a state in the ground manifold is accompanied by an emitted photon. The average energy difference of the emitted photons, corresponding to the zero-phonon line (ZPL), is given by:
\begin{equation}
    E_{\rm ZPL} = \frac{1}{2} \left(E_{\rm C} + E_{\rm A} - E_{\rm 3} - E_{\rm 1}\right),
\end{equation}
where \(E_{i}\) is the eigenenergy of the \(\ket{i}\) state, \(\ket{A}\) and \(\ket{C}\) are the first and second spin-up states in the excited manifold, and \(\ket{1}\) and \(\ket{3}\) are the first and second spin-up states in the ground manifold. The ZPL of an unstrained sample can thus be measured to determine this energy difference. Using this information, along with the formulas in Secs.~\ref{subsec:stress}--\ref{sec:strain_Hami}, the spectrum of the XV emitter under arbitrary stress can be computed.

\begin{figure}[H]
    \centering
    \includegraphics[width=0.98\linewidth]{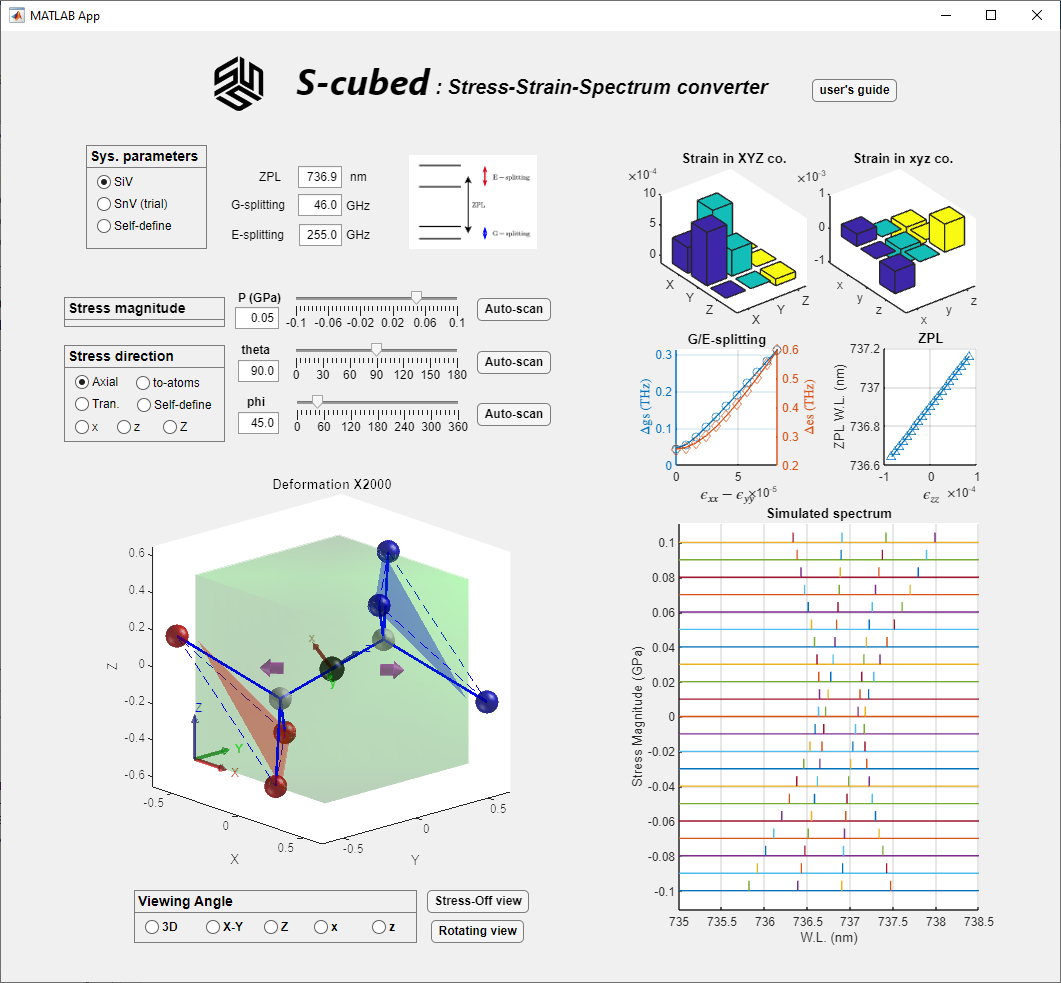}
    \caption{User interface of S-cubed, the stress-strain-spectrum converter. The system parameter input panel (top left) accepts the ZPL wavelength and the splittings of the lowest two spin-up states in the ground and first-excited manifolds under zero stress. The stress input panel (middle left) allows users to specify the magnitude and direction of the stress. The XV atomic system deformation visualization panel (bottom left) displays the structural response. The panels on the right show the strain tensor in the diamond and XV coordinate systems (top right), the splitting changes in the ground and excited manifolds and the ZPL wavelength (middle right), and the spectrum evolution (bottom right). The spectrum calculations are based on the quench-factor model [see Eq.~(\ref{eq:H_tot_Ham})] but can be updated with the phonon-electron interaction model [see Eq.~(\ref{eq:H_0})] in the future. The design of the right-side panels is inspired by Meesala \textit{et al.}~\cite{Meesala2018Strain}. The MATLAB code for this application is available in the 4TU.ResearchData repository~\cite{gu2024scubed}.}
    \label{fig:UI_S_cubed}
\end{figure}

To facilitate these calculations, we developed a user-friendly calculator, S-cubed, which performs stress-strain-spectrum conversions, as shown in Fig.~\ref{fig:UI_S_cubed}. This MATLAB-based tool features a convenient user interface (UI) that allows users to input the ZPL wavelength, the splittings of the lowest two spin-up states in the ground and excited manifolds under zero stress, and the magnitude and direction of uniaxial stress. The tool provides the following outputs:
\begin{enumerate}
    \item Visualization of the deformation of the XV atomic system.
    \item Strain tensor in both the diamond and XV coordinate systems.
    \item Variation of the splitting of the two lowest spin-up states in the ground and first-excited manifolds with respect to strain magnitude.
    \item Change in the ZPL wavelength as a function of strain magnitude.
    \item Evolution of the spectrum as a function of stress magnitude.
\end{enumerate}

\newpage

\section{Stark Effect and Light Emission}
\label{sec:stark}

\subsection{Electric Dipole Moment}
In classical physics, the electric potential energy of a particle with charge $Q_{\rm p}$ in a uniform electric field $\mathbf{E}$ is given by
\begin{equation}
    U = Q_{\rm p} \mathbf{r} \cdot \mathbf{E}\;,
\end{equation}
where $\mathbf{r}$ is the displacement of the particle from a reference point with zero potential energy. This leads to the definition of the electric dipole moment as
\begin{equation}
    \mathbf{d} = Q_{\rm p} \mathbf{r}\;.
\end{equation}
In quantum mechanics, the physical quantity $\mathbf{r}$ is represented as an operator $\hat{\mathbf{r}}$. In the basis set $\{\ket{\psi_i}\}$, the dipole moment operator $\hat{\mathbf{d}}$ is represented by a matrix
\begin{equation}
\begin{split}
    \hat{\mathbf{d}} &= Q_{\rm p} \hat{I} \, \hat{\mathbf{r}} \, \hat{I}, \\
    &= Q_{\rm p} \left[ \bra{\psi_i} \hat{\mathbf{r}} \ket{\psi_j} \right],
\end{split}
\end{equation}
where $\hat{I} = \sum_i \ket{\psi_i} \bra{\psi_i}$ is the identity operator in the basis $\{\ket{\psi_i}\}$ and $\left[ \bra{\psi_i} \hat{\mathbf{r}} \ket{\psi_j} \right]$ is a matrix with elements $ \bra{\psi_i} \hat{\mathbf{r}} \ket{\psi_j} $. When the basis states $\{\ket{\psi_i}\}$ possess inversion symmetry, the diagonal elements $\{\bra{\psi_i} \hat{\mathbf{r}} \ket{\psi_i}\}$ are all zero since the position operator is odd under inversion symmetry. Consequently, the external electric field $\mathbf{E}$ only induces transitions between different states of $\{\ket{\psi_i}\}$.

\subsection{Robustness Against Charge Noise}
We can estimate the electric dipole moment based on the eigenstates of the hole orbitals in the XV Coulomb potential, Eq.~(\ref{eq:ele_eigen_xy}), and the size of the XV system as shown in Fig.~\ref{fig:XV_size}. 

Given the estimated geometry of the X-C bond, the expected position of the hole in an X-C bond state $\ket{\sigma_i}$ ($i=1, 2, \ldots, 6$) is approximately at the center of the X-C bond. Based on the dimensions of the XV system (see Fig.~\ref{fig:XV_size}), the operators $\hat{x}^{\rm o}$, $\hat{y}^{\rm o}$, and $\hat{z}^{\rm o}$ in the X-C bond basis $\{\ket{\sigma_i}\}$ are given by
\begin{equation}
    \begin{split}
        \hat{x}^{\rm o} &= \frac{a_0}{4 \sqrt{6}} \mathrm{diag}\left(2, -1, -1, -2, 1, 1\right), \\
        \hat{y}^{\rm o} &= \frac{a_0}{4 \sqrt{2}} \mathrm{diag}\left(0, -1, 1, 0, 1, -1\right), \\
        \hat{z}^{\rm o} &= \frac{5 \sqrt{3} a_0}{48} \mathrm{diag}\left(1, 1, 1, -1, -1, -1\right).
    \end{split}
\end{equation}

Using the expressions in Eq.~(\ref{eq:ele_eigen_xy}), we calculate the electric dipole moments in the subspace $\{\psi_x, \psi_y\}$. However, all matrix elements vanish:
\begin{equation}
    \bra{\psi_x}\hat{x}^{\rm o}\ket{\psi_y} = \bra{\psi_x}\hat{y}^{\rm o}\ket{\psi_y} = \bra{\psi_x}\hat{z}^{\rm o}\ket{\psi_y} = 0\;.
\end{equation}
The vanishing inner products arise from the parity properties of $\ket{\psi_x}$, $\ket{\psi_y}$, and the operators $\hat{x}^{\rm o}$, $\hat{y}^{\rm o}$, and $\hat{z}^{\rm o}$ as functions of spatial coordinates. The state $\ket{\psi_x}$ ($\ket{\psi_y}$) is odd under $\hat{x}^{\rm o}$ ($\hat{y}^{\rm o}$) and even under $\hat{y}^{\rm o}$ ($\hat{x}^{\rm o}$). Both $\ket{\psi_x}$ and $\ket{\psi_y}$ share the same parity with respect to $\hat{z}^{\rm o}$, being odd (even) in the ground (first-excited) subspace. The operators $\hat{x}^{\rm o}$, $\hat{y}^{\rm o}$, and $\hat{z}^{\rm o}$ are odd functions of their respective coordinates. For example, in the first integral above, $\ket{\psi_x}\hat{x}^{\rm o}\ket{\psi_y}$ is even in $\hat{x}^{\rm o}$ but odd in both $\hat{y}^{\rm o}$ and $\hat{z}^{\rm o}$, making the integral zero. Similar reasoning applies to the other two integrals. Thus, the electric dipole moment in this subspace is zero.

Experimental observations confirm that the electric dipole moment is at least four orders of magnitude smaller than that of the nitrogen-vacancy (NV) center in diamond~\cite{Stark_de2021investigation}. This characteristic has both advantages and disadvantages. On the one hand, the reduced dipole moment narrows the linewidths of energy levels, making them less susceptible to charge noise. On the other hand, it limits the feasibility of using electric fields to tune energy level positions.

\subsection{Stark Shift}
\label{sec:Stark_shift}

Although the electric dipole moments in both the ground and the first-excited subspaces are zero, the electric dipole moment in the space combined with these two is not zero. The off-diagonal block matrix of the three coordinates, mapping from the space spanned by $\{\psi_{x-}, \psi_{y-}\}$ to the space spanned by $\{\psi_{x+}, \psi_{y+}\}$, is given by
\begin{equation}
\begin{split}
    \hat{x}^{\rm o} &= \frac{a_0}{4\sqrt{6}}
    \bordermatrix{
    ~ & \ket{\psi_{x-}} & \ket{\psi_{y-}} \cr
    \bra{\psi_{x+}} & 1 & 0 \cr
    \bra{\psi_{y+}} & 0 & -1 \cr
    }\;, \\
    \hat{y}^{\rm o} &= \frac{a_0}{4\sqrt{6}}
    \bordermatrix{
    ~ & \ket{\psi_{x-}} & \ket{\psi_{y-}} \cr
    \bra{\psi_{x+}} & 0 & 1 \cr
    \bra{\psi_{y+}} & 1 & 0 \cr
    }\;, \\
    \hat{z}^{\rm o} &= \frac{5a_0}{16\sqrt{3}}
    \bordermatrix{
    ~ & \ket{\psi_{x-}} & \ket{\psi_{y-}} \cr
    \bra{\psi_{x+}} & 1 & 0 \cr
    \bra{\psi_{y+}} & 0 & 1 \cr
    }\;.
\end{split}
\end{equation}
Hence the full matrix in the combined space spanned by $\{\psi_{x-}, \psi_{y-}, \psi_{x+}, \psi_{y+}\}$ is given by
\begin{equation}
    \begin{split}
        \hat{x}^{\rm o} &= \frac{a_0}{4\sqrt{6}}
        \begin{bmatrix}
            \mathbf{0} & \tau_x^{\rm o} \\
            \tau_x^{\rm o} & \mathbf{0}
        \end{bmatrix}\;, \\
        \hat{y}^{\rm o} &= \frac{a_0}{4\sqrt{6}}
        \begin{bmatrix}
            \mathbf{0} & \tau_y^{\rm o} \\
            \tau_y^{\rm o} & \mathbf{0}
        \end{bmatrix}\;, \\
        \hat{z}^{\rm o} &= \frac{5a_0}{16\sqrt{3}}
        \begin{bmatrix}
            \mathbf{0} & I^{\rm o} \\
            I^{\rm o} & \mathbf{0}
        \end{bmatrix}\;, \\
    \end{split}
\end{equation}
where $\textbf{0}$ is the zero matrix and $I^{\rm 0}$ is the identity matrix. 

Now we define
\begin{equation}
    \hat{d}_\pm = \frac{1}{2}e(\hat{x}^{\rm o} \pm \text{i}\hat{y}^{\rm o})
\end{equation}
then the there components of the electric dipole moment can be presented as
\begin{equation}
    \begin{split}
        \hat{d}_+ &= \frac{ea_0}{4\sqrt{6}}
        \begin{bmatrix}
            \mathbf{0} & \tau_+^{\rm o} \\
            \tau_+^{\rm o} & \mathbf{0}
        \end{bmatrix}\;, \\
        \hat{d}_- &= \frac{ea_0}{4\sqrt{6}}
        \begin{bmatrix}
            \mathbf{0} & \tau_-^{\rm o} \\
            \tau_-^{\rm o} & \mathbf{0}
        \end{bmatrix}\;, \\
        \hat{d}_z &= \frac{5ea_0}{16\sqrt{3}}
        \begin{bmatrix}
            \mathbf{0} & I^{\rm o} \\
            I^{\rm o} & \mathbf{0}
        \end{bmatrix}\;, \\
    \end{split}
\end{equation}
where $\tau_+^{\rm o}$ are the raise and lower operators for the electronic orbital states.

Lastly, we must calculate the electric dipole moment in the bases of the eigenstates of the intrinsic Hamiltonian $H_0$. We do this numerically by adopting the SnV-G and SnV-E parameters based on the \textit{ab initio} calculation~\cite{thiering2018ab} (see Tab.~\ref{tab:Thiering_para_Ham_fac}). The block matrix mapping from the space spanned by the lowest eight spin-up states $\{\ket{\rm A}, \ket{\rm C}, \ket{\rm E}, \cdots, \ket{\rm O}\}$ in the first-excited subspace to the space spanned by the lowest eight spin-up states $\{\ket{1}, \ket{3}, \ket{5}, \cdots, \ket{15}\}$ in the ground subspace is given by
\begin{equation}
\begin{split}
    \hat{d}_+=&1.75 \mathrm{D} \, \cdot \\
    &\begin{scriptsize}
    \bordermatrix{
    ~ & \ket{\rm A}(-\frac{1}{2}) & \ket{\rm C}(\frac{1}{2}) & \ket{\rm E}(-1\frac{1}{2}) & \ket{\rm G}(1\frac{1}{2}) & \ket{\rm I}(-2\frac{1}{2}) & \ket{\rm K}(-\frac{1}{2}) & \ket{\rm M}(2\frac{1}{2}) & \ket{\rm O}(\frac{1}{2}) \cr
    \bra{1}(-\frac{1}{2}) & 0 & 0 & -0.379 & 0 & 0 & 0 & 0 & 0 \cr
    \bra{3}(\frac{1}{2}) & -0.680 & 0 & 0 & 0 & 0 & -0.302 & 0 & 0 \cr
    \bra{5}(-1\frac{1}{2}) & 0 & 0 & 0 & 0 & -0.415 & 0 & 0 & 0 \cr
    \bra{7}(1\frac{1}{2}) & 0 & -0.613 & 0 & 0 & 0 & 0 & 0 & -0.0739 \cr
    \bra{9}(-\frac{1}{2}) & 0 & 0 & 0.499 & 0 & 0 & 0 & 0 & 0 \cr
    \bra{11}(\frac{1}{2}) & -0.127 & 0 & 0 & 0 & 0 & -0.629 & 0 & 0 \cr
    \bra{13}(-2\frac{1}{2}) & 0 & 0 & 0 & 0 & 0 & 0 & 0 & 0 \cr
    \bra{15}(2\frac{1}{2}) & 0 & 0 & 0 & -0.590 & 0 & 0 & 0 & 0 \cr
    } 
    \end{scriptsize}\\
\end{split}\;,
\label{eq:diple_dx}
\end{equation}

\begin{equation}
\begin{split}
    \hat{d}_-=&1.75 \mathrm{D} \, \cdot \\
    &\begin{scriptsize}
    \bordermatrix{
    ~ & \ket{\rm A}(-\frac{1}{2}) & \ket{\rm C}(\frac{1}{2}) & \ket{\rm E}(-1\frac{1}{2}) & \ket{\rm G}(1\frac{1}{2}) & \ket{\rm I}(-2\frac{1}{2}) & \ket{\rm K}(-\frac{1}{2}) & \ket{\rm M}(2\frac{1}{2}) & \ket{\rm O}(\frac{1}{2}) \cr
    \bra{1}(-\frac{1}{2}) & 0 & -0.488 & 0 & 0 & 0 & 0 & 0 & -0.286 \cr
    \bra{3}(\frac{1}{2}) & 0 & 0 & 0 & -0.315 & 0 & 0 & 0 & 0 \cr
    \bra{5}(-1\frac{1}{2}) & -0.448 & 0 & 0 & 0 & 0 & 0.116 & 0 & 0 \cr
    \bra{7}(1\frac{1}{2}) & 0 & 0 & 0 & 0 & 0 & 0 & -0.353 & 0 \cr
    \bra{9}(-\frac{1}{2}) & 0 & 0.0487 & 0 & 0 & 0 & 0 & 0 & -0.436 \cr
    \bra{11}(\frac{1}{2}) & 0 & 0 & 0 & 0.409 & 0 & 0 & 0 & 0 \cr
    \bra{13}(-2\frac{1}{2}) & 0 & 0 & -0.465 & 0 & 0 & 0 & 0 & 0 \cr
    \bra{15}(2\frac{1}{2}) & 0 & 0 & 0 & 0 & 0 & 0 & 0 & 0 \cr
    }
    \end{scriptsize}
\end{split}\;,
\label{eq:diple_dy}
\end{equation}

\begin{equation}
\begin{split}
    \hat{d}_z=&3.09 \mathrm{D} \, \cdot \\
    &\begin{scriptsize}
    \bordermatrix{
    ~ & \ket{\rm A}(-\frac{1}{2}) & \ket{\rm C}(\frac{1}{2}) & \ket{\rm E}(-1\frac{1}{2}) & \ket{\rm G}(1\frac{1}{2}) & \ket{\rm I}(-2\frac{1}{2}) & \ket{\rm K}(-\frac{1}{2}) & \ket{\rm M}(2\frac{1}{2}) & \ket{\rm O}(\frac{1}{2}) \cr
    \bra{1}(-\frac{1}{2}) & 0.958 & 0 & 0 & 0 & 0 & 0.281 & 0 & 0 \cr
    \bra{3}(\frac{1}{2}) & 0 & 0.877 & 0 & 0 & 0 & 0 & 0 & 0.361 \cr
    \bra{5}(-1\frac{1}{2}) & 0 & 0 & 0.966 & 0 & 0 & 0 & 0 & 0 \cr
    \bra{7}(1\frac{1}{2}) & 0 & 0 & 0 & 0.922 & 0 & 0 & 0 & 0 \cr
    \bra{9}(-\frac{1}{2}) & -0.250 & 0 & 0 & 0 & 0 & 0.820 & 0 & 0 \cr
    \bra{11}(\frac{1}{2}) & 0 & -0.470 & 0 & 0 & 0 & 0 & 0 & 0.546 \cr
    \bra{13}(-2\frac{1}{2}) & 0 & 0 & 0 & 0 & 0.969 & 0 & 0 & 0 \cr
    \bra{15}(2\frac{1}{2}) & 0 & 0 & 0 & 0 & 0 & 0 & 0.939 & 0 \cr
    } 
    \end{scriptsize}\\
\end{split}\;.
\label{eq:diple_dz}
\end{equation}
Here, the symbol D denotes the unit of the Debye (1 D $= 3.336 \times 10^{-30}$ C$\cdot$m). The fraction in parentheses beside the basis states on the border of the matrix represents the conserved quantity $\left\langle L_z^{\rm vo}\right\rangle / \hbar$ (see Sec.~\ref{sec:cons_quan}). It is evident that the matrix element $\hat{d}_\pm[i, j]$ ($\hat{d}_z[i, j]$) is nonzero only when the transition satisfies $\Delta\left\langle L_z^{\rm vo}\right\rangle / \hbar = \pm 1$ (0).

Using the nonzero matrix elements of the electric dipole moment, we estimate the Stark shifts induced by a static electric field $\mathbf{E}$. According to perturbation theory~\cite{SOC_griffiths2018introduction}, the second-order energy shift (the first-order contribution, corresponding to the diagonal elements, vanishes) for a SnV system is given by
\begin{equation}
    \Delta E_i \approx \sum_{f \ne i} \frac{\left|\bra{f}\mathbf{d} \cdot \mathbf{E} \ket{i}\right|^2}{E_i - E_f},
\end{equation}
where $E_i$ and $E_f$ are the energies of the initial and final states, respectively.

\begin{figure}
    \centering
    \includegraphics[width=0.8\linewidth]{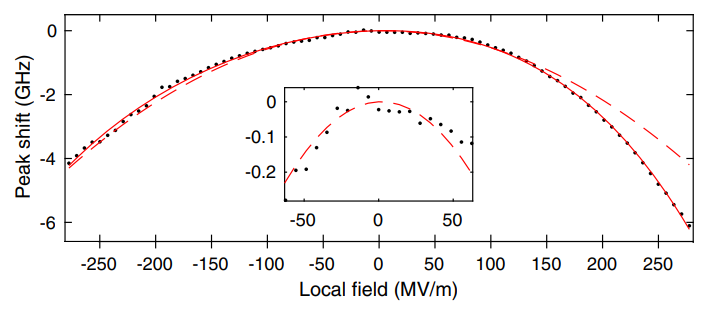}
    \caption{Measured Stark shift of the SnV color center. The $y$ ($x$) axis represents $\Delta_{\rm spl}$ ($E$) as defined in this work. The black dots denote experimental data, while the dashed and solid red lines correspond to fits using second- and fourth-order polynomials, respectively. The inset provides a detailed view of the shift at small applied fields, illustrating its quadratic behavior. Adapted from Ref.~\cite{Stark_de2021investigation}.}
    \label{fig:stark_shift}
\end{figure}

The change in the splitting between $E_{\rm A}$ and $E_1$ is expressed as 
\begin{equation}
    \Delta_{\rm spl} = \Delta E_{\rm A} - \Delta E_1 \;.
\end{equation}
For both $\Delta E_{\rm A}$ and $\Delta E_1$, we consider the two largest terms in their respective expansions, yielding
\begin{equation}
\begin{split}
    \Delta E_{\rm A} &\approx \frac{\left|\bra{1}\hat{d}_z\ket{\rm A}\right|^2 E^2 \cos^2 \theta_z}{E_{\rm A} - E_1} + \frac{\left|\bra{3}\hat{d}_+\ket{\rm A}\right|^2 E^2 \sin^2 \theta_z}{E_{\rm A} - E_3}, \\
    \Delta E_1 &\approx \frac{\left|\bra{\rm A}\hat{d}_z\ket{1}\right|^2 E^2 \cos^2 \theta_z}{E_1 - E_{\rm A}} + \frac{\left|\bra{\rm B}\hat{d}_-\ket{1}\right|^2 E^2 \sin^2 \theta_z}{E_1 - E_{\rm B}}.
\end{split}
\end{equation}
Here, $\theta_z$ is the angle between the applied electric field $\mathbf{E}$ and the $z$ axis of the XV system coordinate, and $E = |\mathbf{E}|$. Substituting the values $E_{\rm A} - E_1 = 484.1 \, \mathrm{THz}$ (corresponding to 619.3 nm), $E_{\rm A} - E_3 = 483.2 \, \mathrm{THz}$, and $E_{\rm B} - E_1 = 486.9 \, \mathrm{THz}$ into the equations, we obtain
\begin{equation}
    \Delta_{\rm spl} \in [1.1, 9.1] \times 10^{-16} E^2 \; \mathrm{GHz},
\end{equation}
where the minimum (maximum) value corresponds to $\theta_z = \frac{\pi}{2}$ ($0$). This result agrees with experimental observations~\cite{Stark_de2021investigation, linear_shift_aghaeimeibodi2021electrical}, as shown in Fig.~\ref{fig:stark_shift}.

\subsection{Light Emission}
\label{sec:light_emission}

The nonzero elements of the electric dipole moment indicate that the XVs can couple to photons, making them photon-addressable quantum devices. Different spatial components of the electric dipole moment couple to photons with specific polarizations. Specifically, the $\hat{d}_z$ component couples to photons propagating in the $x$-$y$ plane with an electric field component along the $z$ axis. In contrast, the $\hat{d}_{+(-)}$ components couple to right (left) circularly polarized photons propagating along the $z$ axis, with their electric fields oscillating in the $x$-$y$ plane.

In experiments, the photon emission and absorption profiles of XV emitters can be characterized using Photoluminescence (PL) and Photoluminescence Excitation (PLE) techniques, respectively. 

In a PL experiment, a laser with a fixed wavelength excites the emitter to an energy level above the first-excited manifold. The emitter first relaxes to states within the first-excited manifold, populating them according to a specific distribution, and subsequently transitions from the first-excited manifold to the ground manifold, accompanied by photon emission.

In contrast, a PLE experiment begins by initializing the XV emitters in the ground manifold with a defined population distribution. A wavelength-tunable laser is then used to scan the spectrum, identifying frequencies that can excite the emitter. The excited emitter subsequently decays, emitting a photon at the same frequency as the excitation laser, which is detected by a photon detector.

Thus, the PL spectrum provides information about the emission characteristics of the emitter, while the PLE spectrum reveals its absorption characteristics.

To simulate the PL and PLE spectra, it is necessary to calculate the relaxed population distribution in the first-excited and ground manifolds, respectively. Both distributions follow the Boltzmann distribution, which depends on the temperature:
\begin{equation}
    n_i = \text{e}^{-\frac{E_i-E_i}{k_B T}},
\end{equation}
where $n_i$ represents the population of the $i$-th energy level of energy $E_i$, $E_1$ is the the lowest-energy within the respective manifold, $k_B$ is the Boltzmann constant, and $T$ is the absolute temperature. The normalization factor for the distribution is omitted here, as we are concerned only with the relative strengths of the photon peaks in the PL or PLE spectrums.

Table~\ref{tab:Boltzmann} provides the calculated energy differences $\Delta E_i$ and the corresponding populations $n$ for the lowest eight energy levels in the ground and excited manifolds at various absolute temperatures. These calculations are based on the SnV parameters obtained from \textit{ab initio} calculations~\cite{thiering2018ab}, as listed in Table~\ref{tab:Thiering_para_Ham_fac}.

\begin{table}[]
    \centering
    \begin{tabular}{l|cccccccc}
        (a) &  $\ket{1}$ & $\ket{3}$ & $\ket{5}$ & $\ket{7}$ & $\ket{9}$ & $\ket{11}$ & $\ket{13}$ & $\ket{15}$  \\
        \hline
        $\Delta E$ (meV) & 0 & 3.82 & 57.9 & 60.9 & 100 & 101 & 120 & 123 \\
        \hline
        $n$ (5 K) & 1 & 1e-4 & 0 & 0 & 0 & 0 & 0 & 0\\
        $n$ (50 K) & 1 & 0.412 & 0 & 0 & 0 & 0 & 0 & 0\\
        $n$ (100 K) & 1 & 0.642 & 1.2e-3 & 9e-4 & 0 & 0 & 0 & 0\\
    \end{tabular}
    \begin{tabular}{l|cccccccc}
        (b) &  $\ket{\rm A}$ & $\ket{\rm C}$ & $\ket{\rm E}$ & $\ket{\rm G}$ & $\ket{\rm I}$ & $\ket{\rm K}$ & $\ket{\rm M}$ & $\ket{\rm O}$  \\
        \hline
        $\Delta E$ (meV) & 0 & 11.6 & 34.6 & 45.5 & 77.6 & 84.1 & 87.7 & 106.8 \\
        \hline
        $n$ (5 K) & 1 & 0 & 0 & 0 & 0 & 0 & 0 & 0\\
        $n$ (50 K) & 1 & 0.0676 & 3e-4 & 0 & 0 & 0 & 0 & 0\\
        $n$ (100 K) & 1 & 0.260 & 0.0181 & 5.1e-3 & 1e-4 & 1e-4 & 0 & 0\\
    \end{tabular}
    \caption{Relative energy $\Delta E$ and (unnormalized) Boltzmann distribution $n(\Delta E)$ of the population in the ground (a) and excited manifolds (b). Calculations are based on the SnV parameters obtained from \textit{ab initio} calculations~\cite{thiering2018ab}, as listed in Table~\ref{tab:Thiering_para_Ham_fac}. The notation "$A$e$B$" denotes number $A\times10^{B}$.}
    \label{tab:Boltzmann}
\end{table}

Both the population distribution and the magnitude of the electric dipole matrix elements determine the intensity of the collected light in PL and PLE experiments. According to Fermi’s golden rule~\cite{Fermi_golden_dirac1927quantum}, the transition intensity from an initial state \(\ket{i}\) to a final state \(\ket{j}\), associated with the \(k\)-component (\(k = +, -, z\)) of the electric dipole moment, \(\hat{d}_k\), is given by
\begin{equation}
    I_k^{ij} = n_i \left|\bra{j}\hat{d}_k\ket{i}\right|^2,
\end{equation}
where \(n_i\) is the population of the initial state \(\ket{i}\).

\begin{figure}[h]
    \centering
    \includegraphics[width=0.9\linewidth]{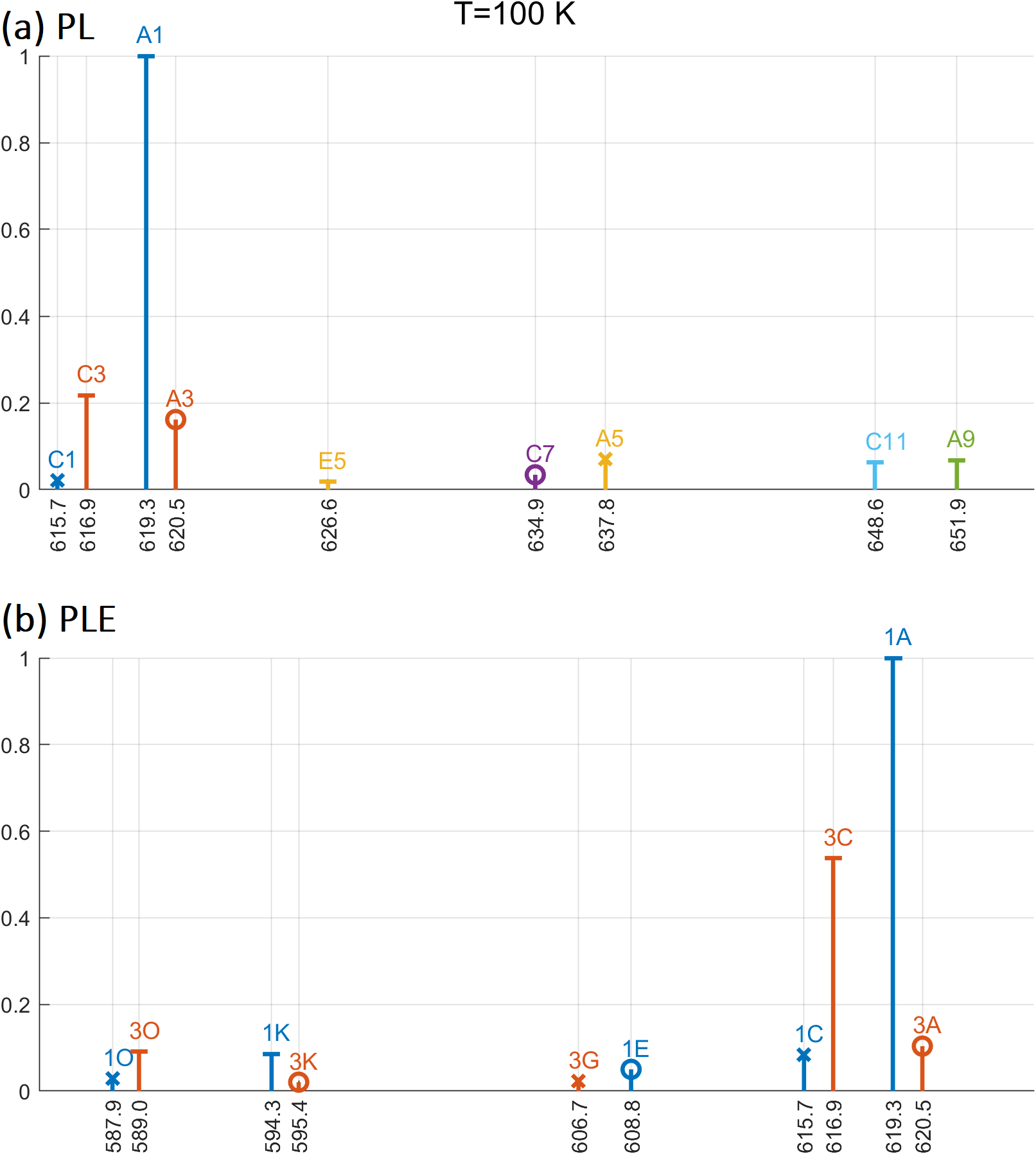}
    \caption{Simulated normalized (a) PL and (b) PLE spectra of the SnV emitter at 100 K. The label \(ij\) on each line denotes the transition from \(\ket{i}\) to \(\ket{j}\). Line colors correspond to the ground manifold states. The endpoints of each line indicate the polarization of the emitted light: a bar end for linear polarization (from \(\hat{d}_z\)), and a circle (cross) end for right (left) circular polarization (from \(\hat{d}_+\) and \(\hat{d}_-\)). Calculations are based on the SnV parameters obtained from \textit{ab initio} calculations~\cite{thiering2018ab}, as listed in Table~\ref{tab:Thiering_para_Ham_fac}.}
    \label{fig:spectrum}
\end{figure}

Figure~\ref{fig:spectrum} presents the calculated magnitudes for transitions in the PL and PLE spectra for a SnV center at 100 K.

From Fig.~\ref{fig:spectrum}(a), the ratio between the intensities of the \(A1\) and \(A3\) lines, \(\frac{I_+^{A3}}{I_z^{A1}}\), is approximately 0.16. This can be understood as follows: due to the similarity between the eigenstates in the ground and first-excited manifolds,
\begin{equation}
\begin{split}
    \bra{1}\hat{d}_z\ket{A} &\approx (3.09 \text{D}) \cdot \bra{1} I^o \ket{1} = 3.09\ \mathrm{D} \\
    \bra{3}\hat{d}_+\ket{A} &\approx (1.75\ \mathrm{D}) \cdot \bra{1}\tau_+^o\ket{3} \approx q_{\rm SnV-G} \times 1.75\ \mathrm{D} = 0.73 \times 1.75\ \mathrm{D},
\end{split}
\end{equation}
where \(q_{\rm SnV-G}\) is the quench factor for the \(\tau_x^o\) and \(\tau_y^o\) operators in the ground manifold. Thus,
\begin{equation}
    \frac{I_+^{A3}}{I_z^{A1}} \approx \left(\frac{0.73 \times 1.75\ \mathrm{D}}{3.09\ \mathrm{D}}\right)^2 = 0.17,
\end{equation}
which closely matches the simulated result of 0.16.

Next, we consider the proportion of light emitted into zero-phonon lines (ZPLs). Since the states \(\ket{A}, \ket{C}, \ket{3},\) and \(\ket{1}\) all reside in the phononic ground states of their respective manifolds, the transitions \(\ket{A}, \ket{C} \to \ket{1}, \ket{3}\) do not involve phonon excitation. Hence, the four lines A1, A3, C1, and C3 are all ZPLs.

For a temperature of 50 K, the calculated proportions of light emitted in the ZPLs for SiV, GeV, SnV, and PbV are 0.84, 0.79, 0.83, and 0.85, respectively.


\section{Zeeman Effect and Addressability Manipulation}
\label{sec:Zeeman}

\subsection{Zeeman Hamiltonian}
The interaction between the external magnetic field and the XV system, known as the Zeeman effect, affects both the hole spin and the hole's orbital angular momentum. In Sec.~\ref{sec:SOC}, we established that the hole in the XV system has only a \(z\)-component of angular momentum. Consequently, the Zeeman Hamiltonian is given by~\cite{HeppThesis}:
\begin{align}
    H_{\rm B} =& \frac{g_S\mu_{\rm B}}{\hbar} \mathbf{B}\cdot \mathbf{S} + \frac{g_L\mu_{\rm B}}{\hbar} B_{z} L_z^{\rm o} \\
    =& \mu_{\rm B} \left[B_{x} \sigma_{x}^{\rm s} + B_{y} \sigma_{y}^{\rm s} + B_{z} \left(-\frac{1}{2}\sigma_{y}^{\rm o}+\sigma_{z}^{\rm s}\right)\right],
\end{align}
where \(\mu_{\rm B} = 14.1 \, \text{GHz/T}\) is the Bohr magneton, and \(B_i\) (\(i = x, y, z\)) is the \(i\)-component of the magnetic field. Here the electron $g$-factor $g_L=1$ and we set the spin $g$-factor $g_s=2$. The hole orbital angular momentum is defined as $L_z^{\rm o}=-\frac{\hbar}{2}\sigma_{y}^{\rm o}$ [see Eqs.~(\ref{eq:Lz_mz}), (\ref{eq:m_z}), and (\ref{eq:L_z_sigma_y})]. Note that the factor $-\frac{1}{2}$ is not apparently shown in Hepp's thesis~\cite{HeppThesis}.

\begin{figure}
    \centering
    \includegraphics[width=0.96\linewidth]{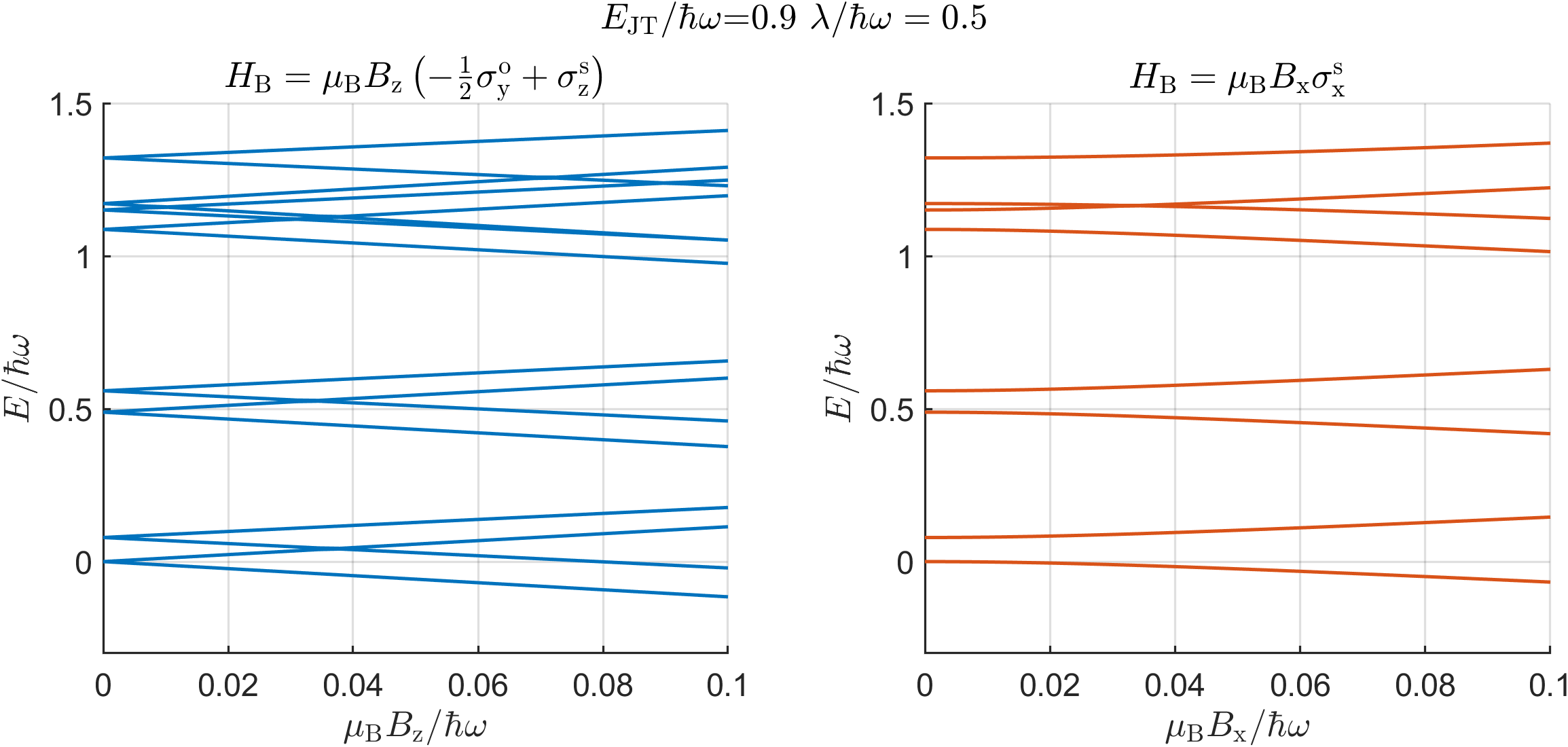}
    \caption{Variation of energy levels with the strength of the external magnetic field applied along the (a) \(z\)-direction and (b) \(x\)-direction.}
    \label{fig:level_mag}
\end{figure}

The Zeeman effect induces energy splitting; however, as discussed in Sec.~\ref{sec:symmetries}, the \(x\)- and \(y\)-components of the magnetic field do not break the degeneracy of the XV system. Thus, only the \(z\)-component of the magnetic field causes energy splitting. Fig.~\ref{fig:level_mag} illustrates the energy levels within an XV manifold under various magnetic field strengths applied along the \(z\)- and \(x\)-directions, respectively.

\subsection{Addressability Manipulation}

\begin{figure}
    \centering
    \includegraphics[width=0.8\linewidth]{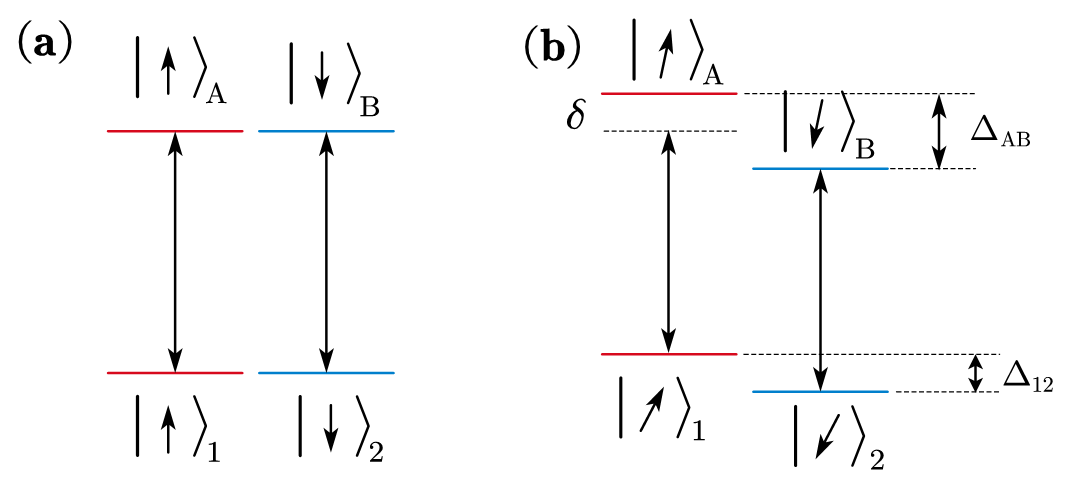}
    \caption{Energy levels of the two ground states in the excited manifold (\( \ket{\rm A} \) and \( \ket{\rm B} \)) and the ground manifold (\( \ket{1} \) and \( \ket{2} \)). (a) Without an external magnetic field, the transitions are symmetric. (b) With non-zero magnetic fields along the \(x\)-, \(y\)-, and \(z\)-directions, the symmetry is broken, resulting in distinct transition energies. The arrow in the Dirac ket symbol indicates the orientation of the hole spin, while the subscript denotes the eigenstate of the XV system.}
    \label{fig:coupling} 
\end{figure}

The Zeeman effect is a crucial mechanism for selectively addressing transitions between specific quantum states in the XV system. Without an external magnetic field, the spin-up and spin-down states in the ground manifold (\(\ket{\uparrow}_1\) and \(\ket{\downarrow}_2\)) are degenerate, as are the spin-up and spin-down states in the first-excited manifold (\(\ket{\uparrow}_{\rm A}\) and \(\ket{\downarrow}_{\rm B}\)). Consequently, coupling light to the \( \ket{\uparrow}_1 \leftrightarrow \ket{\uparrow}_{\rm A} \) transition unavoidably also couples the \( \ket{\downarrow}_2 \leftrightarrow \ket{\downarrow}_{\rm B} \) transition with equal strength, as illustrated in Fig.~\ref{fig:coupling}(a).

This degeneracy presents challenges for utilizing the XV system in quantum information processing. For example, several protocols require asymmetric driving of the transitions between \(\ket{\uparrow}_1 \leftrightarrow \ket{\uparrow}_{\rm A}\) and \(\ket{\downarrow}_2 \leftrightarrow \ket{\downarrow}_{\rm B}\), such as:
\begin{itemize}
    \item \textsc{cnot} gates between two XV qubits driven optically~\cite{cnot_gate_Edwin}.
    \item \textsc{cnot} gates between a photon and an XV qubit~\cite{cnot_pho_spin_Rempe}.
    \item Single-photon generation~\cite{Raman_drive_Tin}.
    \item Single-qubit gates with optical driving~\cite{GeoGate_SiV, Spin_rotation_takou2021optical, micro_control_pieplow2024efficient}.
\end{itemize}

To break the symmetry between the two transition energies, both a \(z\)-component and a combination of \(x\)- and \(y\)-components of the external magnetic field are required. The \(x\)- and/or \(y\)-components create spin orientations that differ between the ground and first-excited manifolds. When the \(z\)-component magnetic field is applied, the spin splitting \(\Delta_{\rm AB}\) in the excited manifold differs from the spin splitting \(\Delta_{12}\) in the ground manifold, as shown in Fig.~\ref{fig:coupling}(b) hence break the symmetry in the two transitions.

The difference in spin orientation between the first-excited manifold and the ground manifold arises from the differing spin-orbit coupling strengths in the two manifolds. As explained in Sec.~\ref{sec:SOC}, the hole experiences an intrinsic magnetic field along the \(z\)-axis due to spin-orbit coupling [see Eq.~(\ref{eq:B_vE})]. This intrinsic magnetic field differs between the two manifolds because of their distinct spin-orbit coupling strengths. Consequently, when identical \(x\)- and/or \(y\)-components of the external magnetic field are applied to both manifolds, the combined magnetic fields in the two manifolds adopt different orientations, resulting in distinct spin orientations.

\begin{figure}
    \centering
    \includegraphics[width=0.96\linewidth]{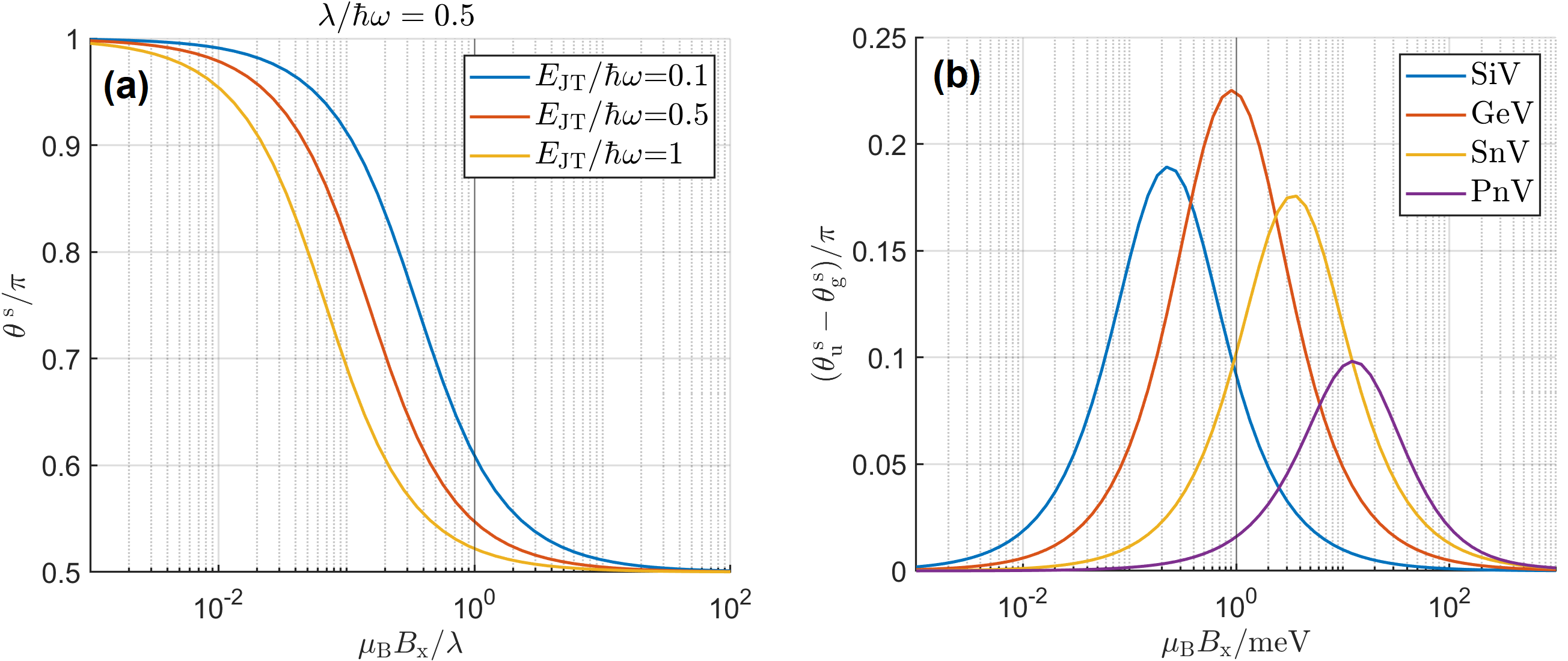}
    \caption{Investigation of hole spin orientations. (a) Variation of the hole spin's polar angle \(\theta^{\rm s}\) for the lowest spin-down state as a function of the applied magnetic field along the \(x\)-axis (\(B_x\)). (b) Variation in the angle difference between the hole spin orientations in the first-excited and ground manifolds (\(\theta^{\rm s}_{\rm e} - \theta^{\rm s}_{\rm g}\)) as a function of \(B_x\). Parameters for the first-excited and ground manifolds are adopted from the \textit{ab initio} calculations by Thiering \textit{et al.}~\cite{thiering2018ab}, as presented in Tab.~\ref{tab:Thiering_para_Ham_fac}.}
    \label{fig:spin_dir}
\end{figure}

Fig.~\ref{fig:spin_dir}(a) shows the spin orientation angle
\begin{equation}
    \theta^{\rm s} = \arccos\left(\langle {\rm G_\downarrow} | \sigma_{z}^{\rm s} | {\rm G_\downarrow} \rangle\right),
\end{equation}
where \(|{\rm G_\downarrow}\rangle\) represents the lowest spin-down state in a single manifold, as a function of the \(x\)-component of the magnetic field strength (\(B_x\)). Fig.~\ref{fig:spin_dir}(b) illustrates the orientation difference \(\theta^{\rm s}_{\rm e} - \theta^{\rm s}_{\rm g}\) between the first-excited and ground manifolds as a function of \(B_x\). The calculations reveal that the SnV system can exhibit a maximum spin orientation difference of \(0.18\pi\) at \(\mu B_x = 4 \, \mathrm{meV}\).

\section{Approximation Methods}

The phonon, whose Hilbert space is infinitely large, introduces coupling that makes the XV center system difficult to solve exactly. Two popular approximation methods are used to simplify the treatment of electron-phonon coupling.

The Born-Oppenheimer (B-O) approximation allows one to treat the phonon motion and hole motion separately, but the approximation is valid only when the phonon energy is far lower than the electron-phonon coupling strength. This prerequisite is not fulfilled in the XV systems, hence the B-O approximation is not applicable. However, this method offers a valuable perspective for XV systems with well-developed concepts, such as the Jahn-Teller energy.

The quench-factor method treats the electron-phonon eigenstates as black boxes and approximates the electron-phonon coupling effect as merely quenching the strength of the other interactions involving the hole. The original method assumes that the strengths of all other hole-related interactions are far smaller compared to the electron-phonon interaction. However, this condition often fails, particularly in systems with strong spin-orbit coupling. In this section, we introduce an improved definition of the quench factors, enabling a more accurate description of XV systems.

\subsection{Born-Oppenheimer Approximation}
\label{sec:B-O_approx}

The Born-Oppenheimer (B-O) approximation is inspired by the assumption that the hole, which has a light effective mass and strong bonding to the atomic core, can adapt almost instantaneously to the atomic motion. Therefore, the atomic displacements are approximately constant from the perspective of the hole, \textit{i.e.}, the state of the hole depends only on the atomic displacements and is independent of the atomic momenta.

A rigorous derivation of the B-O approximation is provided by the adiabatic theorem~\cite{adiabatic_born1928beweis}. It states that if a quantum system with a time-dependent Hamiltonian \(H(t)\) is initially in an eigenstate \(\ket{\psi_i(t=0)}\) of \(H(0)\), it will remain in the corresponding eigenstate of \(H(t)\), \(\ket{\psi_i(t)}\), provided that the following condition holds~\cite{adiabatic_proof}:
\begin{equation}
    \frac{h \left|\bra{\psi_j} \dot{H} \ket{\psi_i}\right|}{\left|E_j - E_i\right|^2} \ll 1,
    \label{eq:adia_condition}
\end{equation}
for all other eigenstates \(\left\{\ket{\psi_j}\right\}\) (\(j \ne i\)) with corresponding eigenenergies \(\{E_j\}\). Here, \(h = 6.626 \times 10^{-34}\ \text{J}\cdot\text{s}\) is the Planck constant.

\begin{figure}
    \centering
    \includegraphics[width=0.8\linewidth]{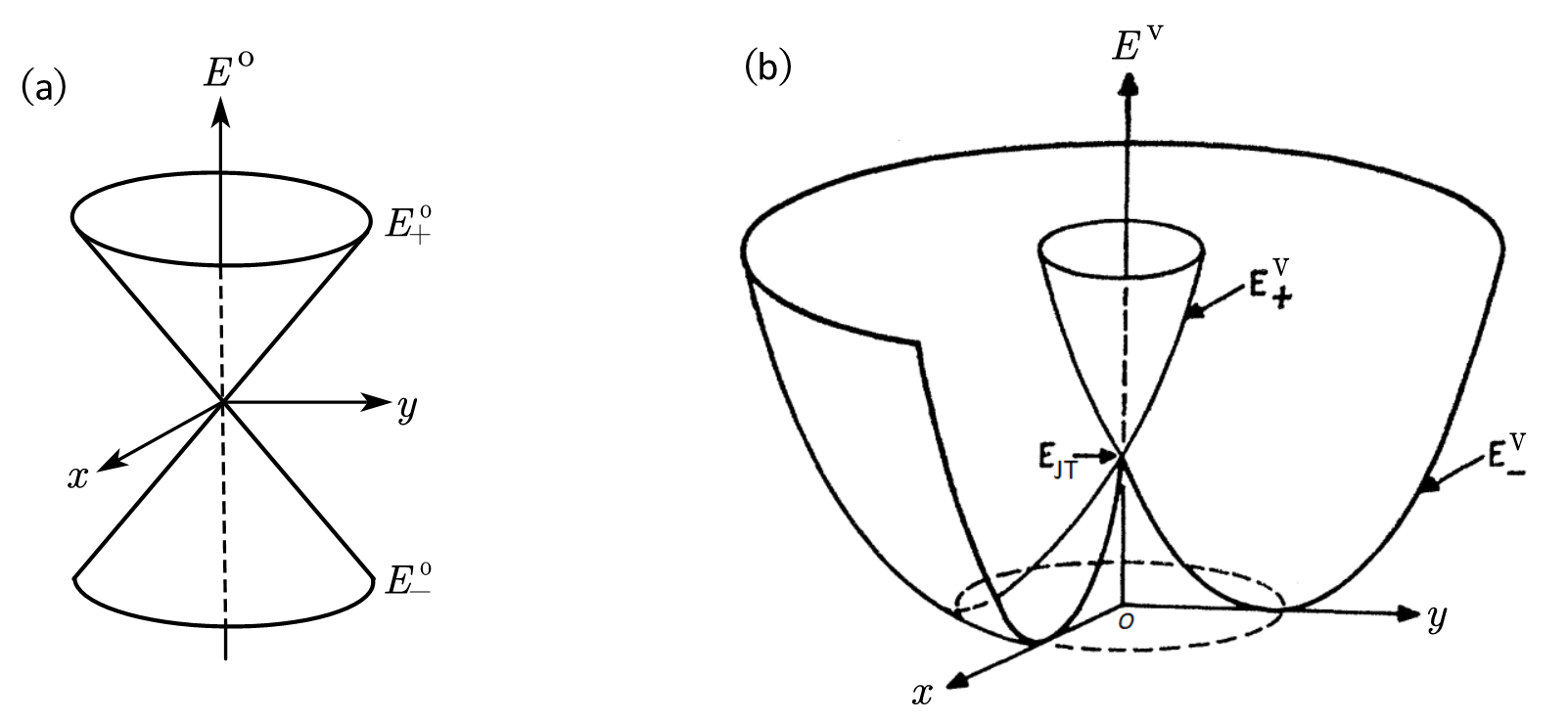}
    \caption{(a) Landscape of the eigenenergies of \(H_{\rm vo}\). (b) Adiabatic potential energy surface for the atomic motion. Adapted from Ref.~\cite{ham1968effect}.}
    \label{fig:landscape}
\end{figure}

Based on the adiabatic theorem, the hole orbital will remain in its eigenstate if the atomic displacement changes slowly enough and the gap between the hole's eigenenergies is sufficiently large. When these conditions are fulfilled, the eigenenergy of the hole orbitals, obtained by diagonalizing the Hamiltonian \(H_{\rm vo}\) (see Eq.~\ref{eq:H_vo}) while treating \(Q_x\) and \(Q_y\) as constants, is given by
\begin{equation}
    E_\pm^{\rm o} = \pm F \sqrt{Q_x^2 + Q_y^2}.
\end{equation}
The eigenenergies \(E_\pm^{\rm o}\) form surfaces in the \(Q_x\)-\(Q_y\) space, taking the shape of cones as shown in Fig.~\ref{fig:landscape}(a). We observe that the two eigenenergies touch at the point \((Q_x, Q_y) = (0, 0)\). The condition~(\ref{eq:adia_condition}) can only be met when the radial atomic displacement \(\sqrt{Q_x^2 + Q_y^2}\) is sufficiently large.

Assuming that this condition is satisfied, the harmonic potential in which the atoms move is modified by the addition of the term \(E_\pm^{\rm o}\), resulting in the potential
\begin{equation}
    E_\pm^{\rm v} = \frac{\mu}{2} \omega^2 \left(Q_x^2 + Q_y^2\right) \pm F \sqrt{Q_x^2 + Q_y^2}.
\end{equation}
This potential is referred to as the adiabatic potential energy surface, as shown in Fig.~\ref{fig:landscape}(b). It has an upper sheet and a lower sheet. The lower sheet \(E_-^{\rm v}\) has minima forming a ring at a distance
\begin{equation}
    \rho_0 = \frac{F}{\mu \omega^2}
\end{equation}
from the origin. Additionally, there is a spike at the origin with a height
\begin{equation}
    E_{\rm JT} = \frac{F^2}{2 \mu \omega^2}.
    \label{eq:Jahn-Teller_ene}
\end{equation}
This height \(E_{\rm JT}\) is called the Jahn-Teller energy.

We can infer that if the phonon energy is much smaller than the spike height:
\begin{equation}
    \hbar \omega \ll E_{\rm JT},
    \label{eq:adia_cond2}
\end{equation}
the atomic displacement vector \((Q_x, Q_y)\) is unlikely to appear near the origin, ensuring that the adiabatic condition~(\ref{eq:adia_condition}) is fulfilled. Consequently, Eq.~(\ref{eq:adia_cond2}) is the alternative prerequisite condition for the B-O approximation.

Lastly, Note that
\begin{equation}
    E_{\rm JT} = \frac{J^2}{\hbar \omega},
\end{equation}
[see Eq.~(\ref{eq:J_esti})]. The condition~(\ref{eq:adia_cond2}) is equivalent to
\begin{equation}
    \hbar \omega \ll J,
    \label{eq:adia_cond3}
\end{equation}
which indicates that the electron-phonon coupling strength is much larger than the phonon energy. This is consistent with the assumption that the atoms move slowly compared to the hole orbitals.

The condition~(\ref{eq:adia_cond3}) implies strong electron-phonon coupling, meaning that the electron can almost instantaneously adapt to the motion of the phonons. However, this assumption does not always hold. The reason is straightforward: although a single hole generally exhibits high mobility and adaptability, in the case of the XV system, the hole is also bonded to the central X atom. This bonding increases its effective inertia, making it less responsive to the motion of the surrounding carbon atoms. As a result, the Born-Oppenheimer approximation no longer applies. Nevertheless, the concept of the Jahn-Teller energy~(\ref{eq:Jahn-Teller_ene}) remains useful as a qualitative reference.
\subsection{Quench Factors}
\label{sec:quench_f}

Quench factors, or Ham factors, originally proposed by Ham~\cite{ham1968effect}, are used to simplify the characterization of the influence of the phonon on the interactions related to the electronic orbital states. The name of quench factors refers to two factors both less than one. It reflects the nature of the phonon effect that it effectively reduces the interaction strengths related to the electronic orbital states. 

This approximation technique reduces the complexity of representing the effect of phonons, which situate in an infinite-dimensional Hilbert space, into two scalar parameters. As a result of this convenience, this method was initially utilized in the SiV system~\cite{HeppThesis} and subsequently in other XV systems~\cite{thiering2018ab, Meesala2018Strain}. Although widely used, few have examined its validity and precision. As shown in Sec.~\ref{sec:SOC} and DFT simulations~\cite{thiering2018ab}, the spin-orbit coupling strength increases along with the atomic number of the X atom. Therefore, there could be potential problems for the XV systems of larger X atomic number to inherit this approximation from the SiV system.

This section aims to establish a foundation for applying the approximate quench factor model. Firstly, we show that the quench factors based on original definitions are less precise for XV systems with strong spin-orbit coupling. Moreover, the accuracy of this approach depends on the relative electron-phonon coupling strength compared to the phonon energy. To address these issues, we introduce refined definitions of the quench factors, which are well-suited to XV systems containing elements with large atomic numbers. While these improved quench factors offer enhanced accuracy over the original formulation, their precision still depends on the ratio of the electron-phonon coupling strength to the phonon energy.

\subsubsection{Original Definition}

In the initial proposal of the quench factor~\cite{ham1968effect}, Ham considers the following Hamiltonian
\begin{equation}
    H=H_{\rm vib}+H_{\rm vo}+H_{\rm int} \left( \tau_z^{\rm o}, \tau_x^{\rm o}, \tau_y^{\rm o} \right)\;,
\end{equation}
where $H_{\rm vib}$ and $H_{\rm vo}$ are as the ones defined in this thesis. And, $H_{\rm int}$ is the interaction Hamiltonian excluding the electron-phonon interaction. It may incorporate the electronic orbital operators $\tau_z^{\rm o}$, $\tau_x^{\rm o}$, and $\tau_y^{\rm o}$.

Without considering the electron spin, the Hamiltonian \(H_{\rm vib} + H_{\rm vo}\) is two-fold degenerate in energy. If the interaction strength of $H_{\rm int}$ is far less than the ground splitting of \(H_{\rm vib} + H_{\rm vo}\), the two degenerate ground states $\{\ket{1}, \ket{2}\}$ effectively form a nearly closed subspace of the Hamiltonian $H$. Therefore, if limiting our scope to the two energetically lowest eigenstates of $H$, it is sufficient to examine the projected Hamiltonian $PHP$, where 
\begin{equation}
    P = \ket{1} \bra{1} + \ket{2} \bra{2}
\end{equation}
is the projection operator onto the subspace $\{\ket{1}, \ket{2}\}$.

Since the states $\ket{1}$ and $\ket{2}$ are degenerate, the term $P(H_{\rm vib} + H_{\rm vo})P$ vanishes under a suitable energy gauge. The effective Hamiltonian becomes
\begin{equation}
    H_{\rm Ham} = H_{\rm int} \left( P\tau_z^{\rm o}P,\, P\tau_x^{\rm o}P,\, P\tau_y^{\rm o}P \right).
\end{equation}

When evaluating the operators $\left\{\; P \, \tau_i^{\rm o} \, P\;\right\}$ ($i=x,y,z$), we find the following relationships:
\begin{gather}
    \bra{1} \tau_z^{\rm o} \ket{1} = - \bra{2} \tau_z^{\rm o} \ket{2} \coloneqq p, \label{eq:tau_z_p} \\
    \bra{1} \tau_x^{\rm o} \ket{2} = \bra{2} \tau_x^{\rm o} \ket{1} = \mathrm{i} \bra{1} \tau_y^{\rm o} \ket{2} = - \mathrm{i} \bra{2} \tau_y^{\rm o} \ket{1} \coloneqq q, \label{eq:tau_xy_q} \\
    \bra{1} \tau_z^{\rm o} \ket{2} = \bra{2} \tau_z^{\rm o} \ket{1}=\bra{1} \tau_{x(y)}^{\rm o} \ket{1} = \bra{2} \tau_{x(y)}^{\rm o} \ket{2} = 0. \label{eq:tau_eq0}
\end{gather}
They can be easily proved using the symmetries as discussed in Sec.~\ref{sec:symmetries}. As a result, the effective Hamiltonian can be expressed as
\begin{equation}
    H_{\rm Ham} = H_{\rm int} \left( p\tau_z^{\rm v}, q\tau_x^{\rm v}, q\tau_y^{\rm v} \right).
    \label{eq:H_tot_Ham}
\end{equation}
where
\begin{equation}
    \begin{split}
        \tau_z^{\rm v} &= \ket{2} \bra{2} - \ket{1} \bra{1}, \\
        \tau_x^{\rm v} &= \ket{1} \bra{2} + \ket{2} \bra{1}, \\
        \tau_y^{\rm v} &= \text{i} \ket{2} \bra{1} - \text{i} \ket{1} \bra{2}.
    \end{split}
\end{equation}
These are Pauli operators in the bases $\{\ket{1}, \ket{2}\}$, which are entangled states involving electron orbitals and phonons. As an entity, the combination of them is also referred to as a vibron.

The effective Hamiltonian $H_{\rm Ham}$ (excluding electron spin) is two-dimensional, hence, this approximation method significantly simplifies the calculation of the ground splitting of $H$. As a concrete example, when $H_{\rm int}=H_{\rm os}$, which is the spin-orbit coupling Hamiltonian, the ground splitting is approximately given by
\begin{equation}
    \Delta_{\rm GS} \approx p\lambda\;.
\end{equation}

\begin{figure}
    \centering
    \includegraphics[width=0.6\linewidth]{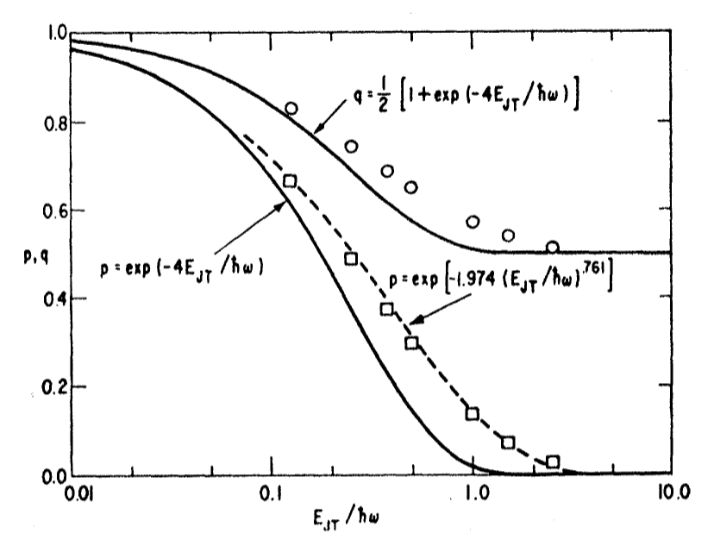}
    \caption{Ham factors \( p \) and \( q \) of the two ground states \( \left\{\ket{1}, \ket{2}\right\} \) as functions of the ratio of the Jahn-Teller energy \( E_{\rm JT} \) to the phonon energy \( \hbar\omega \). The points represent exact values derived from numerical methods, while the solid curves are approximations. The dashed curves fit the data in the range \( 0.1 \leq E_{\rm JT}/\hbar\omega \leq 3.0 \). Adapted from Ref.~\cite{ham1968effect}.}
    \label{fig:p_q}
\end{figure}

Due to the inner product nature, $0\leq p, q \leq 1$, which gives rise to their name as quench factors. This approximation tells us that the effect of electron-phonon coupling is to quench the coupling strengths of other interactions associated with the electronic orbital states. For the concrete values of $p$ and $q$, Ham pointed out~\cite{ham1968effect} that
\begin{equation}
    q = \frac{1}{2}(1 + p),
    \label{eq:p_q_relation}
\end{equation} and $p$ can be expressed as functions of \( E_{\rm JT}/\hbar\omega \) as shown in Fig.~\ref{fig:p_q}.

This approximation achieves adequate precision for the SiV system~\cite{thiering2018ab}, where the spin-orbit coupling is much smaller than the electron-phonon coupling strength and phonon energy. However, the error of this approximation increases for the XV systems with larger spin-orbit coupling strengths. To mitigate this problem, a solution could be to separate the spin-orbit coupling term out from $H_{\rm int}$; thereby, the lowest four energy eigenstates (incorporating electron spin states) of \( H_0=H_{\rm vib} + H_{\rm vo}+ H_{\rm os} \) may form a quasi-independent subspace regarding other interaction terms associated with the electronic orbital states. In this new subspace, quench factors $p'$ and $q'$ can be defined similarly.

\subsubsection{Refined Definition}
\label{sec:refined_def}

We consider the XV system Hamiltonian in the form
\begin{equation}
    H=H_0 +  H'_{\rm int}\left( \tau_z^{\rm o}, \tau_x^{\rm o}, \tau_y^{\rm o} \right)\;,
    \label{eq:H_0+int2}
\end{equation}
where $H_0=H_{\rm vib}+H_{\rm vo}+H_{\rm os}$, and $H'_{\rm int}$ is the interaction Hamiltonian excluding the electron-phonon interaction and spin-orbit interaction, which can also include the electronic orbital operators $\tau_z^{\rm o}$, $\tau_x^{\rm o}$, and $\tau_y^{\rm o}$.

As discussed in Sec.~\ref{sec:degeneracy}, $H_0$ is two-fold degenerate. Since it commutes with the $z$-component spin operator $\sigma_z^{\rm s}$, the four lowest energy eigenstates can be presented as $\left\{\ket{1'}, \ket{2'}\right\}\otimes\{\ket{\uparrow}, \ket{\downarrow}\}$, where $\left\{\ket{1'}, \ket{2'}\right\}$ are the vibronic components and $\{\ket{\uparrow}, \ket{\downarrow}\}$ are the spin components, specifically, the eigenstates of $\sigma_z^{\rm s}$. If the energy difference between the energy of the fifth and fourth lowest energy eigenstate $\Delta_{5, 4}$ is much larger than the interaction strength of $H'_{\rm int}$, the four lowest energy eigenstates form a quasi-independent subspace.

In this subspace, we can define
\begin{align}
    p' &\coloneqq \frac{1}{2}\left(\bra{2', \uparrow} \tau_z^{\rm o} \ket{2', \uparrow} - \bra{1', \uparrow} \tau_z^{\rm o} \ket{1', \uparrow}\right)\;, \\
    q' &\coloneqq \bra{1', \uparrow} \tau_x^{\rm o} \ket{2', \uparrow} \;.
\end{align}
Then, in a suitable energy gauge, the Hamiltonian~\ref{eq:H_0+int2} can be approximated with 
\begin{equation}
    H \approx -\frac{\Delta_0}{2} \tau_z^{\rm m}\sigma_z^{\rm s} + H_{\rm int} \left( p'\tau_z^{\rm m}, q'\tau_x^{\rm m}, q'\tau_y^{\rm m} \right).
    \label{eq:H_improved_Ham}
\end{equation}
where $\Delta_0$ is the ground splitting of $H_0$ and  \(\{\tau_i^{\rm m}\}\) ($i=x, y, z$) are the Pauli matrices in the basis \(\left\{ \ket{1'}, \ket{2'} \right\}\). Note that under this refined definition, $p'$ and $q'$ do not obey the relationship similar to Eq.~\eqref{eq:p_q_relation}.

The quantity $\Delta_0$ can be determined in experiments. It corresponds to the ground splitting of an unstrained sample under zero magnetic field. In some literature~\cite{Meesala2018Strain}, when no need to be specified, the quench factors can be absorbed into the effective interaction strengths and not shown.

To show that the quench factors \( p' \) and \( q' \) are of higher precision than \( p \) and \( q \) for XV species with larger spin-orbit coupling strengths, we harness the parameters of $\hbar\omega$, $E_{\rm JT}$, and $\lambda$ of various XV systems from the reported DFT calculations~\cite{thiering2018ab} as shown in Tab.~\ref{tab:Thiering_para_Ham_fac}). Additionally, we examine two interaction Hamiltonians, (1) $H'_{\rm int}=\gamma_z \tau_z^{\rm o}$ for $p-p'$ comparison and (2) $H'_{\rm int}=\gamma_x \tau_x^{\rm o}$ for $q-q'$ comparison. We limit our scope to
\begin{equation}
    \gamma_z=\gamma_x=1 \text{meV}\;.
\end{equation}

Based on the quench factors $p$ and $q$, the ground splittings for the two types of $H'_{\rm int}$ are given by
\begin{gather}
    \Delta_{\rm GS,qch,z}=\text{min}\left(\Delta_0, 2p\gamma_z\right)\;,\\
    \Delta_{\rm GS,qch,x}=\sqrt{\Delta_0^2+4(q\gamma_x)^2}\;,
\end{gather}
respectively. The corresponding precisions of these evaluations are defined by
\begin{gather}
    \epsilon_{\rm qch,z}=\frac{\Delta_{\rm GS,qch,z}-\Delta_{\rm GS,num,z}}{\Delta_{\rm GS,num,z}}\;,\\
    \epsilon_{\rm qch,x}=\frac{\Delta_{\rm GS,qch,z}-\Delta_{\rm GS,num,z}}{\Delta_{\rm GS,num,z}-\Delta_0}\;,
\end{gather}
where $\Delta_{\rm GS,num,x}$ and $\Delta_{\rm GS,num,z}$ are from the numerical diagonalization of the Hamiltonian $H_0+H'_{\rm int}$, respectively. The precisions of the ground splitting evaluations based on $p'$ and $q'$ are defined in the same fashion and denoted by $\epsilon'_{\rm qch,z}$ and $\epsilon'_{\rm qch,x}$, respectively.

\begin{table}[h]
    \centering
    \begin{tabular}{ccccc}
        \toprule
         & \(\hbar\omega\) & \(E_{\text{JT}}\)  & \(\lambda\) & $\Delta_{5, 4}$\\
         & (meV) & (meV) & (meV) & (meV) \\
         \hline
        SiV-G & 85.2 & 42.3  & 0.82 & 52.0 \\
        GeV-G & 82.2 & 30.1  & 2.20 & 54.5 \\
        SnV-G & 79.4 & 21.6  & 8.28 & 54.1 \\
        PbV-G & 74.9 & 15.6  & 34.6 & 41.6 \\
        \hline
        SiV-E & 73.5 & 78.5  & 6.96 & 31.4 \\
        GeV-E & 73.0 & 85.7  & 36.1 & 26.7 \\
        SnV-E & 75.6 & 83.1  & 96.8 & 23.0 \\
        PbV-E & 78.6 & 91.6  & 245  & 14.5 \\
        \bottomrule
    \end{tabular}
    \caption{XV system parameters \( \hbar\omega \), \( E_{\text{JT}} \), and \( \lambda \) reported in DFT calculations~\cite{thiering2018ab}. The suffix "-G"("-E") denotes the ground and first-excited manifold. The variable $\Delta_{5, 4}$ represents the energy difference between the fifth and fourth lowest eigenenergies.}
    \label{tab:Thiering_para_Ham_fac}
\end{table}

\begin{table}[h]
    \centering
    \begin{tabular}{ccccccccccc}
        \toprule
          & $p$ & $p'$ & $\epsilon_{\rm qch,z}$ & $\epsilon'_{\rm qch,z}$ & $q$ & $q'$ & $\epsilon_{\rm qch,x}$ & $\epsilon'_{\rm qch,x}$\\
         \hline
        SiV-G &0.30&0.30&-0.029&-0.029&0.65&0.65&1.7e-4&1.5e-4\\
        GeV-G &0.38&0.38&-0.029&-0.029&0.69&0.69&3.7e-4&2.1e-4\\
        SnV-G &0.46&0.46&-0.089&-0.092&0.73&0.73&4.5e-3&7.0e-4\\
        PbV-G &0.53&0.49&-0.24&-0.030&0.77&0.73&0.10&9.8e-3\\
        \hline
        SiV-E &0.12&0.12&-0.16&-0.15&0.56&0.56&1.1e-3&5.6e-4\\
        GeV-E &0.11&0.11&-0.50&-0.50&0.55&0.55&1.9e-2&3.2e-3\\
        SnV-E &0.12&0.12&-0.71&-0.71&0.56&0.53&0.15&2.4e-2\\
        PbV-E &0.11&0.10&-0.84&-0.85&0.55&0.42&0.99&0.16\\
        \bottomrule
    \end{tabular}
    \caption{Precisions of the quench factors based on the original and refined definitions for various XV systems. The suffix "-G"("-E") denotes the ground (first-excited) manifold. The notion $AeB$ means $A\times10^{B}$.}
    \label{tab:quench_epsilon}
\end{table}

The comparison between the precisions of the quench factors based on the original and refined definitions for various XV system species are shown in Tab.~\ref{tab:quench_epsilon}. From the comparison, we see that the quench factors $p$ and $p'$ give similar precisions. However, the precision of the quench factor $q'$ is an order of magnitude greater than that of $q$, demonstrating an improved fit of the refined definition compared to the original one. Additionally, the precision of the quench factors $p'$ and $q'$ still decreases as the atomic number of X increases. This is because of the decrease of $\Delta_{5, 4}$, the energy difference between the fifth and fourth lowest eigenenergies of $H_0$, as shown in Tab.~\ref{tab:Thiering_para_Ham_fac}.

\section{Previous Studies}
\label{sec:comparison}

\subsection{Results of this article}
In Secs.~\ref{sec:ele-vib} and~\ref{sec:SOC}, we provided order-of-magnitude estimations for the XV system parameters $\hbar\omega$, $F_0$, $F$, and $\lambda$, with undetermined partition factor $0<\alpha_{\rm X}<1$ and screening factor $0< \beta_{\rm X}<1$ and a relation
\begin{equation}
    F_0=\sqrt{2}F\;.
    \label{eq:F0_F_sum}
\end{equation}
These parameters are shown in Tab.~\ref{tab:XV_para_esti}.
\begin{table}[h]
    \centering
    \begin{tabular}{ccccc}
        \toprule
         & \(\hbar\omega\) & $F_0$ & \(F\)  & \(\lambda\)\\
         & (meV) & (eV/\AA) & (eV/\AA) & (meV)  \\
         \hline
        SiV & 136 & 39$\alpha_{\rm Si}$ & 27$\alpha_{\rm Si}$ & 52$\beta_{\rm Si}$ \\
        GeV & 120 & 35$\alpha_{\rm Ge}$ & 25$\alpha_{\rm Ge}$ & 603$\beta_{\rm Ge}$ \\
        SnV & 114 & 28$\alpha_{\rm Sn}$ & 20$\alpha_{\rm Sn}$ & 1841$\beta_{\rm Sn}$ \\
        PbV & 111 & 25$\alpha_{\rm Pb}$ & 18$\alpha_{\rm Pb}$ & 7706$\beta_{\rm Pb}$ \\
        \hline
    \end{tabular}
    \caption{XV system parameter estimations from Secs.~\ref{sec:ele-vib} and~\ref{sec:SOC}.}
    \label{tab:XV_para_esti}
\end{table}

In Sec.~\ref{sec:strain_Hami}, we discover relations between the strain susceptibilities and the electron-displacement coupling strengths $F_0$ and $F$ as follows
\begin{equation}
    \begin{split}
        t_{\parallel} &= -\frac{25}{4\sqrt{114}}F_0a_0, \\
        t_{\perp} &= -\frac{4}{\sqrt{114}}F_0a_0, \\
        d &= \frac{2}{\sqrt{57}}Fa_0, \\
        f &= \frac{10}{\sqrt{114}}Fa_0,
    \end{split}
    \label{eq:d_F_relation_sum}
\end{equation}
where $a_0 = 3.567 \, \text{\r{A}}$ is the diamond lattice constant. 

In Sec.~\ref{sec:B-O_approx}, we show that the Jahn-Teller energy is defined as 
\begin{equation}
    E_{\rm JT} = \frac{F^2}{2 \mu \omega^2}.
    \label{eq:Jahn-Teller_ene_sum}
\end{equation}
where $\mu \in \{28.0, 72.6, 118.7, 207.2\}\,\text{Da}$ (1 Da = \(1.66 \times 10^{-27}\) kg) is the mass of the X nucleus.

In Sec.~\ref{sec:interaction_diag}, we show that the electron-phonon coupling strength $J$ is given by
\begin{equation}
    J = F\sqrt{\frac{\hbar}{2\mu\omega}}\;.
    \label{eq:J_esti_sum}
\end{equation}

\subsection{Comparison with Hepp's model}

Hepp has pioneered modeling the SiV system~\cite{HeppThesis}. His model was later extended to apply to the SnV system by Trusheim \textit{et al.}~\cite{SnV_model_trusheim2020transform}. This article functions as both a rectification and an expansion of Hepp's model. Building upon first-principle reasoning, the primary accomplishment of this article is threefold.

Firstly, using group theory, Hepp analyzes the symmetries of the electronic orbital states and nuclear vibrations \underline{separately} and concludes that there are four nuclear vibration modes, two sets of couples, each with $\epsilon_g$ symmetry, coupled to the electronic orbital doublet. Nevertheless, this article found it problematic to analyze the symmetries of the nuclear vibrations independent of the electronic orbital configurations. Instead, we classify the nuclear vibration modes according to the spatial configuration of the electronic orbital states, as presented by Eq.~\eqref{eq:vib_ele_coef}, and derive the conclusion that, in the simplest form, there are only two nuclear vibration modes dominantly coupled to the electronic orbital doublet. The four nuclear vibration modes identified by Hepp all have non-zero projections onto the two modes identified in this article, which is why they all have significant coupling to the electronic orbital doublet.

Secondly, within Hepp's framework, the spin-orbit coupling is considered the primary interaction in terms of magnitude, while the Jahn-Teller interaction (also known as the electron-phonon interaction) is regarded as secondary. Based on his logic, the interaction Hamiltonian is expressed in the bare basis of electronic orbital and spin states, and the Jahn-Teller is presented as an interaction term. Additionally, the quench factors, as derived from the Jahn-Teller effect, only apply to terms weaker than the spin-orbit couplings, for example, the orbital gyromagnetic ratio. However, this logic can be potentially wrong, as the presumption that the spin-orbit coupling is much stronger than the Jahn-Teller interaction is not justified, and the Jahn-Teller interaction term presented in his framework is not proven to be non-zero. 

What's more, Hepp's modeling of the Jahn-Teller effect directly contradicts the model used in Theiring's calculation~\cite{thiering2018ab}. In Hepp's model, in the presence of both the spin-orbit coupling and Jahn-Teller effect, the ground splitting of the SiV system is given by
\begin{equation}
    \Delta_{\rm gs}^{\rm Hep}=\sqrt{\lambda^2+4\Upsilon_x^2+4\Upsilon_y^2}\;,
\end{equation}
where $\Upsilon_x$ and $\Upsilon_y$ are the parameters that represent the strength of the Jahn-Teller effect. Thus, the presence of the Jahn-Teller effect increases the ground-state splitting beyond that induced by spin-orbit coupling. However, in the framework used in Theiring's calculation, it leads to a reduction, since the Jahn-Teller effect is characterized by quench factors, which are applied to the spin-orbit coupling strengths.

Through meticulous reasoning, this article supports the model utilized in Theiring's calculation. And, as presented in Sec.~\ref{sec:refined_def}, we also propose a refinement on that approximate model with improved precision. In this refined model, the spin-orbit coupling and Jahn-Teller effect are considered simultaneously, and we propose refined quench factors that apply to all the other terms associated with electronic orbital states except for these two. Coincidentally, the data processing methods used in Meesala's paper~\cite{Meesala2018Strain} align with the refined quench-factor model.

Thirdly, in Hepp's study, he experimentally observed a branching ratio of about $\frac{1}{2}$ in the SiV emission spectrum but did not explain it. This article accomplishes this explanation, as presented in Sec.~\ref{sec:light_emission}. We show that this factor originates from the Jahn-Teller effect, or the quench factors.

\subsection{Partition factor and screening factor based on Thiering's calculation}

Thiering \textit{et al.}~\cite{thiering2018ab} have calculated the absolute values of $\hbar\omega$, $E_{\rm JT}$, and $\lambda$ for the four XV systems with DFT methods. Tab.~\ref{tab:p_s_factors} shows the partition factors ($\alpha_{\rm X}$) and screening factors ($\beta_{\rm X}$) based on their calculation results.

\begin{table}[]
    \centering
    \begin{tabular}{ccccc}
        \toprule
         & $\alpha_{\rm X}$ & $\beta_{\rm X}$\\
         \hline
        SiV-G & 0.049 & 0.016   \\
        GeV-G & 0.043 & 0.0036   \\
        SnV-G & 0.044 & 0.0045   \\
        PbV-G & 0.039 & 0.0045   \\
        \hline
        SiV-E & 0.058 & 0.13   \\
        GeV-E & 0.065 & 0.060   \\
        SnV-E & 0.082 & 0.053   \\
        PbV-E & 0.10 & 0.032   \\
        \bottomrule
    \end{tabular}
    \caption{Partition factors ($\alpha_{\rm X}$) and screening factors ($\beta_{\rm X}$) of the XV systems in their ground (-G) and first-excited (-E) manifolds. The calculation is based on Thiering's DFT simulations~\cite{thiering2018ab} as shown in Tab.~\ref{tab:Thiering_para_Ham_fac} and relation~\eqref{eq:Jahn-Teller_ene_sum}.}
    \label{tab:p_s_factors}
\end{table}

Based on the Tab.~\ref{tab:p_s_factors}, we observe larger partition factors and screening factors in the first-excited manifolds than the ground manifolds for all XV systems. This can be explained by the presence and absence of the central node in the ground and first-excited manifolds, respectively. Additionally, we note that, within the same manifold, partition factors and screening factors exhibit minimal variation among the XV species.

\subsection{Lowest splitting measurement}

In the Tab.~\ref{tab:lowest_spltting} below, we show the measured lowest splittings in the ground and first-excited manifolds for all the XV species along with the ones calculated with the DFT method by Thiering \textit{et al.}~\cite{thiering2018ab}.

\begin{table}[h]
    \centering
    \begin{tabular}{ccc}
        \toprule
         & \(\Delta_{0, \rm X, mea}^{g/e}\) (GHz) & \(\Delta_{0, \rm X, Thi}^{g/e}\) (GHz)~\cite{thiering2018ab} \\
         \hline
        SiV-G &  50~\cite{GS_SiV_hepp2014electronic} & 61 \\
        GeV-G &  170~\cite{GS_GeV_ekimov2015germanium} & 207 \\
        SnV-G &  850~\cite{GS_SnV_iwasaki2017tin} & 946 \\
        PbV-G &  5700~\cite{GS_PbV_trusheim2019lead} & 4514 \\
        \hline
        SiV-E & 260~\cite{GS_SiV_hepp2014electronic} & 215 \\
        GeV-E & 1120~\cite{GS_GeV_ekimov2015germanium} & 987 \\
        SnV-E & 3000~\cite{GS_SnV_iwasaki2017tin} & 2897 \\
        PbV-E &  no data  &  7051 \\
        \bottomrule
    \end{tabular}
    \caption{Lowest splittings in the ground and first-excited manifolds for all the XV species measured in experiments ($\Delta_{0, \rm X, mea}^{g/e}$) along with the ones calculated by Thiering \textit{et al.} (\(\Delta_{0, \rm X, Thi}^{g/e}\))~\cite{thiering2018ab}. One can derive these values with the Hamiltonian~\eqref{eq:H_0} along with Thiering's DFT simulated parameters as shown in Tab.~\ref{tab:Thiering_para_Ham_fac}. Note that the unit is in GHz, and 1 meV = 242 GHz.}
    \label{tab:lowest_spltting}
\end{table}

\subsection{Strain Susceptibilities}

Meesala \textit{et al.} have claimed measuring the quenched SiV strain susceptibility parameters $d$, $f$ for both the ground and first-excited manifolds, along with two other parameters $t_{\parallel, e}-t_{\parallel, g}$ and $t_{\perp, e}-t_{\perp, g}$, the differences in parameter $t_{\parallel}$ and $t_{\perp}$ between the first-excited and ground manifolds~\cite{Meesala2018Strain}. In the data analysis section, Meesala \textit{et at} haven't taken the Jahn-Teller effect into account. However, the measured values of $d$ and $f$ can be interpreted as the quenched parameters $p'd$, $p'f$ based on the refined definition of the quench factors as introduced in Sec.~\ref{sec:refined_def} while the values of $t_{\parallel, e}-t_{\parallel, g}$ and $t_{\perp, e}-t_{\perp, g}$ do not experience any quenching effects hence being measured faithfully. The method of deriving the strain susceptibility from measuring the splitting under a controlled strain is also present in Sec.~\ref{sec:predictions}.

On the other hand, Guo \textit{et al.} have calculated the original (un-quenched) SnV strain susceptibility parameters $d$ and $f$ with DFT methods using both PBE and SCAN functionals~\cite{strain_guo2023microwave}.

In the meantime, based on the relationships [Eq.~\eqref{eq:d_F_relation_sum}] between the Jahn-Teller energy and the strain susceptibilities we found in this article, we are allowed to convert Thiering's DFT calculation results of the Jahn-Teller energies and phonon energies into the strain susceptibilities and compare them with Meesala's and Guo's results. We summarize these three data sets in Tab.~\ref{tab:strain_comp} and Tab.~\ref{tab:strain_comp_2}.

\begin{table}[]
    \centering
    \begin{tabular}{|c|c|c|c|c||c|c|c|}
    \hline
        \multicolumn{2}{|c|}{} & \multicolumn{3}{c||}{Ground Manifold} & \multicolumn{3}{c|}{first-excited Manifold} \\
        \hline
        & data set &  \(d\) & \(f\) & \(d/f\) & \(d\) & \(f\) & \(d/f\) \\
      \hline
      \multirow{2}{*}{SiV} & T-conv & 0.30 & 1.07 & 0.28 & 0.36 & 1.26 & 0.28 \\
      & M-res & 1.3/\(q_{\rm g}'\) & -1.7/\(q_{\rm g}'\) & -0.76 & 1.8/\(q_{\rm e}'\) & -3.4/\(q_{\rm e}'\) & -0.53 \\
      \hline
      \multirow{3}{*}{SnV} & T-conv & 0.20 & 0.71 & 0.28 & 0.38 & 1.33 & 0.28 \\
      & G-PBE & 1.57 & -1.12 & -1.40 & 1.91 & -5.11 & -0.37 \\
      & G-SCAN & 1.67 & -1.12 & -1.49 & 1.84 & -5.18 & -0.35 \\
      \hline
    \end{tabular}
    \caption{Strain susceptibility parameters from 1) T-conv: Thiering's DFT calculation for the SiV and SnV system parameters~\cite{thiering2018ab} (shown in Tab.~\ref{tab:Thiering_para_Ham_fac}) converted with relations~\eqref{eq:Jahn-Teller_ene_sum} and~\eqref{eq:d_F_relation_sum}; 2) M-res: restored parameters based on the Meesala's experimental measurement for the quenched (See Sec.~\ref{sec:quench_f}) SiV strain susceptibility~\cite{Meesala2018Strain}; Based on Thiering's DFT calculation as shown in Tab.~\ref{tab:quench_epsilon}, the quench factors $q_{\rm g}'=0.65$, $q_{\rm e}'=0.56$; 3) G-PBE/G-SCAN: Guo's strain susceptibility DFT calculation for the SnV strain susceptibilities with the PBE/SCAN functionals~\cite{strain_guo2023microwave}. The unit for the strain susceptibility parameters $d$ and $f$ is PHz/strain.}
    \label{tab:strain_comp}
\end{table}

\begin{table}[]
    \centering
    \begin{tabular}{|c|c|c|c|}
        \hline
        & data set &  $t_{\parallel,\rm e}-t_{\parallel,\rm g}$ & $t_{\perp,\rm e}-t_{\perp,\rm g}$  \\
      \hline
      \multirow{2}{*}{SiV} & T-conv & -0.16 & -0.11  \\
      & M-res & -1.7 & 0.078 \\
      \hline
    \end{tabular}
    \caption{Strain susceptibility parameters from 1) T-conv: Thiering's DFT calculation for the SiV system parameters~\cite{thiering2018ab} (shown in Tab.~\ref{tab:Thiering_para_Ham_fac}) converted with relations~\eqref{eq:Jahn-Teller_ene_sum},~\eqref{eq:d_F_relation_sum}, and~\eqref{eq:F0_F_sum}; 2) M-res: Meesala's experimental measurement for the quenched (See Sec.~\ref{sec:quench_f}) SiV strain susceptibility~\cite{Meesala2018Strain}. Unlike $d$ or $f$, no quench factors are associated with $t_{\parallel}$ and $t_{\perp}$; hence, the restoration to the original values is exempted. The unit for the strain susceptibility parameters $t_{\parallel}$ and $t_{\perp}$ is PHz/strain.}
    \label{tab:strain_comp_2}
\end{table}

From Tab.~\ref{tab:strain_comp} and Tab.~\ref{tab:strain_comp_2}, we see that the results obtained by Meesala and Guo exhibit some agreement in both signs and the order of magnitudes, even though the XV species involved are distinct. Nonetheless, the parameters converted from Thiering's calculation possess discrepancies with Meesala's and Guo's results in both signs and magnitude. Additional research endeavors are required to address this conflict.

If Thiering's results and the connections between the Jahn-Teller energy and the strain susceptibilities outlined in this article are accurate, one possible explanation for this discrepancy lies in the intrinsic strain of the diamond sample. To explain this, when there is intrinsic strain $\gamma_0$ in the XV sample, the original ground splitting of the XV system is given by
\begin{equation}
    \Delta_{0, \rm s}=\sqrt{\lambda^2+4\gamma_0^2}\;.
\end{equation}
The artificial stress, when in certain directions, may counteract the intrinsic strain $\gamma_0$, resulting in diminishing the ground splitting and giving rise to opposite signs in the strain susceptibilities. Given that Meesala \textit{et al.} conducted measurements on a single sample in their study, this possibility is not excluded. Furthermore, in section~\ref{sec:predictions}, we propose a method to evaluate the sample's inherent strain, potentially helping to eliminate its effect.

\section{Intrinsic strain measurement}
\label{sec:predictions}

Based on the refined definition of the quench factors, the Hamiltonian of an XV system under strain is given by
\begin{equation}
    H_{\rm S} = \frac{\Delta_0}{2} \tau_z^{\rm m}\sigma_z^{\rm s}+\gamma \tau_{\perp}^{\rm m} \;,
\end{equation}
where
\begin{equation}
    \gamma=q'\sqrt{\left[d \left(\epsilon_{xx} - \epsilon_{yy}\right) + f\epsilon_{xz}\right]^2+\left(-2d \epsilon_{xy} + f\epsilon_{yz}\right)^2}
\end{equation}
and $\tau_{\perp}^{\rm m}=\cos\theta\tau_{x}^{\rm m}+\sin\theta\tau_{y}^{\rm m}$ with 
\begin{equation}
    \tan\theta=\frac{-2d \epsilon_{xy} + f\epsilon_{yz}}{d \left(\epsilon_{xx} - \epsilon_{yy}\right) + f\epsilon_{xz}}\;.
\end{equation}
This Hamiltonian has four dimensions and is two-fold degenerate. The splitting between the two sets of doublets is given by
\begin{equation}
    \Delta_{\rm S}=\sqrt{\Delta_0^2+4\gamma^2}\;.
\end{equation}
Additionally, when $\gamma \ll \Delta_0$,
\begin{equation}
    \Delta_{\rm S} \approx \Delta_0 + \frac{8\gamma}{\Delta_0}.
    \label{eq:Del_M}
\end{equation}
Based on this relation and varying the magnitude and direction of the applied stress, one may measure the quenched strain susceptibilities of the SiV system, which is what Meesala \textit{et al.} obtained. 

However, there is a drawback to this method. The strain of an XV system can originate from two primary sources: the intrinsic strain, which arises from adjacent lattice imperfections, and the artificial strain resulting from externally applied stress, \textit{e.g.}
\begin{equation}
    \gamma_{x/z}=\gamma_{x/z, 0}+\gamma_{x/z, \mathrm{art}}\,,
\end{equation}
where $\gamma_0$ represents the intrinsic strain and $\gamma_{\rm art}$ the artificial strain. Meesala's calculation integrates the intrinsic strain $\gamma_0$ into the intrinsic splitting $\Delta_0$, hence not being measured. 

Nevertheless, here we propose a method to measure the intrinsic strain by harnessing a magnetic field along the $y$-direction. Based on the refined quench-factor model, the Hamiltonian of an XV system in the presence of a magnetic field in the $y$-direction and strain is given by
\begin{equation}
    H_{\rm SM} = \frac{\Delta_0}{2} \tau_z^{\rm m}\sigma_z^{\rm s} + \mu_{\rm B}B_y \sigma_y^{\rm s}+\gamma \tau_{\perp}^{\rm m}
\end{equation}
where $B_y$ is the strength of the magnetic field in the $y$-direction.

With a magnetic field added, the double degeneration in $H_{\rm S}$ is lifted. The four eigenenergies of $H_{\rm SM}$ are
\begin{equation}
    E_{\rm SM}^{\pm\pm}=\pm\sqrt{\left(\frac{\Delta_0}{2}\right)^2+\left(\mu_{\rm B}B_y \pm \gamma\right)^2}
\end{equation}
Hence, the splitting of those doublets is given by
\begin{equation}
    \Delta_{\rm SM}=\sqrt{\left(\frac{\Delta_0}{2}\right)^2+\left(\mu_{\rm B}B_y + \gamma\right)^2}-\sqrt{\left(\frac{\Delta_0}{2}\right)^2+\left(\mu_{\rm B}B_y - \gamma\right)^2}
\end{equation}
when $\gamma \ll \mu_{\rm B}B_y$,
\begin{equation}
    \Delta_{\rm SM}\approx \frac{4\gamma}{\sqrt{\Delta_0^2+4\left(\mu_{\rm B}B_y\right)^2}}\;.
    \label{eq:Del_MS}
\end{equation}

When no artificial strain is applied, $\gamma_{\rm art}=0$, the splitting $\Delta_{\rm SM}$ is contributed solely from the intrinsic strain $\gamma_0$, hence it can be measured.

\section{Conclusion}

We refine the theoretical framework upon previous endeavors of the XV center quantum system. Specifically, we give thorough explanations for the origin of the spin-orbit and electron-phonon interactions, and the mechanism of how the strain, electric field, and magnetic field influence the potential energy of the XV system. From those explorations, the readers may understand why the XV centers possess properties such as large ground-state splitting, robust light coupling without phonon excitation, strain tunability, resilience to charge noise, and the ability to be manipulated via magnetic fields. Original insights of this article include
\begin{itemize}
    \item only two photonic modes dominantly couple to the XV orbital states,
    \item a novel symmetry, the joint-reflection symmetry, that explains the system degeneracy under a magnetic field perpendicular to the principal symmetry axis,
    \item the relationship between the electron-displacement coupling strength and the strain susceptibility,
    \item the explanation of the stark shift,
    \item the explanation of the branching ratio of light emission intensity,
    \item the examination of the quantum state addressability with the aid of the magnetic field,
    \item an improved quench-factor model with higher precision for large spin-orbit coupling strengths, and
    \item a potential proposal to measure the intrinsic strain with the aid of a magnetic field.
\end{itemize}
Besides, this article compares the existing theoretical and experimental studies, pointing out the discrepancies among them and suggesting the potential reasons for the conflicts. In summary, our studies deepen the understanding of the XV center systems and provide useful guidance to manipulate them, expanding the set of tools available for quantum information processing.

\backmatter

\bmhead{Acknowledgements}

Parts of the research presented in Secs.~\ref{sec:interaction_diag}, \ref{sec:num_diag}, and~\ref{sec:light_emission} were conceptualized and supervised by Dr.~V.~V.~Dobrovitski and Dr.~J.~Borregaard, who also participated in the development of the methodology, analysis, investigation, and validation of the results. 
The author thanks Enli Chen for providing the proof of the lemma in Sec.~\ref{sec:degeneracy}. 
The author gratefully acknowledges that this work was partially supported by the collaboration between Fujitsu~Limited and Delft University of Technology, co-funded by the Netherlands Enterprise Agency under project number PPS2007. 
The author also acknowledges funding from the NWO Gravitation Program Quantum Software Consortium (Project QSC No.~024.003.037).

\noindent

\begin{appendices}

\section[List of Used Approximations]{List of Used Approximations} 
\label{app:list_of_approx}

\subsection{Approximations Used in the Electron-phonon Interaction Modeling}
\label{sec:approx_phonon}

\begin{itemize}
    \item \textbf{Localized Phonon Approximation}: We neglect the mutual coupling between the seven nuclei within the XV system and their coupling to the remaining nuclei in the diamond bulk, resulting in seven localized independent nuclear vibrations. This approximation is based on the significant difference between the phonon frequencies of the XV system and those of the bulk, which minimizes the coupling.
    
    \item \textbf{Imaginary Bonding Structures}: The imaginary bonding structures as shown in Fig.~\ref{fig:X-C-X} are simple models used to approximate the potential that a nucleus experiences in a certain direction. Their applicability is based on the similarity of the potential stiffness in certain directions.

    \item \textbf{X-C Bond Length}: The X-C bond lengths listed in Tab.~\ref{tab:bond_data} are based on measurements or DFT calculations for molecules or perfect crystals. Those values may differ from those in the XV color center as a defect in the diamond.
    
    \item \textbf{Potential-Width Parameters}: The potential-width parameters for the X-C bonds, $\alpha_{\rm X}$, are estimated using a guess formula (Eq.~\ref{eq:stiff_guess}). This approximation assumes that the width parameter $\alpha$ for each X-C bond and C-C bond scales proportionally with its bond dissociation energy $E_{\rm 0}$ and inversely with its bond length $d_{\rm 0}$.

    \item \textbf{Nearest Neighbor Interactions}: We neglect the interaction of the $i$-th X-C bond orbital with atomic cores other than the $i$-th carbon atomic core. This approximation is justified by the larger distances between the $i$-th X-C bond and the other atomic cores, as well as the shielding effect provided by the electronic clouds between them.
    
    \item \textbf{Point-Charge Model}: This model approximates the middle portion of the spatially extended X--C bond orbital and the carbon atomic core as point charges with effective charges and positions. The validity of this approximation relies on the fact that the middle portion of the X--C bond orbital is distant enough from both the X and carbon nuclei. Here, the middle portion refers to the portion of the hole population that is not stationary with respect to its nearest nucleus.

    \item \textbf{Small nuclear vibration amplitudes} This approximation is based on the calculation presented in Sec.~\ref{sec:nuclear_vib}, where it shows that the typical value of the nuclear vibration amplitudes is more than 10 times smaller than the inverse of the potential-width parameter $\alpha$ and the bond length. This approximation is the basis for approximating the movement of the nuclei as harmonic oscillators. And, combined with the adaptability of the hole orbital, it is also the basis to neglect the effect caused by the nuclear displacement in the \(x_i\) and \(y_i\) directions because the net effect is approximately a rotation of the X--C bond.

    \item \textbf{Linear Coulomb Coupling}: We assume that the Coulomb energy is proportional to the atomic displacement, with the proportionality factor given by the first-order derivative of the Coulomb energy, evaluated at the equilibrium nuclear configuration, with respect to the displacements of nuclei. The validity of this approximation relies on the "Small nuclear vibration amplitudes" approximation.

    \item \textbf{Isolated X--C bond}: When determining the effective position of the middle portion of the hole cloud, as approximated as a point charge, we assume the X--C bond orbital takes the shape as in an isolated X--C bond, neglecting the influence on the shape of the orbital from the other five X--C bonds in the XV system. Specifically, the presence (absence) of a node in the location of the X nucleus when the XV system is in its ground (first-excited) manifold changes the shape of the X--C bond orbital, hence the effective position of the middle portion of the hole cloud. Consequently, our model gives orders-of-magnitude estimations of the electron-phonon coupling strength for both the ground and first-excited manifolds without specifying the difference between the two cases.
\end{itemize}

\subsection{Approximations Used in the Spin-Orbit Interaction Modeling}
\label{sec:list_approx_soc}

\begin{itemize}
    \item \textbf{Isolated X atom}: We assume an isolated X atom when calculating the spin-orbit coupling strength of the hole in the XV system. This approximation is valid because the majority of the hole population is confined within the region enclosed by the six carbon nuclei of the XV system. Adopting the classical picture, the hole orbits around only the X nucleus and its inner-shell electrons. The electric fields generated by the charge outside the hole's trajectory have no divergence within the area enclosed by the hole's trajectory, hence contribute zero net spin-orbit coupling strength. 

    \item \textbf{Neglected contribution from the carbon \(sp^3\) orbitals}: The hole is distributed across the six X--C bonds, where each bond involves an \(sp^3d^2\) orbital on the X atom and an \(sp^3\) orbital on the carbon atom. However, we neglect the contribution of the spin-orbit coupling strength from the population located in the \(sp^3\) orbitals for two reasons: 1) the spin-orbit coupling strength is reciprocal cubic decay in distance from the X nucleus and (2) the population in the \(sp^3\) orbitals suffers a more substantial screening effect on the electric field of the X nucleus than that in the \(sp^3d^2\) orbitals. Regarding the ratio between the population in the \(sp^3d^2\) and \(sp^3\) orbitals, it is smaller in the ground manifold than in the first-excited manifold because of the presence and absence of the node in the electronic orbital eigenstates, respectively. However, we assume this ratio to be one for both cases as a rough estimation.

\end{itemize}

\section[Deduction of the Matrix Elements of the displacement operator]{Deduction of the Matrix Elements of the displacement operator\raisebox{.3\baselineskip}{\normalsize\footnotemark}} 
\label{app:matrix_ele}
\footnotetext{The method in this appendix is based on Proc. R. Soc. Lond. A 1958 244, 1-16\cite{longuet1958studies}.}

This appendix aims to project the nuclear displacement operators \(\hat{Q}_x\) and \(\hat{Q}_y\) onto the basis set \(\{|nm\rangle\}\), as introduced in Sec.~\ref{sec:interaction_diag}. 

\textbf{To improve the readability, we replace the nuclear displacement operators \(\hat{Q}_x\) and \(\hat{Q}_y\) with simpler notations \(\hat{x}\) and \(\hat{y}\) in the following. Please don't mix them with the hole orbital displacements in the main text.}

To achieve the goal, we need to calculate the matrix elements \(\langle n'm' |x| nm\rangle\) and \(\langle n'm' |y| nm\rangle\). Instead of performing cumbersome integrals with the analytical expressions of \(|nm\rangle\), we adopt an algebraic approach, deducing the result from three fundamental facts:
\begin{gather}
    \langle nm|H_{\textrm{v}}|nm\rangle = (n+1)\hbar\omega, \label{Hv} \\
    \langle nm|\frac{\mu}{2}\omega^{2}\left(x^{2}+y^{2}\right)|nm\rangle = \frac{1}{2}(n+1)\hbar\omega, \label{halfEnrg} \\
    \langle nm|L^\mathrm{v}_{z}|nm\rangle = \hbar m. \label{Lz}
\end{gather}

In Eq.~(\ref{halfEnrg}), the potential energy of the harmonic oscillator is known to account for half of the total energy. At first glance, the connection between these equations and the desired matrix elements might seem indirect. However, by carefully applying logical reasoning, we bridge this gap and derive the desired results. Let the logic do the magic.

First, we aim to identify which matrix elements are non-zero. To achieve this, we begin by examining a seemingly unrelated calculation:

By evaluating the commutator \(\left[x, H_{\textrm{v}}\right]\), we derive the following relation:
\begin{equation}
    H_{\textrm{v}}^{2}x - 2H_{\textrm{v}}xH_{\textrm{v}} + xH_{\textrm{v}}^{2} = \left(\hbar\omega\right)^{2}x \;.
\end{equation}
If we replace \(x\) in the equation above with 
\begin{equation}
    Q_\pm \coloneqq x \pm \mathrm{i}y \;,
\end{equation}
the relation remains valid. By sandwiching this relation with \(\langle n'm'|\) and \(|nm\rangle\) and defining 
\begin{equation}
    Q_{\pm \:nm}^{\ \ n'm'} \coloneqq \langle n'm'|Q_\pm|nm\rangle \;,
\end{equation}
we arrive at:
\begin{equation}
    Q_{\pm \:nm}^{\ \ n'm'} \left(n' - n\right)^2 = Q_{\pm \:nm}^{\ \ n'm'} \;. 
\end{equation}
Since this relation must always hold, it implies that for \(Q_{\pm \:nm}^{\ \ n'm'}\) to be non-zero, the following condition must be satisfied:
\begin{equation}
    n' = n \pm 1 \;.
\end{equation}
This result indicates that we only need to compute two types of matrix elements: \(Q_{\pm \:nm}^{\ \ n+1\;m'}\) and \(Q_{\pm \:nm}^{\ \ n-1\;m'}\).

Next, by calculating the commutators \([x, L_z]\) and \([y, L_z]\), we derive the following relation:
\begin{equation}
    L^\mathrm{v}_{z} Q_\pm = Q_\pm \left(L^\mathrm{v}_{z} \pm \hbar\right) \;.
\end{equation}
Sandwiching this relation with \(\langle n'm'|\) and \(|nm\rangle\) and applying similar reasoning as above, we obtain:
\begin{equation}
    \begin{cases}
        m' = m + 1 & \text{for } Q_+, \\
        m' = m - 1 & \text{for } Q_- \;.
    \end{cases}
\end{equation}

In summary, only the following four types of matrix elements can potentially be non-zero:
\begin{equation}
    Q_{+ \:nm}^{\ \ n\pm1\;m+1} \quad \text{and} \quad Q_{- \:nm}^{\ \ n\pm1\;m-1} \;.
\end{equation}

Furthermore, noting that, based on the definition of the sandwiching operation, 
\begin{equation}
    Q_{+ \:nm}^{\ \ n\pm1\;m+1} = Q_{- \:n\pm1\;m+1}^{* \ nm},
\end{equation}
only two of the elements, \(Q_{+ \:nm}^{\ \ n\pm1\;m+1}\), need to be calculated. In fact, we can write:
\begin{equation}
    Q_{+ \:nm}^{\ \ n\pm1\;m+1} = Q_{- \:n\pm1\;m+1}^{\ \ nm},
\end{equation}
because, based on the analytical expression of the eigenwavefunctions of \(|nm\rangle\)~\cite{pauling2012introduction}, the matrix elements are all real. This property will be used again in subsequent derivations.

To proceed, we utilize the relation \(x^2 + y^2 = Q_- Q_+\). Sandwiching this relation with \(\langle nm|\) and \(|nm\rangle\), we find:
\begin{align}
    \langle nm|x^{2}+y^{2}|nm\rangle &= \sum_{n'm'}\langle nm|Q_-|n'm'\rangle \langle n'm'|Q_+|nm\rangle \notag \\
    &= Q_{- \:n+1\;m+1}^{\ \ nm} Q_{+ \:nm}^{\ \ n+1\;m+1} + Q_{- \:n-1\;m+1}^{\ \ nm} Q_{+ \:nm}^{\ \ n-1\;m+1} \notag \\
    &= Q_{+ \:nm }^{* \ n+1\;m+1} Q_{+ \:nm}^{\ \ n+1\;m+1} + Q_{+ \:nm }^{* \ n-1\;m+1} Q_{+ \:nm}^{\ \ n-1\;m+1} \notag \\
    &= \left(Q_{+ \:nm }^{\ \ n+1\;m+1}\right)^2 + \left(Q_{+ \:nm }^{\ \ n-1\;m+1}\right)^2.
\end{align}

Recalling Eq.~(\ref{halfEnrg}), we derive the first key relation for determining \(Q_{+ \:nm}^{\ \ n\pm1\;m+1}\):
\begin{equation} \label{plus}
    \left(Q_{+ \:nm }^{\ \ n+1\;m+1}\right)^2 + \left(Q_{+ \:nm }^{\ \ n-1\;m+1}\right)^2 = \frac{\hbar}{\mu\omega}\left(n+1\right).
\end{equation}

Next, we derive a second relation by examining the following:
\begin{equation}
    -\left(x^{2}+y^{2}\right)H_{\textrm{v}} + 2\left(x-\mathrm{i}y\right)H_{\textrm{v}}\left(x+\mathrm{i}y\right) - H_{\textrm{v}}\left(x^{2}+y^{2}\right) = 2\frac{\hbar}{\mu}\left[L^\mathrm{v}_{z} + \hbar\right],
\end{equation}
which can be proven using the commutators employed earlier. The first few essential steps are:
\begin{align}
    \text{Left side} &= (x-\mathrm{i}y)\left[H_{\textrm{v}}(x+\mathrm{i}y) - (x+\mathrm{i}y)H_{\textrm{v}}\right] \notag \\
    &\quad - \left[H_{\textrm{v}}(x-\mathrm{i}y) - (x-\mathrm{i}y)H_{\textrm{v}}\right](x+\mathrm{i}y) \notag \\
    &= (x-\mathrm{i}y)\frac{\hbar^{2}}{\mu}\left(-\frac{\partial}{\partial x} - \mathrm{i}\frac{\partial}{\partial y}\right) + \frac{\hbar^{2}}{\mu}\left(\frac{\partial}{\partial x} - \mathrm{i}\frac{\partial}{\partial y}\right)(x+\mathrm{i}y). \notag
\end{align}

Sandwiching the equation with \(\langle nm|\) and \(|nm\rangle\), we obtain:
\begin{equation} \label{minus}
    \left(Q_{+ \:nm }^{\ \ n+1\;m+1}\right)^2 - \left(Q_{+ \:nm }^{\ \ n-1\;m+1}\right)^2 = \frac{\hbar}{\mu\omega}(m+1).
\end{equation}

Combining Eqs.~(\ref{plus}) and (\ref{minus}), we solve for the matrix elements:
\begin{gather}
    Q_{+ \:nm }^{\ \ n+1\;m+1} = \sqrt{\frac{\hbar}{2\mu\omega}}\sqrt{n+m+2}, \\
    Q_{+ \:nm }^{\ \ n-1\;m+1} = \sqrt{\frac{\hbar}{2\mu\omega}}\sqrt{n-m}.
\end{gather}

For clarity, \(Q_+\) and \(Q_-\) can be expressed as the sum of submatrices with the bases explicitly denoted:
\begin{equation}
    Q_+ = \sqrt{\frac{\hbar}{2\mu\omega}} \sum_{nm} \ 
    \bordermatrix{
        ~                 & \langle nm | \cr
        \ \  |n-1 \  m-1 \rangle & 0 \cr
        \ \  |n-1 \  m+1 \rangle & \sqrt{n-m} \cr
        \ \  |n+1 \  m-1 \rangle & 0 \cr
        \ \  |n+1 \  m+1 \rangle & \sqrt{n+m+2} \cr
    }
\end{equation}
\begin{equation}
    Q_- = \sqrt{\frac{\hbar}{2\mu\omega}} \sum_{nm} \ 
    \bordermatrix{
        ~                 & \langle nm | \cr
        \ \  |n-1 \  m-1 \rangle & \sqrt{n+m} \cr
        \ \  |n-1 \  m+1 \rangle & 0 \cr
        \ \  |n+1 \  m-1 \rangle & \sqrt{n-m+2} \cr
        \ \  |n+1 \  m+1 \rangle & 0 \cr
    }
\end{equation}

Note that for cases where \(m = \pm n\), \(|nm\rangle\) may couple to undefined states, such as \(|n-1 \ n+1\rangle\) or \(|n-1 \ -n-1\rangle\). However, since the corresponding matrix elements are zero, this inclusion does not cause confusion. 

Returning to the original question of expressing \(x\) and \(y\) in matrix form, the operators \(x\) and \(y\) are given by \(x = \frac{1}{2}\left(Q_+ + Q_-\right)\) and \(y = \frac{1}{2\mathrm{i}}\left(Q_+ - Q_-\right)\). Thus, we can write:
\begin{equation}
    x = \frac{1}{2}\sqrt{\frac{\hbar}{2\mu\omega}} \sum_{nm} \ 
    \bordermatrix{
        ~                 & \langle nm | \cr
        \ \  |n-1 \  m-1 \rangle & \sqrt{n+m} \cr
        \ \  |n-1 \  m+1 \rangle & \sqrt{n-m} \cr
        \ \  |n+1 \  m-1 \rangle & \sqrt{n-m+2} \cr
        \ \  |n+1 \  m+1 \rangle & \sqrt{n+m+2} \cr
    }
\end{equation}
\begin{equation}
    y = \frac{1}{2\mathrm{i}}\sqrt{\frac{\hbar}{2\mu\omega}} \sum_{nm} \ 
    \bordermatrix{
        ~                 & \langle nm | \cr
        \ \  |n-1 \  m-1 \rangle & -\sqrt{n+m} \cr
        \ \  |n-1 \  m+1 \rangle & \sqrt{n-m} \cr
        \ \  |n+1 \  m-1 \rangle & -\sqrt{n-m+2} \cr
        \ \  |n+1 \  m+1 \rangle & \sqrt{n+m+2} \cr
    }
\end{equation}

For the second-order Jahn-Teller effect, we need to compute \(x^2-y^2\) and \(2xy\). The results are as follows:
\begin{equation}
    x^2 - y^2 = \frac{\hbar}{4\mu\omega} \sum_{nm} \ 
    \bordermatrix{
        ~                 & \langle nm | \cr
        \ \  |n-2 \  m-2 \rangle & \sqrt{n+m-2} \sqrt{n+m} \cr
        \ \  |n-2 \  m+2 \rangle & \sqrt{n-m-2} \sqrt{n-m} \cr
        \ \  |n \  m-2 \rangle   & 2\sqrt{n-m+2} \sqrt{n+m} \cr
        \ \  |n \  m+2 \rangle   & 2\sqrt{n+m+2} \sqrt{n-m} \cr
        \ \  |n+2 \  m-2 \rangle & \sqrt{n-m+4} \sqrt{n-m+2} \cr
        \ \  |n+2 \  m+2 \rangle & \sqrt{n+m+4} \sqrt{n+m+2} \cr
    }
\end{equation}
\begin{equation}
    2xy = -i\frac{\hbar}{4\mu\omega} \sum_{nm} \ 
    \bordermatrix{
        ~                 & \langle nm | \cr
        \ \  |n-2 \  m-2 \rangle & \sqrt{n+m-2} \sqrt{n+m} \cr
        \ \  |n-2 \  m+2 \rangle & -\sqrt{n-m-2} \sqrt{n-m} \cr
        \ \  |n \  m-2 \rangle   & 2\sqrt{n-m+2} \sqrt{n+m} \cr
        \ \  |n \  m+2 \rangle   & -2\sqrt{n+m+2} \sqrt{n-m} \cr
        \ \  |n+2 \  m-2 \rangle & \sqrt{n-m+4} \sqrt{n-m+2} \cr
        \ \  |n+2 \  m+2 \rangle & -\sqrt{n+m+4} \sqrt{n+m+2} \cr
    }
\end{equation}

For the Jahn-Teller operators \(Q_{\pm\pm} \coloneqq \frac{1}{\sqrt{2}}\left[\left(x^2 - y^2\right) \mp \mathrm{i}2xy\right]\), the expressions are:
\begin{equation}
    Q_{++} = \frac{\hbar}{\sqrt{8}\mu\omega} \sum_{nm} \ 
    \bordermatrix{
        ~                 & \langle nm | \cr
        \ \  |n-2 \  m-2 \rangle & 0 \cr
        \ \  |n-2 \  m+2 \rangle & \sqrt{n-m-2} \sqrt{n-m} \cr
        \ \  |n \  m-2 \rangle   & 0 \cr
        \ \  |n \  m+2 \rangle   & 2\sqrt{n+m+2} \sqrt{n-m} \cr
        \ \  |n+2 \  m-2 \rangle & 0 \cr
        \ \  |n+2 \  m+2 \rangle & \sqrt{n+m+4} \sqrt{n+m+2} \cr
    }
\end{equation}
\begin{equation}
    Q_{--} = \frac{\hbar}{\sqrt{8}\mu\omega} \sum_{nm} \ 
    \bordermatrix{
        ~                 & \langle nm | \cr
        \ \  |n-2 \  m-2 \rangle & \sqrt{n+m-2} \sqrt{n+m} \cr
        \ \  |n-2 \  m+2 \rangle & 0 \cr
        \ \  |n \  m-2 \rangle   & 2\sqrt{n-m+2} \sqrt{n+m} \cr
        \ \  |n \  m+2 \rangle   & 0 \cr
        \ \  |n+2 \  m-2 \rangle & \sqrt{n-m+4} \sqrt{n-m+2} \cr
        \ \  |n+2 \  m+2 \rangle & 0 \cr
    }
\end{equation}

With these results, all matrix elements required for this thesis have been derived.

\end{appendices}


\bibliography{sn-bibliography}

\end{document}